\documentclass[12pt,a4paper]{article}

\usepackage{amsmath,amssymb,amsfonts,bm,graphicx,subfiles,xspace}

\newlength{\abstwidth}
\renewcommand{\b}{\mathrm{b}}
\renewcommand{\c}{\mathrm{c}}
\renewcommand{\d}{\mathrm{d}}
\newcommand{\e}{\mathrm{e}}
\newcommand{\hrm}{\mathrm{h}}
\newcommand{\n}{\mathrm{n}}
\newcommand{\p}{\mathrm{p}}
\newcommand{\q}{\mathrm{q}}
\newcommand{\s}{\mathrm{s}}
\renewcommand{\u}{\mathrm{u}}
\newcommand{\A}{\mathrm{A}}
\newcommand{\B}{\mathrm{B}}
\newcommand{\D}{\mathrm{D}}
\renewcommand{\H}{\mathrm{H}}
\newcommand{\J}{\mathrm{J}}
\newcommand{\K}{\mathrm{K}}
\newcommand{\N}{\mathrm{N}}

\newcommand{\cbar}{\overline{\mathrm{c}}}
\newcommand{\dbar}{\overline{\mathrm{d}}}

\newcommand{\pbar}{\overline{\mathrm{p}}}
\newcommand{\qbar}{\overline{\mathrm{q}}}

\newcommand{\ubar}{\overline{\mathrm{u}}}
\newcommand{\Kbar}{\overline{\mathrm{K}}}
\newcommand{\Pb}{\mathrm{Pb}}
\newcommand{\PbPb}{\mathrm{PbPb}}
\newcommand{\XeXe}{\mathrm{XeXe}}

\newcommand{\pT}{p_{\perp}}
\newcommand{\pTo}{p_{\perp 0}}
\newcommand{\Pom}{\mathbb{P}}
\newcommand{\CM}{\mathrm{CM}}

\renewcommand{\eqref}[1]{eq.~(\ref{#1})}
\newcommand{\figref}[1]{Figure~\ref{#1}}
\newcommand{\secref}[1]{Section~\ref{#1}}
\newcommand{\tabref}[1]{Table~\ref{#1}}

\newcommand{\eg}{\textit{eg.}\xspace}
\newcommand{\ie}{\textit{ie.}\xspace}

\begin{document}
\sloppy
 
\pagestyle{empty}
 
\begin{flushright}
LU TP 21--08\\
MCnet--21--03\\
March 2021
\end{flushright}

\vspace{\fill}

\begin{center}
{\Huge\bf Hadronic Rescattering}\\[4mm]
{\Huge\bf in pA and AA Collisions}\\[10mm]
{\Large Christian Bierlich, Torbj\"orn Sj\"ostrand and Marius Utheim} \\[3mm]
{\it Theoretical Particle Physics,}\\[1mm]
{\it Department of Astronomy and Theoretical Physics,}\\[1mm]
{\it Lund University,}\\[1mm]
{\it S\"olvegatan 14A,}\\[1mm]
{\it SE-223 62 Lund, Sweden}
\end{center}

\vspace{\fill}

\begin{center}
\begin{minipage}{\abstwidth}
{\bf Abstract}\\[2mm]
In a recent article we presented a model for hadronic rescattering,
and some results were shown for $\p\p$ collisions at LHC energies.
In order to extend the studies to $\p\A$ and $\A\A$ collisions, the
\textsc{Angantyr} model for heavy-ion collisions is taken as the
starting point.  Both these models are implemented within the
general-purpose Monte Carlo event generator \textsc{Pythia}, which makes the 
matching reasonably straightforward, and allows for detailed studies
of the full space--time evolution. The rescattering rate is significantly
higher than in $\p\p$, especially for central $\A\A$ collisions,
where the typical primary hadron rescatters several times. We study
the impact of rescattering on a number of distributions, such as
$\pT$ and $\eta$ spectra, and the space--time evolution of the
whole collision process. Notably rescattering is shown to give a
significant contribution to elliptic flow in $\XeXe$ and $\PbPb$, and to give  
a nontrivial impact on charm production.

\end{minipage}
\end{center}

\vspace{\fill}

\phantom{dummy}

\clearpage

\pagestyle{plain}
\setcounter{page}{1}


\section{Introduction}

Heavy-ion experiments at RHIC and LHC have produced convincing
evidence that a Quark-Gluon Plasma (QGP) is formed in high-energy 
nucleus-nucleus ($\A\A$) collisions. The discussion therefore has developed
into one of understanding the underlying detailed mechanisms, such
as the nature of the initial state, the early thermalization,  
the subsequent hydrodynamical expansion, and the transition back to
a hadronic state. Numerous models have been and are being developed
to study such issues.

The standard picture of heavy ion collisions, separates the evolution of
the QGP phase into three or four stages, outlined in the following. 

The first $<1$ fm after the collision, is denoted the ``initial state''.
It consists of dense matter, highly out of equilibrium. Most QGP-based
models seek to calculate an energy density (or a full energy-momentum
tensor) from a model of the evolution of the initial stage. The simplest
approaches are based purely on geometry, and are denoted Glauber models \cite{Miller:2007ri}.
Here, the energy density in the transverse plane is determined purely from
the distributions of nucleons in the incoming nuclei. Going beyond nucleonic
degrees of freedom, some of the more popular choices includes either
introducing constituent quarks \cite{Bozek:2016kpf}, or invoking the more involved
formalism known the Colour Glass Condensate \cite{Gelis:2010nm}. In the latter case, the
so-called IP-Glasma \cite{Schenke:2012wb} program is often used, as it allows for computations
with realistic boundary conditions. 

The initial state, glasma or not, will then transition into a plasma. Recently,
progress has been made to describe the transition from an out-of-equilibrium initial state 
to a hydrodynamized plasma, using kinetic theory \cite{Kurkela:2018wud}. In such cases,
the pre-equilibration will describe the dynamics between $\approx 0.1-1$ fm. 

Between $1-10$ fm, the plasma evolves
according to relativistic viscous hydrodynamics \cite{Heinz:2013th,Luzum:2013yya,Gale:2013da}.
Hydrodynamics is a long wavelength effective theory, able to describe interactions at low momentum,
when the mean free path of particles is much smaller than the characteristic size of the system. As such,
its use has been criticised in small collision systems, but nevertheless seems to be able to describe
flow observables reasonably well even there \cite{Weller:2017tsr}.

Finally, after 10 fm, the QGP freezes out to hadronic degrees of freedom. The physics involved 
after this freeze-out is the main topic of this paper, though with the large difference to traditional
approaches, that it happens much sooner. 

Paradoxically, one of the key problems is that the QGP picture has
been too successful. QGP formation was supposed to be unique to 
$\A\A$ collisions, while $\p\A$ and $\p\p$ collisions would not involve
volumes and time scales large enough for it. And yet QGP-like
signals have been found in these as well. One key example is the observation 
of a non-isotropic particle flow, in the form of a ``ridge'' at the same 
azimuthal angle as a trigger jet
\cite{Khachatryan:2010gv,Aad:2015gqa,Khachatryan:2016txc}
or of non-vanishing $v_2$ azimuthal flow coefficients 
\cite{Aad:2015gqa,Khachatryan:2016txc,Acharya:2019vdf}. Another example is
that the fraction of strange hadrons, and notably multi-strange baryons,
is smoothly increasing from low-multiplicity to high-multiplicity
$\p\p$, on through $\p\A$ to saturate for $\A\A$ multiplicities \cite{ALICE:2017jyt}.

The most obvious way out is to relax the large-volume requirement,
and accept that a QGP, or at least a close-to-QGP-like state, can
be created in smaller systems. An excellent example of this approach
is the core--corona model \cite{Werner:2007bf}, implemented in the EPOS event 
generator \cite{Pierog:2013ria}, wherein the high-density core of a system
hadronizes like a plasma, while the outer lower-density corona
does not. The evolution from low-multiplicity $\p\p$ to $\A\A$ is then
a consequence of an increasing core fraction.

Another approach is to ask what physics mechanisms, not normally
modelled in $\p\p$ collisions, would be needed to understand $\p\p$ 
data without invoking QGP formation. And, once having such a model, one
could ask what consequences that would imply for $\p\A$ and $\A\A$ collisions.
More specifically, could some of the signals attributed to QGP
formation have alternative explanations? If nothing else, exploring these
questions could help sharpen experimental tests, by providing a 
straw-man model. At best, we may actually gain new insights.

This is the road taken by the \textsc{Angantyr} model
\cite{Bierlich:2016smv,Bierlich:2018xfw}. It is based on and
contained in the \textsc{Pythia} event generator
\cite{Sjostrand:2006za,Sjostrand:2014zea},
which successfully describes many/most features of LHC $\p\p$ events.
\textsc{Angantyr} adds a framework wherein $\p\A$ and $\A\A$
collisions can be constructed as a superposition of simpler binary
collisions, in the spirit of the old \textsc{Fritiof} model
\cite{Andersson:1986gw,Andersson:1992iq}. Such a framework is already
sufficient to describe many simple $\p\A$ and $\A\A$ distributions,
such as $\d n_{\mathrm{charged}}/\d \eta$. Beyond that, it also offers
a platform on top of which various collective non-QGP phenomena
can be added. One example is shoving
\cite{Bierlich:2016vgw,Bierlich:2017vhg,Bierlich:2020naj}, whereby
closely overlapping colour fields repel each other, to give a
collective flow. Another is colour rope formation
\cite{Bierlich:2014xba}, wherein overlapping colour fields can
combine to give a higher field strength, thus enhancing strangeness
production relative to the no-overlap default. 

In this article we will study a third mechanism, that of hadronic
rescattering. The basic idea here is that the standard fragmentation
process produces a region of closely overlapping hadrons, that then
can collide with each other as the system expands. Each single such
collision on its own will give negligible effects, but if there are
many of them then together they may give rise to visible physics
signals. Rescattering is often used as an afterburner to the
hadronization of the QGP, commonly making use of the UrQMD
\cite{Bass:1998ca} or SMASH \cite{Weil:2016zrk} programs.
What makes \textsc{Angantyr/Pythia} different is that there is no QGP
phase, so that rescattering can start earlier, and therefore hypothetically
can give larger effects. 
In order to use a rescattering framework as an afterburner 
to \textsc{Angantyr}, a first step is to describe the space--time
structure of hadronization in \textsc{Pythia}, which was worked out in 
\cite{Ferreres-Sole:2018vgo}. This picture can easily be extended from
$\p\p$ to $\p\A$ and $\A\A$ using the nuclear geometry set up in
\textsc{Angantyr}. Thereby the road is open to add rescattering \eg with
UrQMD, which was done by Ref.~\cite{daSilva:2020cyn}.

Using two different programs is cumbersome, however. It requires the user
to learn to use each individual framework, and they have to convert the
output from the first program into a format that can be input to the
second. A related issue arises if the two programs represent event records
differently, so that it might be impossible to trace the full particle
history. A desire for convenience is one of the main motivations behind a
recently developed framework for hadronic rescattering, implemented
natively in \textsc{Pythia} \cite{Sjostrand:2020gyg}.
With this framework, rescattering can be enabled with just a single
additional line of code, which is a trivial task for anyone already familiar
with \textsc{Pythia}. In addition, this framework also introduces physics
features not found in some other frameworks, such as a basic model for
charm and bottom hadrons in rescattering, and with \textsc{Pythia} being
in active development, there is a low threshold for making further
improvements in the future.

In \cite{Sjostrand:2020gyg}, initial studies using the framework were limited
to implications for $\p\p$ collisions, which not unexpectedly were found
to be of moderate size. That is, while visible enough in model studies,
generally they are less easy to pin down experimentally, given all
other uncertainties that also exist. In this article the rescattering
studies are extended to $\p\A$ and $\A\A$ collisions, where effects are
expected to be larger.
Indeed, as we shall see, the outcome confirms this expectation.
The number of rescatterings rises faster than the particle multiplicity,
such that the fraction of not-rescattered hadrons is small in
$\PbPb$ collisions. Rescatterings are especially enhanced at lower
masses, but the process composition at a given mass is universal.
Obviously the primary production volume increases from $\p\p$ and
$\p\A$ to $\A\A$, and thus so does the range of rescatterings. 
Transverse momentum spectra are significantly more deformed by
rescattering in $\A\A$. There is a clear centrality dependence on
particle production rates, \eg a $\J/\psi$ depletion in central
collisions. The most interesting result is a clear signal of
elliptic flow induced by rescatterings, that even matches experimental
$\PbPb$ numbers at large multiplicities, to be contrasted with the
miniscule effects in $\p\p$.

The outline of the article is as follows. In \secref{sec:model} we
describe the main points of the model, from the simulation of the
nuclear collision, through the modelling of individual nucleon--nucleon
sub-collisions and on to the rescattering framework proper.
In \secref{sec:modelTests} effects in this model are tested on its own,
while \secref{sec:comparisonWithData} shows comparisons with data. Some
conclusions and an outlook are presented in \secref{sec:summary}. Finally,
technical aspects related to computation time for rescattering are discussed
in the appendix.

Natural units are assumed throughout the article, \ie $c = \hbar = 1$.
Energy, momentum and mass are given in GeV, space and time in fm, and
cross sections in mb.


\section{The model}
\label{sec:model}

In this section we will review the framework used to simulate nuclear
collisions. Initially the \textsc{Angantyr} framework is used to set
the overall nucleus--nucleus ($\A\A$) collision geometry and select
colliding nucleon-nucleon ($\N\N$) pairs. Then the Multiparton Interactions
(MPI) concept is used to model each single $\N\N$ collision. The resulting
strings are fragmented to provide the primary setup of hadrons, that
then can begin to decay and rescatter. All of these components are
described in separate publications, where further details may be found,
so only the key aspect are collected here is to describe how it all
hangs together.

\subsection{ANGANTYR}
\label{subsec:angantyr}

The \textsc{Angantyr} part of the modelling is responsible for
setting up the $\A\A$ collision geometry, and selecting the number and
nature of the ensuing $\N\N$ collisions \cite{Bierlich:2018xfw}.

Take the incoming high-energy nucleons to be travelling along the
$\pm z$ directions. By Lorentz contraction all the $\N\N$ collisions then
occur in a negligibly small range around $t = z = 0$, and the nucleon
transverse $(x, y)$ positions can be considered frozen during that time.
The nucleon locations inside a nucleus are sampled according to a
two-dimensional Woods-Saxon distribution in the GLISSANDO parametrisation 
\cite{Broniowski:2007nz,Rybczynski:2013yba}, applicable for 
heavy nuclei with $A > 16$, and with a nuclear repulsion effect implemented
algorithmically as a ``hard core'' radius of each nucleon, below which two
nucleons cannot overlap. The $\A\A$ collision impact
parameter provides an offset $\pm b_{\A\A}/2$, \eg along the
$x$ axis. Up to this point, this is a fairly standard Glauber model
treatment, where one would then combine the geometry with measured cross
sections (usually total and/or inelastic non-diffractive), to obtain
the amount of participating or wounded nucleons, and the number of binary
sub-collisions (see \eg Ref. \cite{Miller:2007ri} for a review). In 
\textsc{Angantyr}, a distinction between nucleons wounded inelastic non-diffractively,
diffractively or elastically is desired, along with a dependence on the
nucleon-nucleon impact parameter. To this end, a parametrization of the
nucleon-nucleon elastic amplitude in impact parameter space ($T(\vec{b})$) is used.
It allows for the calculation of the amplitude $T_{kl}(\vec{b})$ for any
combination of projectile and target state, $k$ and $l$ respectively. All parameters
of the parametrization can be estimated from proton-proton total and semi-inclusive
cross sections, and varies with collision energy. The input cross sections used
are the ones available in \textsc{Pythia}, with the SaS model \cite{Schuler:1993wr}
being the default choice. The parametrization of $T(\vec{b})$ thus adds no
new parameters beyond the ones already present in the model for hadronic cross
sections.

Inelastic non-diffractive collisions involve a colour exchange between
two nucleons. In the simplest case, where each incoming nucleon undergoes
at most one collision, the traditional \textsc{Pythia} collision
machinery can be used essentially unchanged. The one difference is
that the nuclear geometry has already fixed the $\N\N$ impact parameter
$b_{\N\N}$, whereas normally this would be set only in conjunction
with the hardest MPI.

The big extension of \textsc{Angantyr} is that it also handles 
situations where a given nucleon $A$ interacts inelastic non-diffractively
with several nucleons $B_1, B_2, \ldots B_n$ from the other nucleus.
Colour fields would then be stretched from $A$ to each $B_i$.
It would be rare for all the fields to stretch all the way out to
$A$, however, but rather matching colour--anticolour pairs would
``short-circuit'' most of the colour flow out to the remnants. Such a
mechanism is already used for MPIs in a single $\N\N$ collision, but here it
is extended to the full set of interconnected nucleons. Therefore only
one $A B_i$ collision is handled as a normal $\N\N$ one, while the other
$A B_j, j \neq i$ ones will produce particles over a smaller rapidity
range. This is analogous to the situation encountered in single
diffraction $A B_j \to A X_j$. If we further assume that the
short-circuiting can occur anywhere in rapidity with approximately 
flat probability distribution, this translates into an excited mass
spectrum like $\d M_{X_j}^2 /  M_{X_j}^2$, again analogous to diffraction.
To this end, $n-1$ carrier particles with vacuum quantum numbers $\Pom_j$ 
(denoted $\Pom$ for the similarity with pomerons) are emitted, with 
fractions $x_j$ of the incoming $A$ (lightcone) momentum picked
according to $\d x_j / x_j$, subject to momentum 
conservation constraints, with a leftover $x_i$ that usually should 
represent the bulk of the $A$ momentum. Thereby the complexity of the 
full problem is reduced to one of describing one regular $A B_i$ collision, 
at a slightly reduced energy, and $n - 1$ $\Pom_j B_j$ collisions, at
significantly reduced energies, similar to diffraction. 
The pomeron-like objects have no net colour or flavour, but they 
do contain partons and the full MPI machinery can be applied to describe also 
these collisions. As the particles are not true pomerons, the PDFs can be different 
from the pomeron ones measured at HERA,
and the transverse size is that of the original nucleon rather than
the smaller one expected for a pomeron. 

In a further step of complexity, the nucleons on side $A$ and $B$
may be involved in multiply interrelated chains of interactions.
Generalizing the principles above, it is possible to reduce even
complex topologies to a set of decoupled $\N\N$, $\N\Pom$, and $\Pom\Pom$
collisions, to be described below. The reduction is not unique, but
may be chosen randomly among the allowed possibilities.

One current limitation is that there is no description of the breakup
of the nuclear remnant. Rather, all non-wounded nucleons of a nucleus
are collected together into a single fictitious new nucleus, that
is not considered any further.

\subsection{Multiparton interaction vertices}
\label{subsec:MPIvertices}

At the end of the \textsc{Angantyr} modelling, a set of separate
hadron--hadron ($\H\H$) interactions have been defined inside an $\A\A$
collision, where the hadron can be either a nucleon or a pomeron-like
object as discussed above. The locations of the $\H\H$ collisions in the
transverse plane is also fixed. 

When two Lorentz-contracted hadrons collide inelastically with each
other, a number of separate (semi-)perturbative parton--parton
interactions can occur. These are modelled in a sequence of falling
transverse momenta $\pT$, as described in detail elsewhere
\cite{Sjostrand:1987su,Sjostrand:2017cdm}. The MPI vertices are spread
over a transverse region of hadronic size, but in the past it was not
necessary to assign an explicit location for every single MPI.
Now it is. The probability for an interaction at a given transverse
coordinate $(x, y)$ can be assumed proportional to the overlap of the
parton densities of the colliding hadrons in that area element. A few
possible overlap function options are available in \textsc{Pythia},
where the Gaussian case is the simplest one. If two Gaussian-profile
hadrons pass with an impact parameter $b_{\H\H}$, then the nice
convolution properties gives a total overlap that is a Gaussian in
$b_{\H\H}$, and the distribution of MPI vertices is a Gaussian
in $(x,y)$. Specifically note that there is no memory of the collision
plane in the vertex distribution. 

This property is unique to Gaussian convolutions, however. In general,
the collision region will be elongated either out of or in to the
collision plane. The former typically occurs for a distribution with a
sharper proton edge, \eg a uniform ball, which gives rise to the
almond-shaped collision region so often depicted for heavy-ion collisions.
The latter shape instead occurs for distributions with a less pronounced
edge, such as an exponential. The default \textsc{Pythia} behaviour is
close to Gaussian, but somewhat leaning towards the latter direction.
Even that is likely to be a simplification. The evolution of the incoming
states by initial-state cascades is likely to lead to ``hot spots'' 
of increased partonic activity, see \eg \cite{Bierlich:2019wld}.
A preliminary study in \cite{Sjostrand:2020gyg} showed that azimuthal
anisotropies in the individual $\H\H$ collision give unambiguous, but miniscule
flow effects, and furthermore the many $\H\H$ event planes of an $\A\A$ collision
point in random directions, further diluting any such effects. In the
end, it is the asymmetries related to the $\A\A$ geometry that matter for
our studies.

Only a fraction of the full nucleon momentum is carried away by the
MPIs of an $\H\H$ collision, leaving behind one or more beam remnants
\cite{Sjostrand:2004pf}. These are initially distributed according to
a Gaussian shape around the center of the respective hadron. By the
random fluctuations, and by the interacting partons primarily being
selected on the side leaning towards the other beam hadron, the
``center of gravity'' will not agree with the originally assumed origin. 
All the beam remnants will therefore be shifted so as to ensure that
the energy-weighted sum of colliding and remnant parton locations
is where it should be. Shifts are capped to be at most a proton radius, 
so as to avoid extreme spatial configurations, at the expense of 
a perfectly aligned center of gravity.

Not all hadronizing partons are created in the collision moment
$t = 0$. Initial-state radiation (ISR) implies that some partons
have branched off already before this, and final-state radiation (FSR)
that others do it afterwards. These partons then can travel some
distance out before hadronization sets in, thereby further complicating  
the space--time picture, even if the average time of parton showers
typically is a factor of five below that of string fragmentation
\cite{Ferreres-Sole:2018vgo}. We do not trace the full shower
evolution, but instead include a smearing of the transverse location
in the collision plane that a parton points back to. No attempt is
made to preserve the center of gravity during these fluctuations.

The partons produced in various stages of the collision process
(MPIs, ISR, FSR) are initially assigned colours according to the
$N_C \to \infty$ approximation, such that different MPI systems are
decoupled from each other. By the beam remnants, which have as one
task to preserve total colour, these systems typically become connected
with each other through the short-circuiting mechanism already mentioned.
Furthermore, colour reconnection (CR) is allowed to swap colours,
partly to compensate for finite-$N_C$ effects, but mainly that it
seems like nature prefers to reduce the total string length drawn out
when two nearby strings overlap each other. When such effects have
been taken into account, what remains to hadronize is one or more
separate colour singlet systems.

\subsection{Hadronization}
\label{subsec:hadronize}

Hadronization is modelled in the context of the Lund string
fragmentation model \cite{Andersson:1983ia}. In it, a linear confinement
is assumed, \ie a string potential of $V = \kappa r$, where the string
tension $\kappa \approx 1$~GeV/fm and $r$ is the separation between a
colour triplet--antitriplet pair. For the simplest possible case, that of
a back-to-back $\q\qbar$ pair, the linearity leads to a straightforward
relationship between the energy--momentum and the space--time pictures:  
\begin{equation}
\left| \frac{\d p_{z,\q/\qbar}}{\d t} \right| =
\left| \frac{\d p_{z,\q/\qbar}}{\d z} \right| =
\left| \frac{\d E_{\q/\qbar}}{\d t} \right| =
\left| \frac{\d E_{\q/\qbar}}{\d z} \right| = \kappa ~.
\label{eq:xplinearity}
\end{equation}
If there is enough energy, the string between an original $\q_0 \qbar_0$
pair may break by producing new $\q_i \qbar_i$ pairs, where the
intermediate $\q_i$ ($\qbar_i$) are pulled towards the $\qbar_0$ ($\q_0$)
end, such that the original colour field is screened. This way the
system breaks up into a set of $n$ colour singlets
$\q_0\qbar_1 - \q_1\qbar_2 - \q_2\qbar_3 - \ldots - \q_{n-1}\qbar_0$,
that we can associate with the primary hadrons. By \eqref{eq:xplinearity}
the location of the breakup vertices in space--time is linearly related
to the energy--momentum of the hadrons produced between such vertices
\cite{Ferreres-Sole:2018vgo}.

When quarks with non-vanishing mass or $\pT$ are created,
they have to tunnel out a distance before they can end up on mass
shell. This tunnelling process gives a suppression of heavier quarks,
like $\s$ relative to $\u$ and $\d$ ones, and an (approximately)
Gaussian distribution of the transverse momenta. Effective
equivalent massless-case production vertices can be defined.
Baryons can be introduced \eg by considering diquark--antidiquark
pair production, where a diquark is a colour antitriplet and thus can
replace an antiquark in the flavour chain.

Having simultaneous knowledge of both the energy--momentum
and the space--time picture of hadron production violates the 
Heisenberg uncertainty relations. In this sense the string model should 
be viewed as a semiclassical one. The random nature of the Monte Carlo
approach will largely mask the issue, and smearing factors are introduced
in several places to further reduce the tension. 

A first hurdle is to go on from a simple straight string to a longer
string system. In the limit where the number of colours is large,
the $N_C \to \infty$ approximation \cite{tHooft:1973alw}, a string
typically will be stretched from a quark end via a number intermediate
gluons to an antiquark end, where each string segment is stretched
between a matching colour-anticolour pair. To first approximation each
segment fragments as a boosted copy of a simple $\q\qbar$ system, but
the full story is more complicated, with respect to what happens 
around each gluon. Firstly, if a gluon has time to lose its energy
before it has hadronized, the string motion becomes more complicated.
And secondly, even if not, a hadron will straddle each gluon kink,
with one string break in each of the two segments it connects. 
A framework to handle energy and momentum sharing in such complicated
topologies was developed in Ref. \cite{Sjostrand:1984ic}, and was then
extended to reconstruct matching space--time production vertices in
\cite{Ferreres-Sole:2018vgo}. This includes many further details not
covered here, such as a transverse smearing of breakup vertices, to
represent a width of the string itself, and various safety checks.

In addition to the main group of open strings stretched between $\q\qbar$
endpoints, there are two other common string topologies. One is a closed 
gluon loop. It can be brought back to the open-string case by a first
break somewhere along the string. The other is the junction topology,
represented by three quarks moving out in a different directions,
each pulling out a string behind itself. These strings meet at a common
junction vertex, to form a Y-shaped topology. This requires a somewhat
more delicate extensions of the basic hadronization machinery.

One complication is that strings can be stretched between partons that
do not originate from the same vertex. In the simplest case, a $\q$
connected with a $\qbar$ from a different MPI, the vertex separation
could be related to a piece of string already at $t = 0$. At the small
distances involved it is doubtful whether the full string tension is
relevant, in particular since the net energy associated with such initial
strings should not realistically exceed the proton mass. Since this energy
is then to be spread over many of the final-state hadrons, the net effect
on each hardly would be noticeable, and is not modelled.

For the space--time picture we do want to be somewhat more careful about
the effects of the transverse size of the original source. Even an
approximate description would help smear the hadron production vertices
in a sensible manner. To begin, consider a simple $\q\qbar$ string, where
the relevant length of each hadron string piece is related to its energy.  
For a given hadron, define $E_{\hrm\q}$ ($E_{\hrm\qbar}$)
as half the energy of the hadron plus the full energy of all hadrons
lying between it and the $\q$ ($\qbar$) end, and use this as a measure
of how closely associated a hadron is with the respective endpoint. 
Also let $\mathbf{r}_{\perp\q}$ ($\mathbf{r}_{\perp\qbar}$) be the 
(anti)quark transverse production coordinates. Then define the hadron 
production vertex offset to be
\begin{equation}
  \Delta \mathbf{r}_{\perp\hrm} = \frac{E_{\hrm\qbar} \, 
    \mathbf{r}_{\perp\q} + E_{\hrm\q} \, \mathbf{r}_{\perp\qbar}}%
    {E_{\hrm\q} + E_{\hrm\qbar}}
  = \frac{(E_{\mathrm{tot}} - E_{\hrm\q}) \, \mathbf{r}_{\perp\q} 
    + E_{\hrm\q} \, \mathbf{r}_{\perp\qbar}}{E_{\mathrm{tot}}} ~,  
\end{equation}
relative to what a string motion started at the origin would have given.

This procedure is then generalized to more complicated string topologies.
Again energy is summed up from one string end, for partons and hadrons
alike, to determine which string segment a given hadron is most closely
associated with, and how the endpoints of that segment should be mixed.
Note that, although energy is not a perfect measure of location along the
string, the comparison between parton and hadron energies is only mildly
Lorentz-frame dependent, which is an advantage. More complicated string
topologies, like junction ones, require further considerations not
discussed here. Again we stress that the main point is not to provide a
perfect location for each individual hadron, but to model the average 
effects. 

\subsection{The hadronic rescattering formalism}
\label{subsec:rescatterxyzt}

By the procedure outlined so far, each primary produced hadron has
been assigned a production vertex $x_0 = (t_0, \mathbf{x}_0)$ and a
four-momentum $p = (E, \mathbf{p})$. The latter 
defines its continued motion along straight trajectories
$\mathbf{x}(t) = \mathbf{x}_0 + (t - t_0) \, \mathbf{p} /m$.
Consider now two particles produced at $x_1$ and $x_2$ with 
momenta $p_1$ and $p_2$. Our objective is to determine whether these
particles will scatter and, if so, when and where. To this end, the
candidate collision is studied in the center-of-momentum frame of the
two particles. If they are not produced at the same time, the position
of the earlier one is offset to the creation time of the later one.
Particles moving away from each other already at this common time
are assumed unable to scatter.

Otherwise, the probability $P$ of an interaction is a function of the
impact parameter $b$, the center-of-mass energy $E_{\CM}$, and the two
particle species $A$ and $B$. There is no solid theory for the $b$
dependence of $P$, so a few different options are implemented, such as
a black disk, a grey disk or a Gaussian. In either case the normalization
is such that $\int P(b) \, \d^2 b = \sigma_{AB}(E_{\CM})$. To first
approximation all options thus give the same interaction rate, but the
drop of hadronic density away from the center in reality means fewer
interactions for a broader distribution.

If it is determined that the two particles will interact, the interaction
time is defined as the time of closest approach in the rest frame.
The spatial component of the interaction vertex depends on the character
of the collision. Elastic and diffractive processes can be viewed as
$t$-channel exchanges of a pomeron (or reggeon), and then it is reasonable
to let each particle continue out from its respective location at the
interaction time. For other processes, where either an intermediate
$s$-channel resonance is formed or strings are stretched between the
remnants of the two incoming hadrons, an effective common interaction
vertex is defined as the average of the two hadron locations at the 
interaction time. In cases where strings are created, be it by $s$-channel
processes or by diffraction, the hadronization starts around this vertex
and is described in space--time as already outlined. This means an
effective delay before the new hadrons are formed and can begin to
interact. For the other processes, such as elastic scattering or an
intermediate resonance decay, there is the option to have effective
formation times before new interactions are allowed. 

In actual events with many hadrons, each hadron pair is checked to see
if it fulfils the interaction criteria and, if it does, the interaction
time for that pair (in the CM frame of the event) is recorded in a
time-ordered list. Furthermore, unstable particles can decay during the
rescattering phase. For these, an invariant lifetime $\tau$ is picked
at random according to an exponential $\exp(-\tau/\tau_0)$, where
$\tau_0 = 1 / \Gamma$ is the inverse of the width. The resulting decay
times are inserted into the same list. Then the scattering or decay that
is first in time order is simulated, unless the particles involved have
already interacted/decayed. This produces new hadrons that are checked
for rescatterings or decays, and any such are inserted into the
time-ordered list. This process is repeated until there are no more
potential interactions.

There are some obvious limitations to the approach as outlined so far:
\begin{itemize}
\item The procedure is not Lorentz invariant, since the time-ordering of 
interactions is defined in the CM frame. We do not expect this to be a
major issue. This has been studied and confirmed within existing
rescattering approaches \cite{Bass:1998ca,Xu:2004mz,Weil:2016zrk}, and
reconfirmed in our $\p\p$ studies. 
\item Currently only collisions between two incoming hadrons are
considered, even though in a dense environment one would also expect
collisions involving three or more hadrons. This is a more relevant
restriction, that may play a role for some observables, and to be
considered in the future. 
\item Since traditional \textsc{Pythia} tunes do not include rescattering
effects, some retuning to $\p\p$ events has to be made before the model
is applied to $\A\A$ ones. For now, only the simplest possible one is used,
wherein the $\pTo$ parameter of the MPI framework is increased slightly
so as to restore the same average charged multiplicity in proton collisions
at LHC energies as without rescattering.
\item All modelled subprocesses are assumed to share the same hadronic
impact-parameter profile. In a more detailed modelling the $t$-channel
elastic and diffractive processes should be more peripheral than the
rest, and display an approximately inverse relationship between the
$t$ and $b$ values.
\item The model only considers the effect of hadrons colliding with
hadrons, not those of strings colliding/overlapping with each other
or with hadrons. An example of the former is the already-introduced
shoving mechanism. Both shoving and rescattering act to correlate the
spatial location of strings/hadrons with a net push outwards, giving
rise to a radial flow. Their effects should be combined, but do not add
linearly since an early shove leads to a more dilute system of strings
and primary hadrons, and thereby less rescattering.
\end{itemize}
 
\subsection{Hadronic rescattering cross sections}
\label{subsec:rescatteringsigma}

A crucial input for deciding whether a scattering can occur is the total
cross section. Once a potential scattering is selected, it also becomes
necessary to subdivide this total cross section into a sum of partial
cross sections, one for each possible process, as these are used to
represent relative abundances for each process to occur. A staggering
amount of details enter in such a description, owing to the multitude
of incoming particle combinations and collision processes. To wit, not
only ``long-lived'' hadrons can collide, \ie $\pi$, $\K$, $\eta$,
$\eta'$, $\p$, $\n$, $\Lambda$, $\Sigma$, $\Xi$, $\Omega$, and their
antiparticles, but also a wide selection of short-lived hadrons,
starting with $\rho$, $\K^*$, $\omega$, $\phi$, $\Delta$, $\Sigma^*$
and $\Xi^*$. Required cross sections are described in detail in Ref. 
\cite{Sjostrand:2020gyg}, and we only provide a summary of the main
concepts here.

Of note is that most rescatterings occur at low invariant masses,
typically only a few GeV. Therefore the descriptions are geared to this
mass range, and cross sections are not necessarily accurate above 10~GeV.
Furthermore event properties are modelled without invoking any perturbative
activity, \ie without MPIs. We will see in \secref{sec:rescRates} that
the number of interactions above 10~GeV is small enough that these
discrepancies can safely be disregarded.

For this low-energy description, the following process types are available:
\begin{itemize}
\item Elastic interactions are ones where the particles do not change
species, \ie $AB \to AB$. In our implementation, these are considered
different from elastic scattering through a resonance, \eg
$\pi^+\pi^- \to \rho^0 \to \pi^+\pi^-$, although the two could be linked
by interference terms. In experiments, usually all $AB \to AB$ events are
called elastic because it is not possible to tell which underlying
mechanism is involved. 
\item Resonance formation typically can be written as $AB \to R \to CD$,
where $R$ is the intermediate resonance. This can only occur when one or
both of $A$ and $B$ are mesons. It is the resonances that drive rapid and
large cross-section variations with energy, since each (well separated)
resonance should induce a Breit-Wigner peak.
\item Annihilation is specifically aimed at baryon--anti\-bar\-yon
collisions where the baryon numbers cancel out and gives a mesonic final
state. It is assumed to require the annihilation of at least one
$\q\qbar$ pair. This is reminiscent of what happens in resonance
formation, but there the final state is a resonance particle, while
annihilation forms strings between the outgoing quarks.
\item Diffraction of two kinds are modelled here: single $AB \to XB$ or 
$AB \to AX$ and double $AB \to X_1 X_2$. Here $X$ represents a massive
excited state of the respective incoming hadron, and there is no net
colour or flavour exchange between the two sides of the event.
\item Excitation can be viewed as the low-mass limit of diffraction,
where either one or both incoming had\-rons are excited to a related
higher resonance. It can be written as $AB \to A^*B$, $AB \to AB^*$ or
$AB \to A^*B^*$. Here $A^*$ and $\B^*$ are modelled with Breit-Wigners,
as opposed to the smooth mass spectra of the $X$ diffractive states.
In our description, this has only been implemented in nucleon-nucleon
interactions.
\item Non-diffractive topologies are assumed to correspond to a net colour 
exchange between the incoming had\-rons, such that colour strings are
stretched out between them after the interaction.
\end{itemize}

Some examples of input used for the modelling of these total and partial
cross sections are as follows.
\begin{itemize}
\item Cross sections are invariant when all particles are replaced by 
their antiparticles.
\item In some cases good enough data exists that interpolation works.
\item $\pi\pi$ and $\K\pi$ cross sections are found using the
calculations of Pel{\'a}ez et al.\
\cite{GarciaMartin:2011cn,Pelaez:2019eqa,Pelaez:2016tgi},
which partly are based on Chiral Perturbation Theory.
\item The neutral Kaon system is nontrivial, with strong interactions
described by the $\K^0/\Kbar^0$ states and weak decays by the 
$\K^0_{\mathrm{S}}/\K^0_{\mathrm{L}}$ ones. Cross sections for a
$\K_{\mathrm{S}}^0/\K_{\mathrm{L}}^0$ with a hadron are given by the mean of the cross
section for $\K^0$ and $\Kbar^0$ with that hadron. When a collision
occurs, the $\K_{\mathrm{S,L}}$ is converted into either $\K^0$ or $\Kbar^0$,
where the probability for each is proportional to the total cross section
for the interaction with that particle.
\item Several total cross sections are described by the $HPR_1R_2$
parameterization \cite{Tanabashi:2018oca}, consisting of one fixed term,
one ``pomeron'' $\ln^2 s$ ($s = E_{\CM}^2$) and two ``reggeon'' $s^{-\eta}$
ones.
\item $\N\N$ and $\N\pi$ elastic cross sections are partly covered by the
CERN/HERA data parameterizations \cite{Montanet:1994xu}.
\item The UrQMD program \cite{Bass:1998ca} has a complete set of total
and partial cross sections for all light hadrons, and in several cases  
we make use of these expressions.
\item Intermediate resonance formation can be modelled in terms of
(non-relativistic) Breit-Wigners, given a knowledge of mass and
(partial) width of the resonance. The widths are made mass-dependent
using the ansatz in UrQMD.
\item The annihilation cross section is the difference between the
total and the elastic ones near threshold, and above the inelastic
threshold it is based on a simple parameterization by Koch and Dover
\cite{Koch:1989zt}.
\item Differential diffractive cross sections are described by the SaS
(Schuler and Sj\"ostrand) ansatz \cite{Schuler:1993wr,Schuler:1996en},
and their integrated cross sections are parameterized with special
attention to achieving the relevant threshold behaviour.
\item Excitation into explicit higher resonances is implemented for $\N\N$
collisions, using the UrQMD expressions. For other collision types the
low-mass diffraction terms of SaS are included instead.
\item Inelastic non-diffractive events are represented by the cross
section part that remains when everything else is removed. Typically it
starts small near the threshold, but then grows to dominate at higher
energies.
\item The Additive Quark Model (AQM) \cite{Levin:1965mi,Lipkin:1973nt}
assumes that total cross sections scales like the product of the number
of valence quarks in the two incoming hadrons. The contribution of heavier
quarks is scaled down relative to that of a $\u$ or $\d$ quark, presumably
by mass effects giving a narrower wave function. Assuming that quarks
contribute inversely proportionally to their constituent masses, this gives
an effective number of interacting quarks in a hadron of approximately
\begin{equation}
    n_{\q,\mathrm{AQM}} = n_{\u} + n_{\d} + 0.6 \, n_{\s} + 0.2 \, n_{\c} +
    0.07 \, n_{\b}~. 
    \label{eq:nqAQM}
\end{equation}
For lack of alternatives, many unmeasured cross sections are assumed
to scale in proportion to this, relative to known ones. For heavier
particles, notably charm and bottom ones, it is also necessary to
correct the collision energy relative to the relevant mass threshold.
\end{itemize}

\subsection{Hadronic rescattering events}
\label{subsec:rescatteringevent}

The choice of subprocess is not enough to specify the resulting final
state. In some cases only a few further variable choices are needed.
For elastic scattering the selection of the Mandelstam $t$ is sufficient,
along with an isotropic $\varphi$ variable. Resonances are assumed to
decay isotropically, as are the low-mass excitations related to diffraction.
For inelastic non-diffractive events, higher-mass diffractive ones,
and annihilation processes, generically one one would expect strings to form
and hadronize. For diffraction these strings would be stretched inside
a diffractively excited hadron, while for the other two cases the strings
would connect the two original hadrons.

To illustrate the necessary steps, consider an inelastic non-diffractive
event. Each of the incoming hadrons first has to be split into a colour
piece, $\q$ or $\qbar\qbar$, and an anticolour ditto, $\qbar$ or $\q\q$.
For a baryon, \textbf{SU(6)} flavour$\times$spin factors are used to pick the
diquark spin. Then the lightcone momentum $p^+ (p^-)$ is split between the
two pieces of incoming hadron $A (B)$ moving along the $+z (-z)$ direction,
in such a way that a diquark is likely to carry the major fraction.  
The pieces also are given a relative $\pT$ kick. Including (di)quark
masses, the transverse masses $m_{\perp A 1}$ and $m_{\perp A 2}$ of the
two $A$ hadron pieces are defined. The $p_{A i}^-$ can now be obtained from 
$p^+ p^- = m_{\perp}^2$, and combined to give an effective mass $m_A^*$,
and similarly an $m_B^*$ is calculated. Together, the criterion
$m_A^* + m_B^* < E_{\CM}$ must be fulfilled, or the whole selection
procedure has to be restarted. Once an acceptable pair $(m_A^*, m_B^*)$
has been found, it is straightforward first to construct the kinematics
of $A^*$ and $B^*$ in the collision rest frame, and thereafter the
kinematics of their two constituents. 

Since the procedure has to work at very small energies, some additional
aspects should be mentioned. At energies very near the threshold, the
phase space for particle production is limited. If the lightest hadrons
that can be formed out of each of the two new singlets together leave
less than a pion mass margin up to the collision CM energy, then a simple
two-body production of those two lightest hadrons is (most likely) the
only option and is thus performed. There is then a risk to end up with an
unintentional elastic-style scattering. For excesses up to two pion masses,
instead an isotropic three-body decay is attempted, where one of the strings
breaks up by the production of an intermediate $\u\ubar$ or $\d\dbar$ pair.
If that does not work, then two hadrons are picked as in the two-body case
and a $\pi^0$ is added as third particle.

Even when the full collision energy is well above threshold, either one
or both of the strings individually may have a small mass, such that only
one or at most two hadrons can be produced from it. It is for cases like
this that the ministring framework has been developed, where it is allowed
for a string to collapse into a single hadron, with liberated excess
momentum shuffled to the other string. In a primary high-energy collisions,
low-mass strings are rare, and typically surrounded by higher-mass ones
that easily can absorb the recoil. At lower energies it is important to
try harder to find working solutions, and several steps of different
kinds have been added to the sequence of tries made. The new setup still
can fail occasionally to find an acceptable final state, but far less
than before the new measures were introduced.


\section{Model tests}
\label{sec:modelTests}

In this section we will study the rescattering model in $\p\p$, $\p\Pb$ and 
$\PbPb$ collisions. All collision energies are set to 5.02~TeV per
nucleon-nucleon system. This includes $\p\p$, for comparison reasons;
results at the more standard 13~TeV $\p\p$ energy have already been
presented elsewhere \cite{Sjostrand:2020gyg}.

\subsection{Multiplicities}
\label{sec:model-multiplicities}

The current lack of $3 \to 2$ processes in our model, to partly balance
the $2 \to 3$ ones, means that rescattering will increase the charged hadron
multiplicity. Effects are modest for $\p\p$ but, to compensate, the $\pTo$
parameter of the MPI framework is increased slightly when rescattering
is included. Thus the number of MPIs is reduced slightly, such that the
$\p\p$ charged multiplicity distribution is restored to be in reasonable
agreement with experimental data. We have used the same value for this
parameter also for the $\p\Pb$ and $\PbPb$ rescattering cases. Then
rescattering increases the final charged multiplicities by about 4~\% and
20~\%, respectively, due to a larger relative amount of rescattering in
larger systems. To simultaneously restore the multiplicity for all cases,
a retune also of \textsc{Angantyr} parameters would be necessary. This is
beyond the scope of the current article, and should rather wait until
$3 \to 2$ has been included. For now we accept some mismatch.

\begin{figure}[t!]
\begin{minipage}[c]{0.49\linewidth}
\centering
\includegraphics[width=\linewidth]{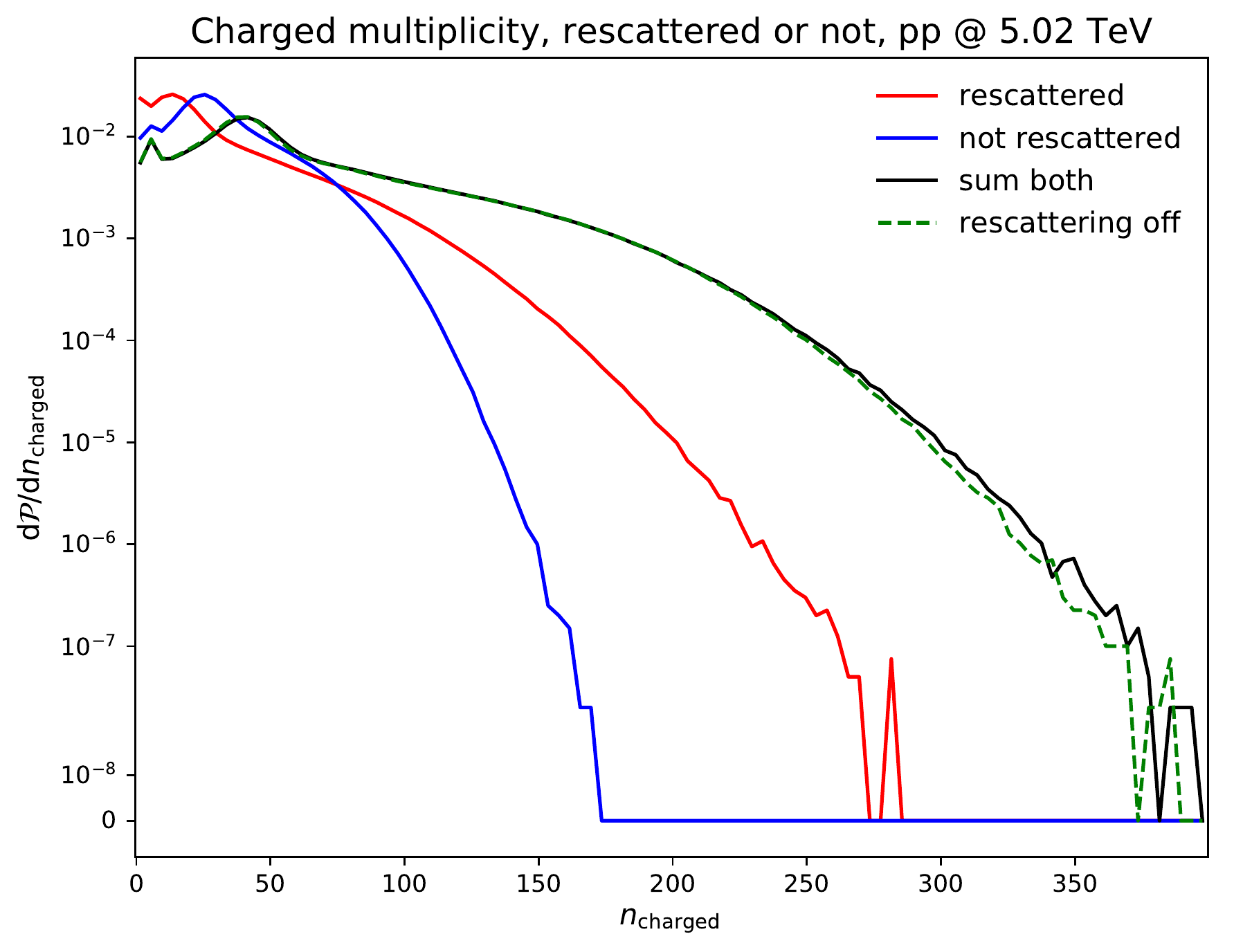}
\end{minipage}
\begin{minipage}[c]{0.49\linewidth}
\centering
\includegraphics[width=\linewidth]{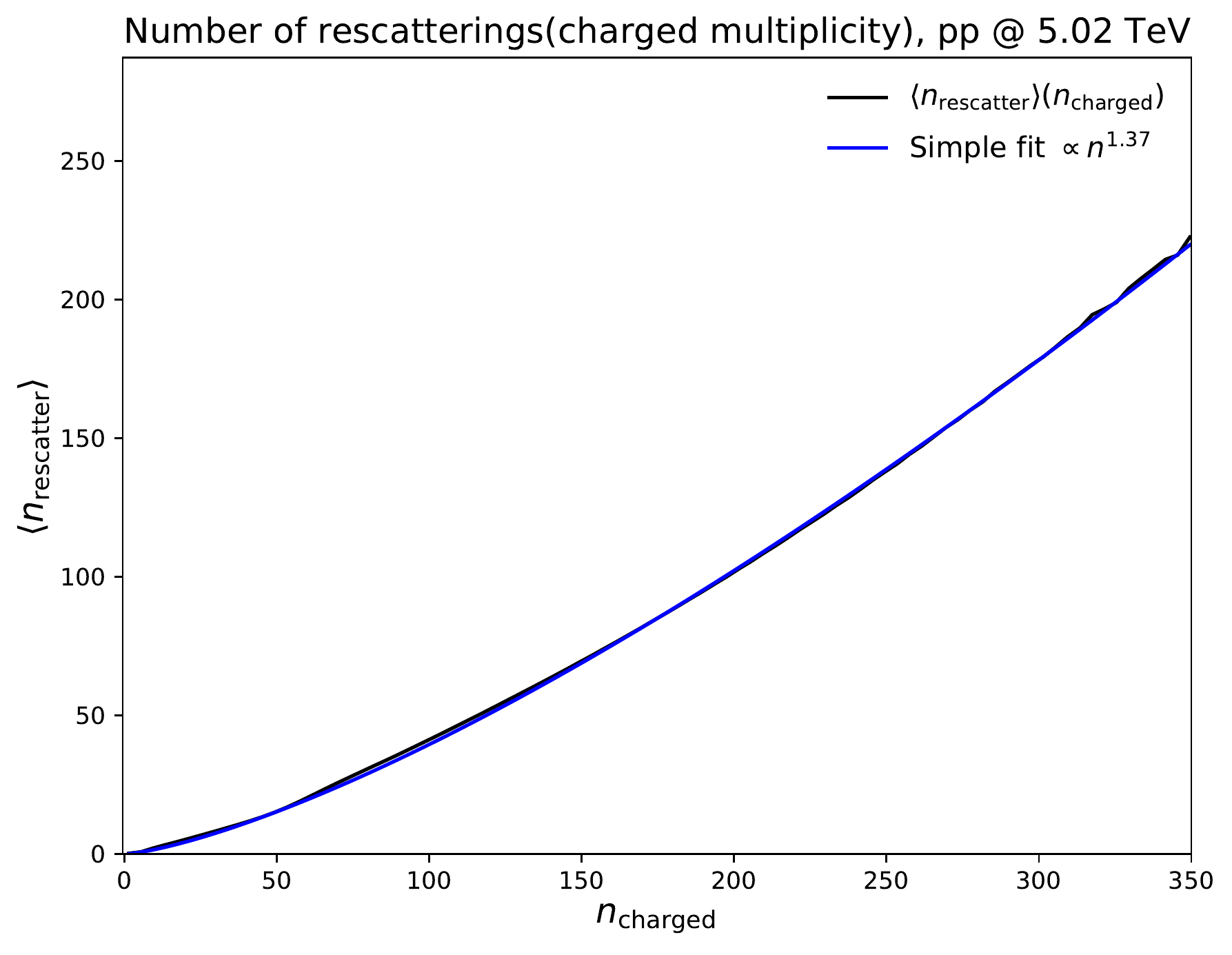}
\end{minipage}\\
\begin{minipage}[c]{0.49\linewidth}
\centering
\includegraphics[width=\linewidth]{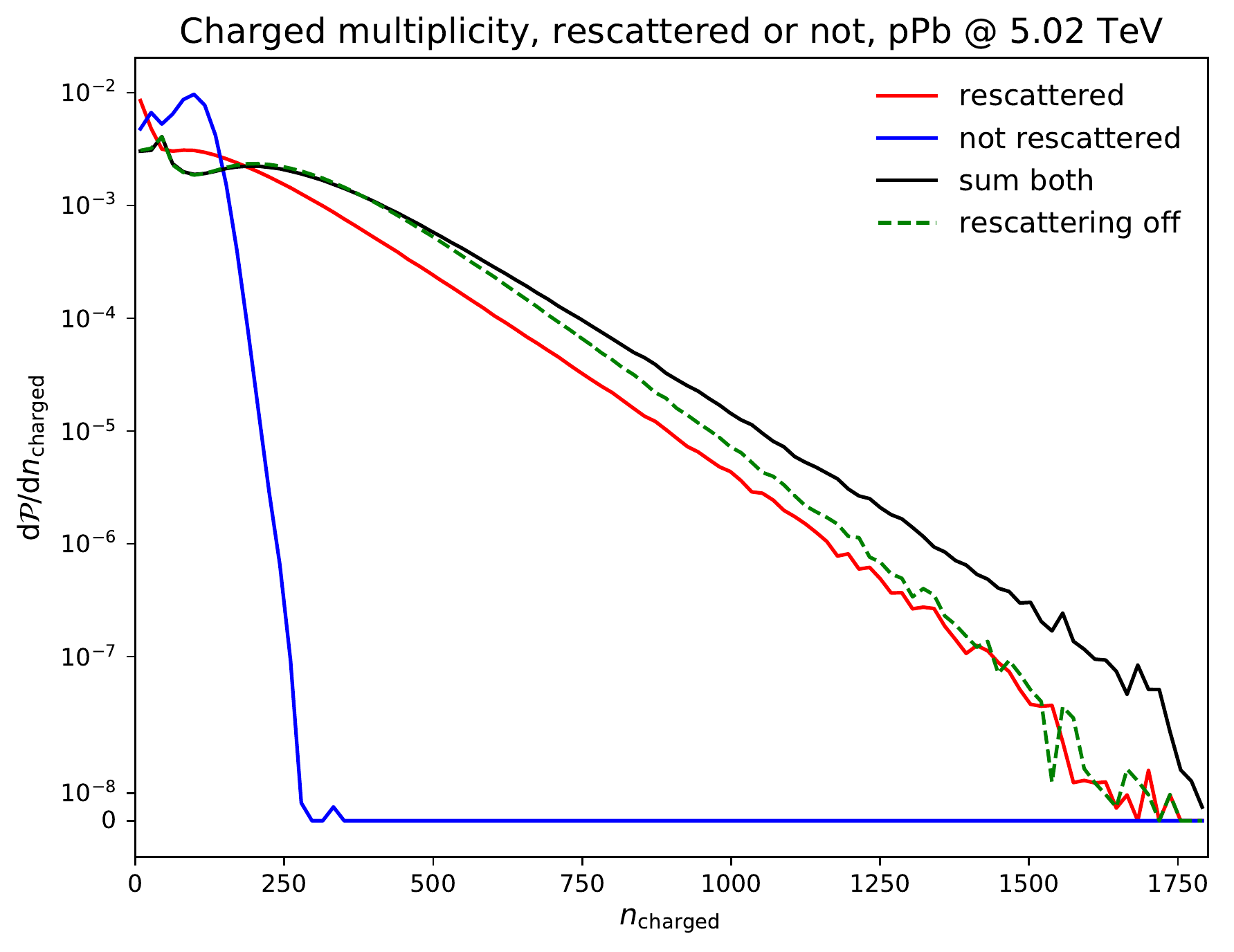}
\end{minipage}
\begin{minipage}[c]{0.49\linewidth}
\centering
\includegraphics[width=\linewidth]{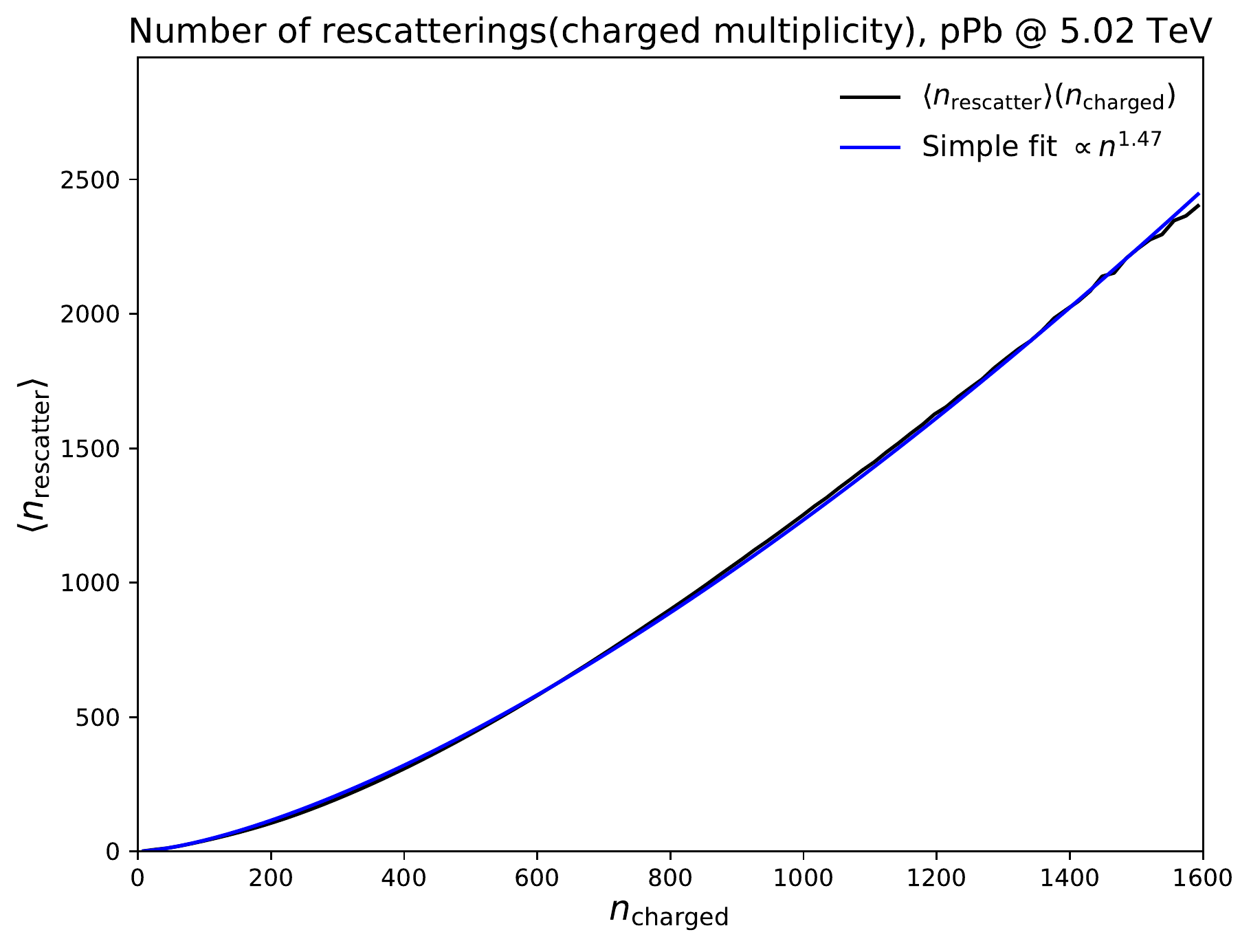}
\end{minipage}\\
\begin{minipage}[c]{0.49\linewidth}
\centering
\includegraphics[width=\linewidth]{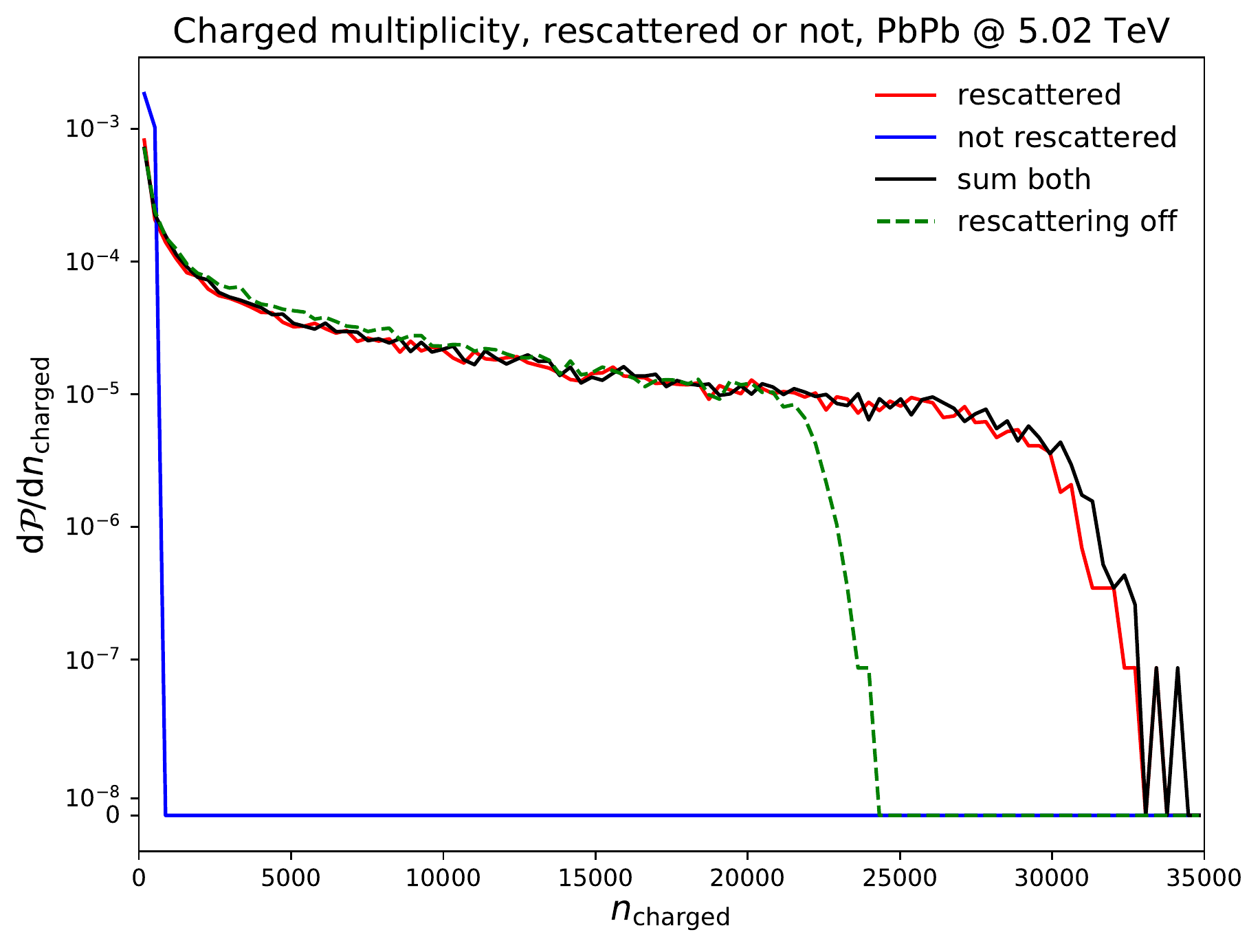}\\
(a)
\end{minipage}
\begin{minipage}[c]{0.49\linewidth}
\centering
\includegraphics[width=\linewidth]{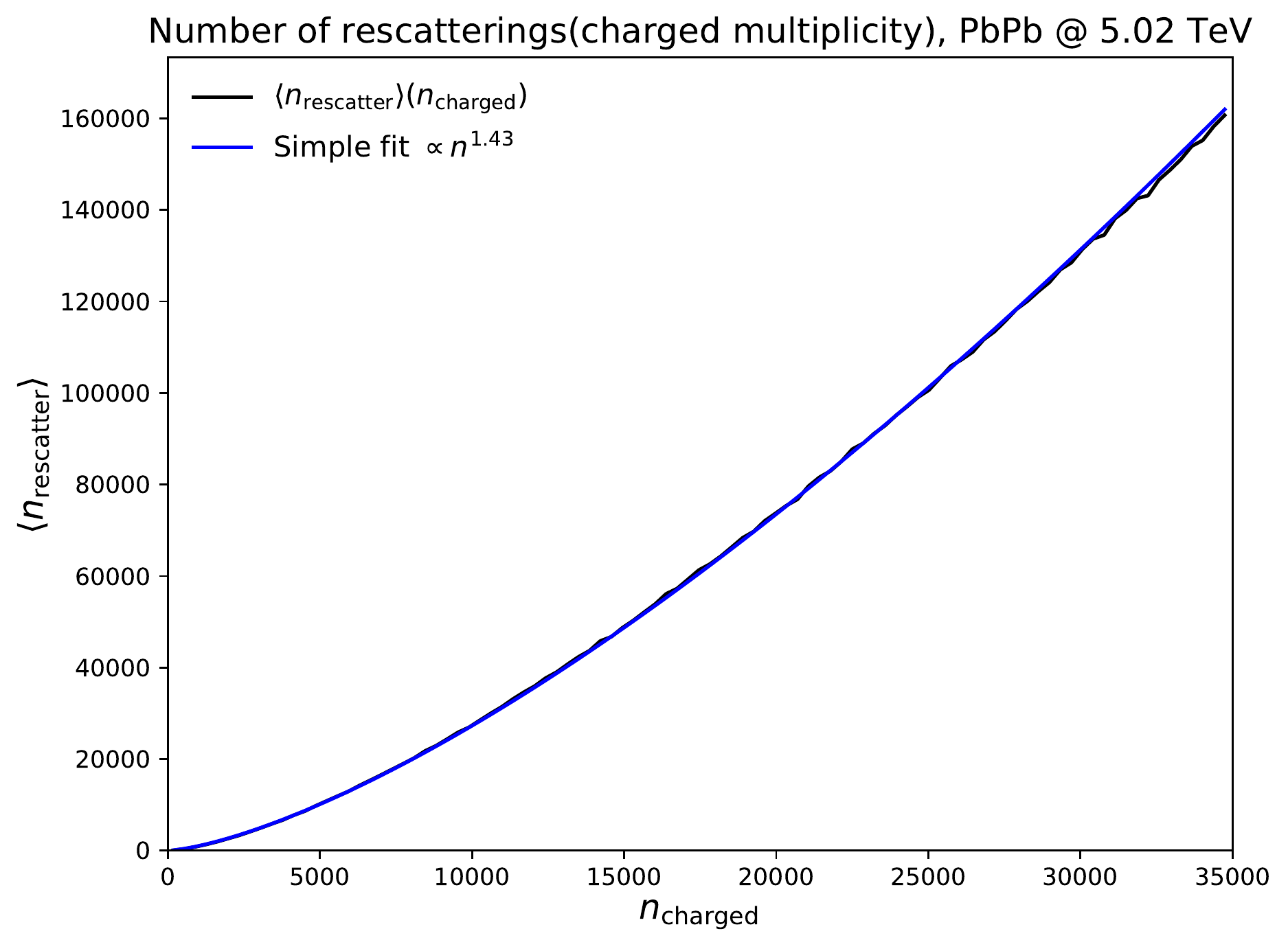}\\
(b)
\end{minipage}
\caption{
(a) Probability distributions for the total number of charged hadrons,
with and without rescattering, as well as the former number split in 
those where the final charged hadrons have been affected (directly or
indirectly) by rescattering and those where not. 
(b) Average number of rescatterings as a function of the charged hadron
multiplicity, together with a simple fit proportional to $n_{\mathrm{ch}}^p$.
}
\label{fig:rescatrate}
\end{figure}

Charged multiplicity distributions are shown in \figref{fig:rescatrate}a,
split into hadrons that have or have not been affected by rescattering.
Particles with a proper lifetime $\tau_0 > 100$~fm have been considered stable,
and multiplicities are reported without any cuts on $\eta$ or $\pT$.
Moving from $\p\p$ to $\p\Pb$ to $\PbPb$ we see how the fraction of
particles that do not rescatter drops dramatically. In absolute numbers
there still are about as many unrescattered in $\p\Pb$ as in $\p\p$,
and about twice as many in $\PbPb$. A likely reason is that many
collisions are peripheral, and even when not there are particles produced
at the periphery.

The total charged multiplicity is also compared with and without rescattering.
As foretold, the  $\p\p$ case has there been tuned to show no difference,
whereas rescattering enhances the high-multiplicity tail in $\p\Pb$ and
$\PbPb$. Rescattering also changes the relative abundances of different
particle types. In particular, baryon-antibaryon annihilation depletes the
baryon rate, by 7.5~\% for $\p\p$, 9.9~\% for $\p\Pb$ and 23.4~\% for
$\PbPb$, compared to the baryon number with a retuned $\pTo$. The retuning
itself gives in all cases a $\sim$2~\% reduction, that should be kept separate
in the physics discussion. The observed strange-baryon enhancement
\cite{Adam:2015qaa,ALICE:2017jyt} thus has to be explained by other
mechanisms, such as the rope model \cite{Bierlich:2014xba} or other
approaches that give an increased string tension \cite{Fischer:2016zzs}.

\subsection{Rescattering rates}
\label{sec:rescRates}

One of the most basic quantities of interest is the number of rescatterings
in an event. The average number of rescatterings as a function of the final
charged multiplicity $n_{\mathrm{ch}}$ is shown in \figref{fig:rescatrate}b.
The number of potential interactions at the beginning of rescattering is
proportional to $n_{\mathrm{primary}}^2$, where the number of primary hadrons
$n_{\mathrm{primary}} \simeq n_{\mathrm{ch}}$. The scaling is different in
practice however, due to the fact that some particles rescatter several times,
while others do not rescatter at all. As a first approximation one might still
expect the number of rescatterings to increase as $n_{\mathrm{ch}}^p$
for some power $p$. As seen in \figref{fig:rescatrate}b, this relation
appears to hold remarkably well, with $p=1.37$ for $\p\p$, $p=1.47$ for
$\p\Pb$, and $p=1.43$ for $\PbPb$. Interestingly, the exponent is highest
for the intermediate case $\p\Pb$, but the rescattering activity as such is
still highest for $\PbPb$. A possible explanation could be that in $\PbPb$,
high multiplicity corresponds to more central events with a larger volume, and
thus higher multiplicity does not necessarily mean higher density in this
case. We have also studied other $\p\A$ and $\A\A$
cases for a wide variety of sizes of $\A$, including Li, O, Cu and
Xe. While there is some $\A$ dependence in the exponent, this variation is
less significant than the overall difference between the $\p\A$ and $\A\A$ cases,
and in all instances the respective $p$ numbers for $\p\Pb$ and $\PbPb$
provide a reasonable description.

\begin{figure}[t!]
\begin{minipage}[c]{\linewidth}
\centering
\includegraphics[width=0.49\linewidth]{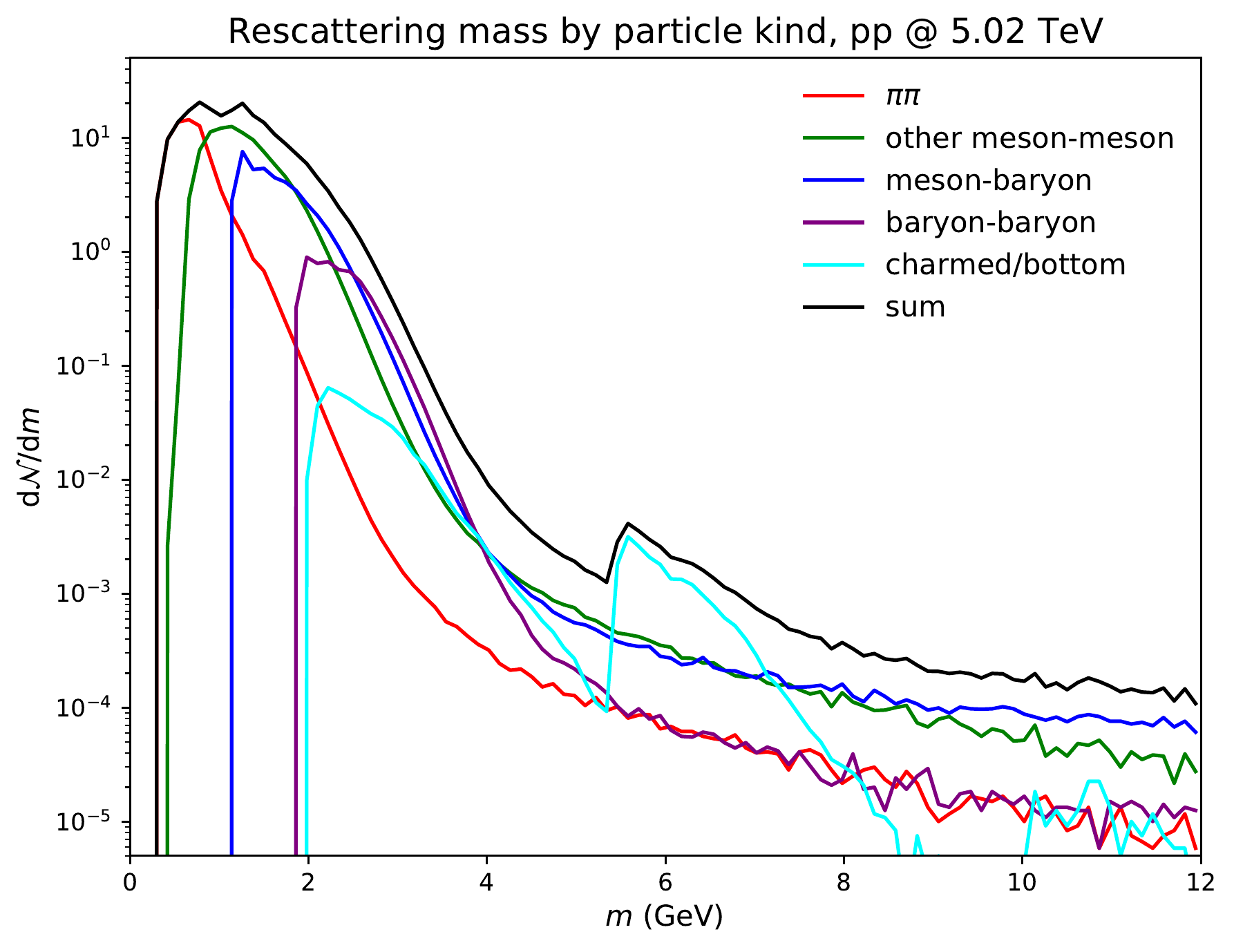}
\includegraphics[width=0.49\linewidth]{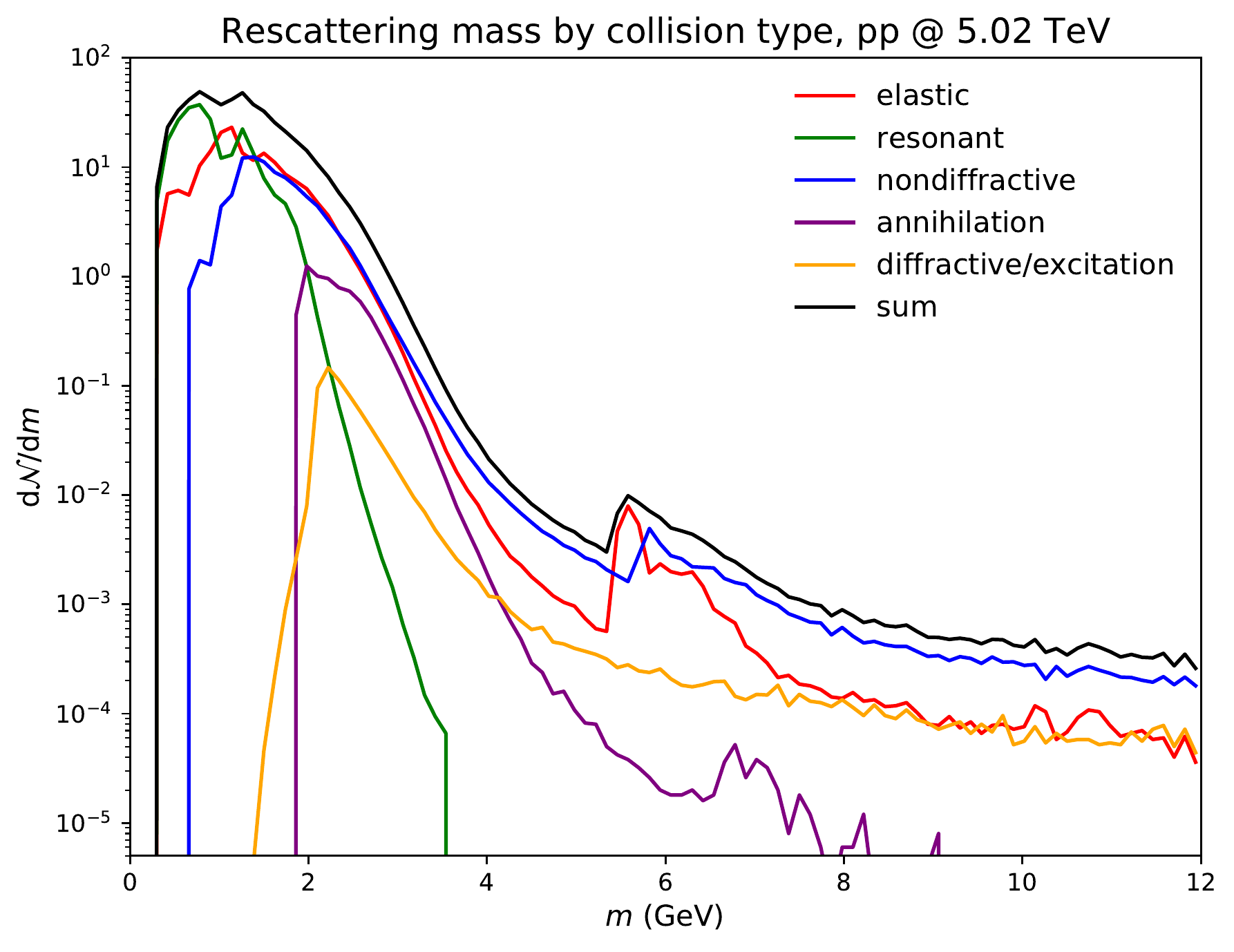}
\end{minipage}\\
\begin{minipage}[c]{\linewidth}
\centering
\includegraphics[width=0.49\linewidth]{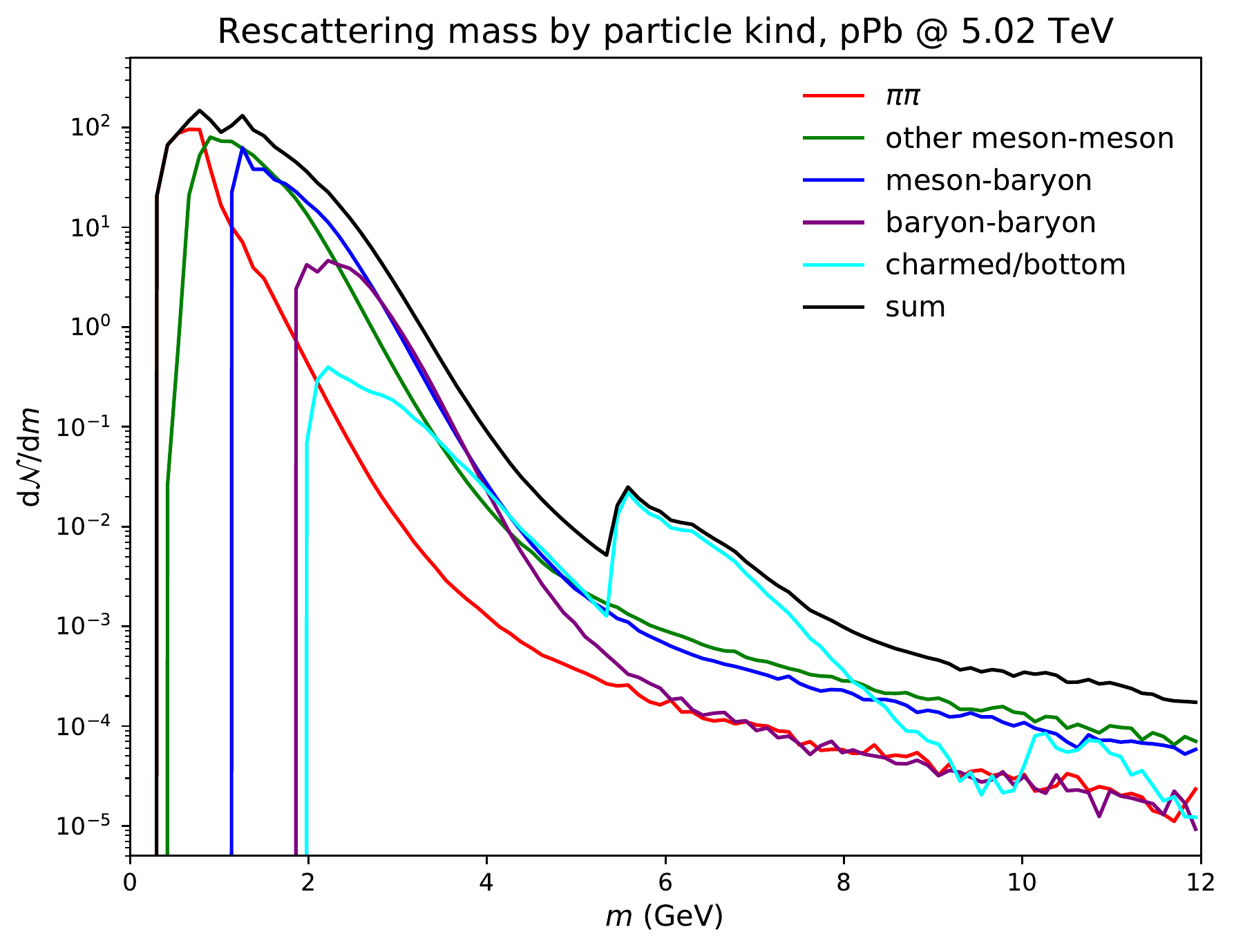}
\includegraphics[width=0.49\linewidth]{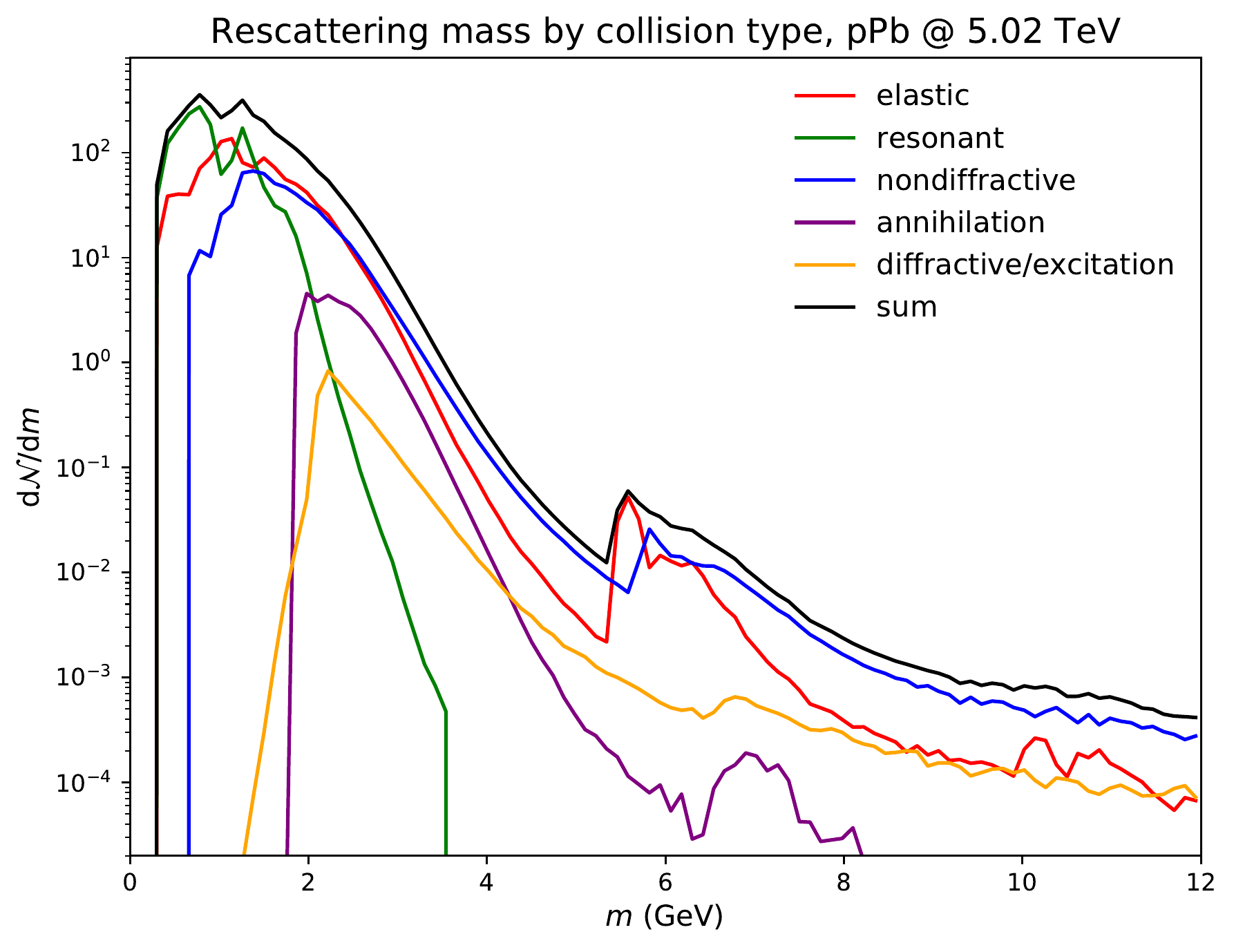}
\end{minipage}\\
\begin{minipage}[c]{\linewidth}
\centering
\includegraphics[width=0.49\linewidth]{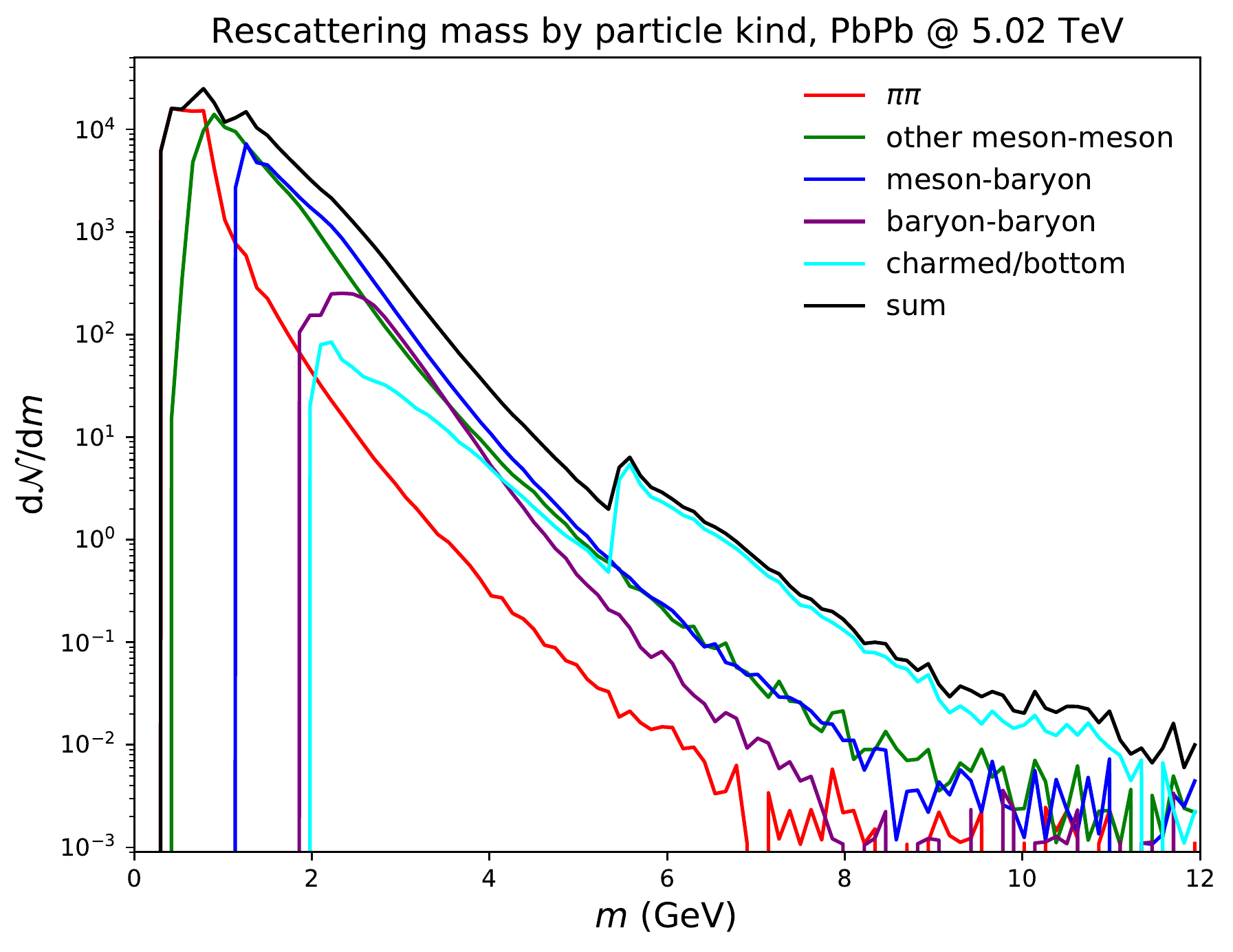}
\includegraphics[width=0.49\linewidth]{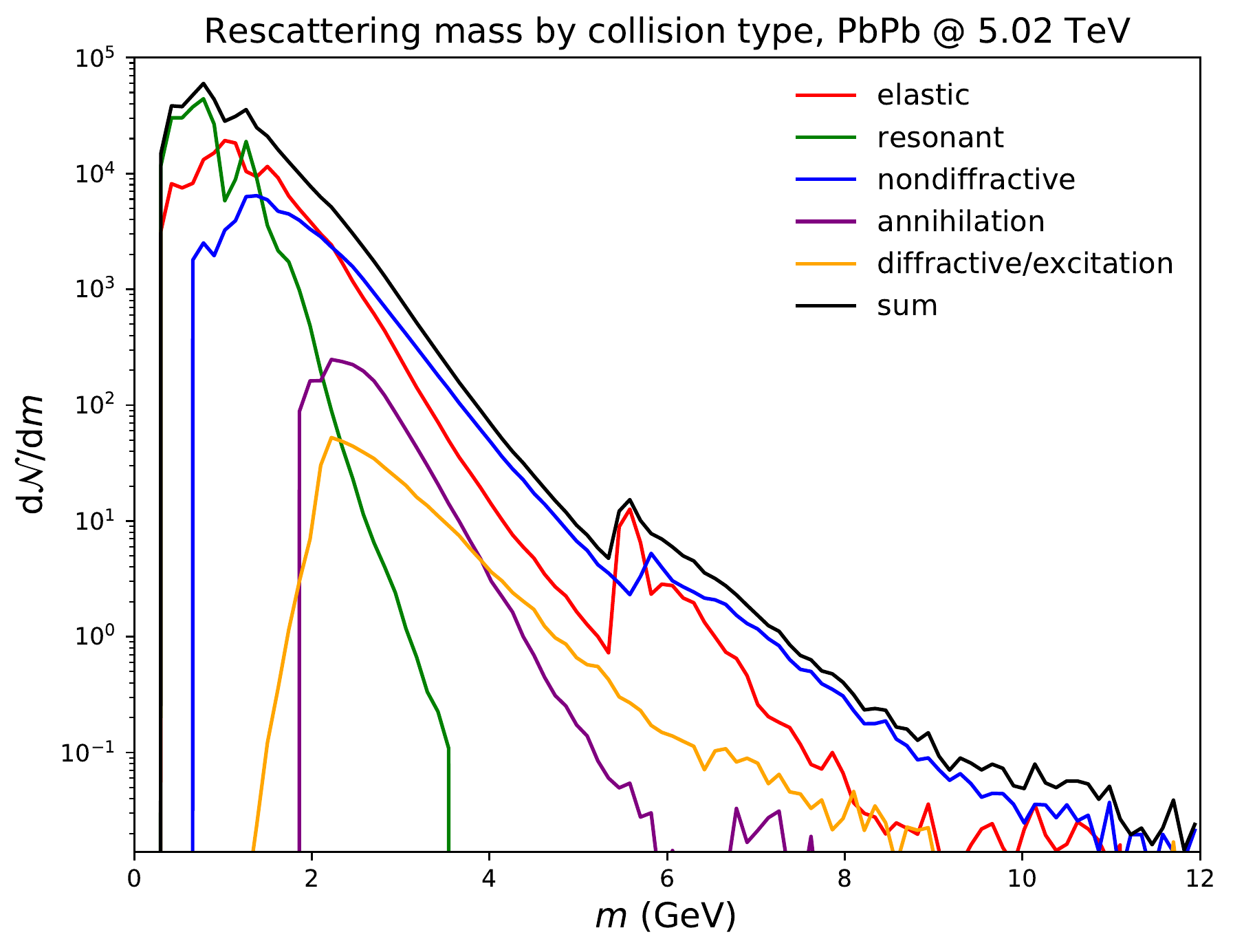}
\end{minipage}\\
\begin{minipage}[c]{0.49\linewidth}
\centering
(a)
\end{minipage}
\begin{minipage}[c]{0.49\linewidth}
\centering
(b)
\end{minipage}
\caption{Invariant masses for rescatterings, (a) by particle kind and
(b) by rescattering process type.
}
\label{fig:mResc}
\end{figure}

The invariant mass distributions of rescatterings are shown in
\figref{fig:mResc}a by incoming particle kind and in \figref{fig:mResc}b
by rescattering type. For increasingly large systems the fraction of
low-mass rescatterings goes up. A likely reason for this is rescattering causes
a greater multiplicity increase in the larger systems, reducing the average
energy of each particle. The composition of collision types
at a given mass is the same (within errors), as could be expected.
Our rescattering model is based on a
non-perturbative framework intended to be reasonably accurate up to
around $\sim$10~GeV. It would have to be supplemented by perturbative
modelling if a significant fraction of the collisions were well above
10~GeV, but clearly that is not the case. As an aside, the bump around
5.5~GeV comes from interactions involving bottom hadrons. 

\subsection{Transverse momentum spectra}

\begin{figure}[t!]
\begin{minipage}[c]{0.49\linewidth}
\centering
\includegraphics[width=\linewidth]{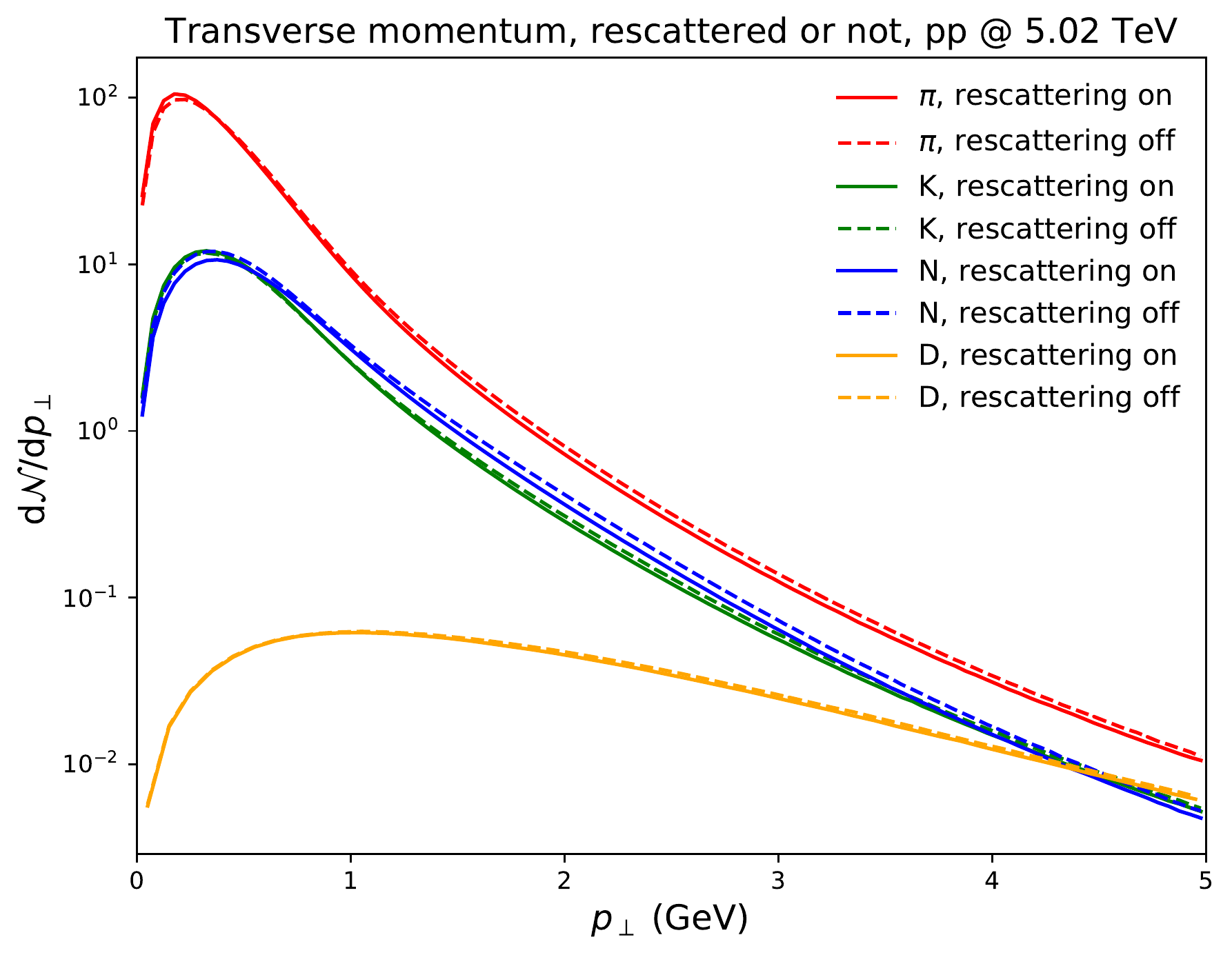}\\
(a)
\end{minipage}
\begin{minipage}[c]{0.49\linewidth}
\centering
\includegraphics[width=\linewidth]{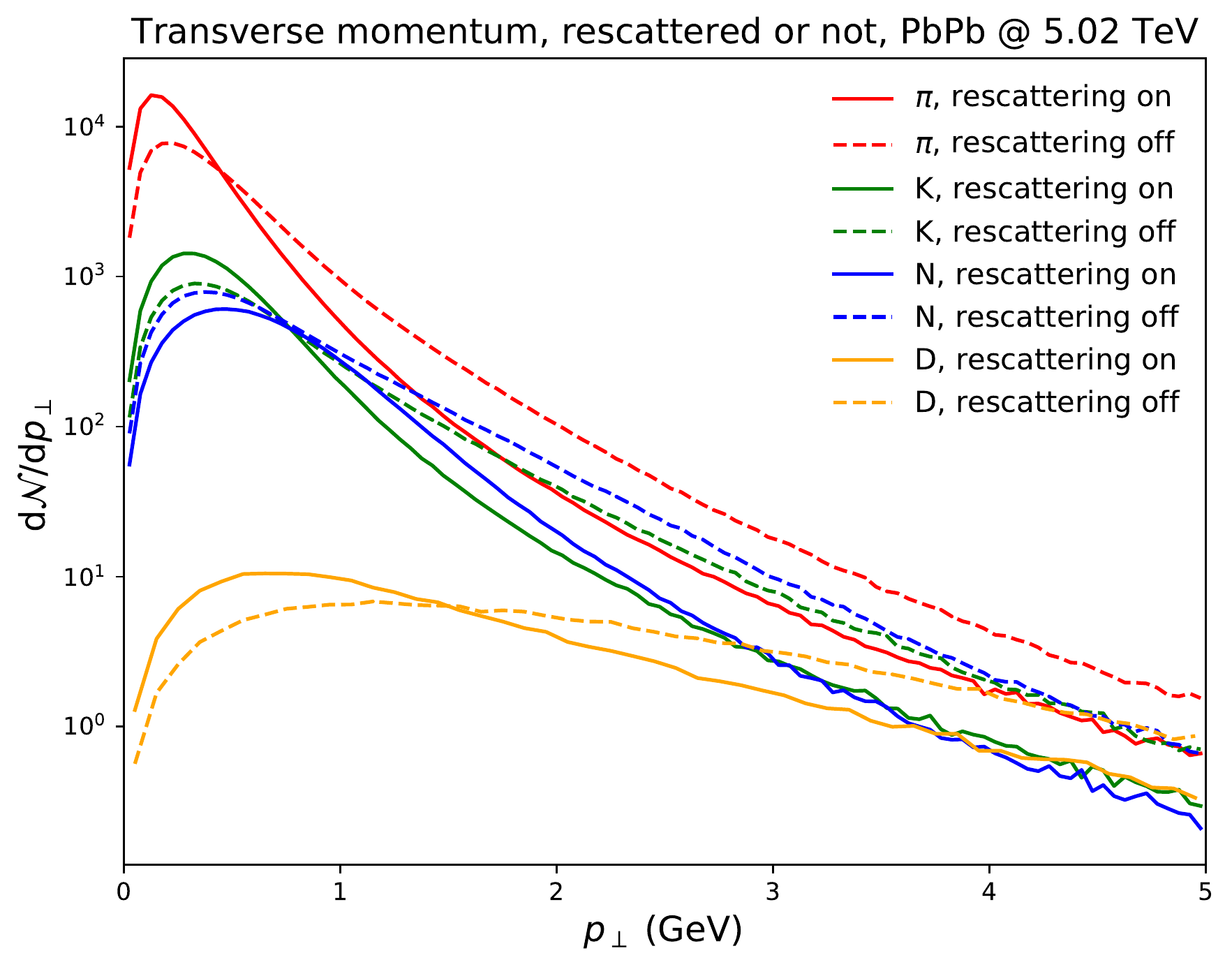}\\
(b)
\end{minipage}\\
\begin{minipage}[c]{0.49\linewidth}
\centering
\includegraphics[width=\linewidth]{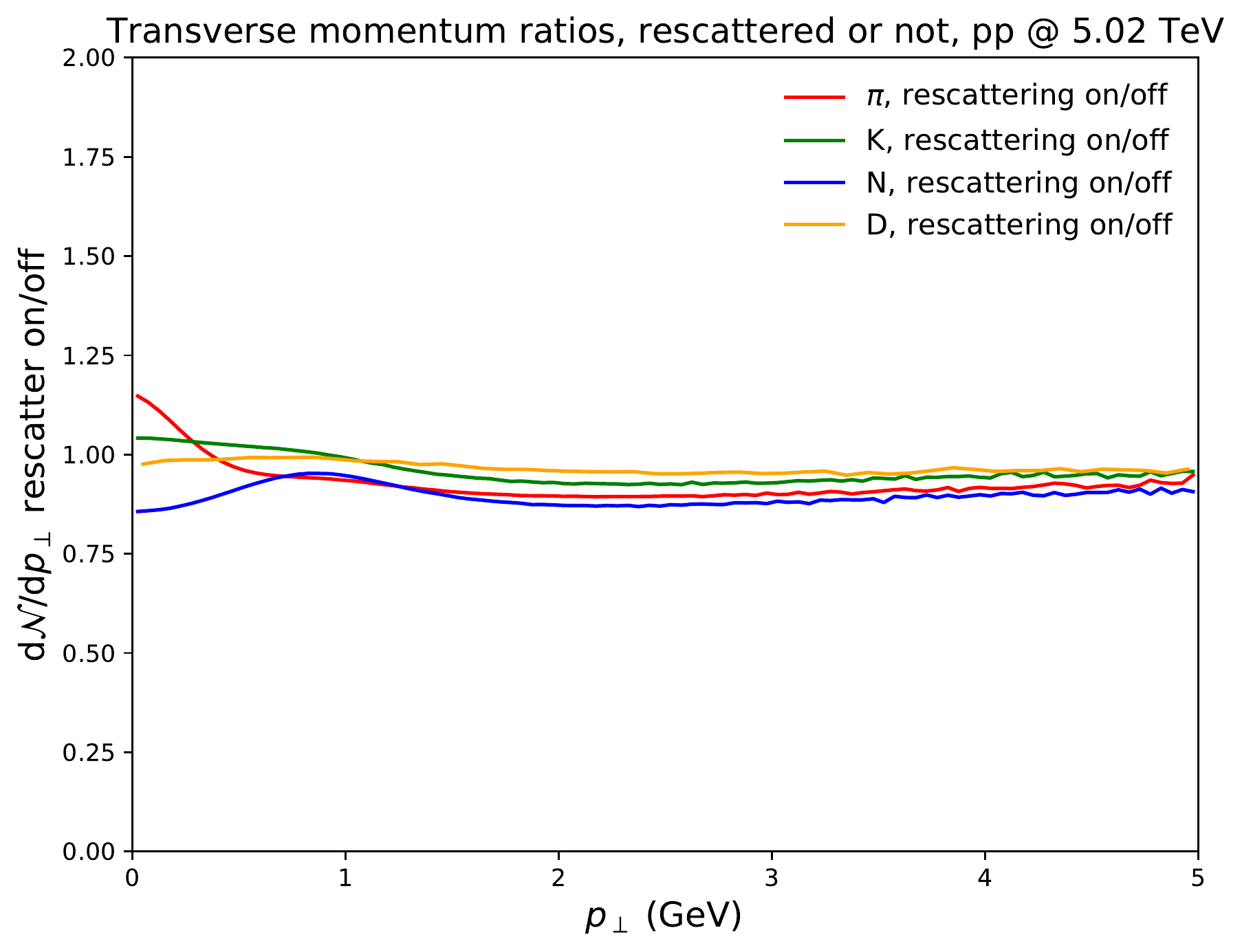}\\
(c)
\end{minipage}
\begin{minipage}[c]{0.49\linewidth}
\centering
\includegraphics[width=\linewidth]{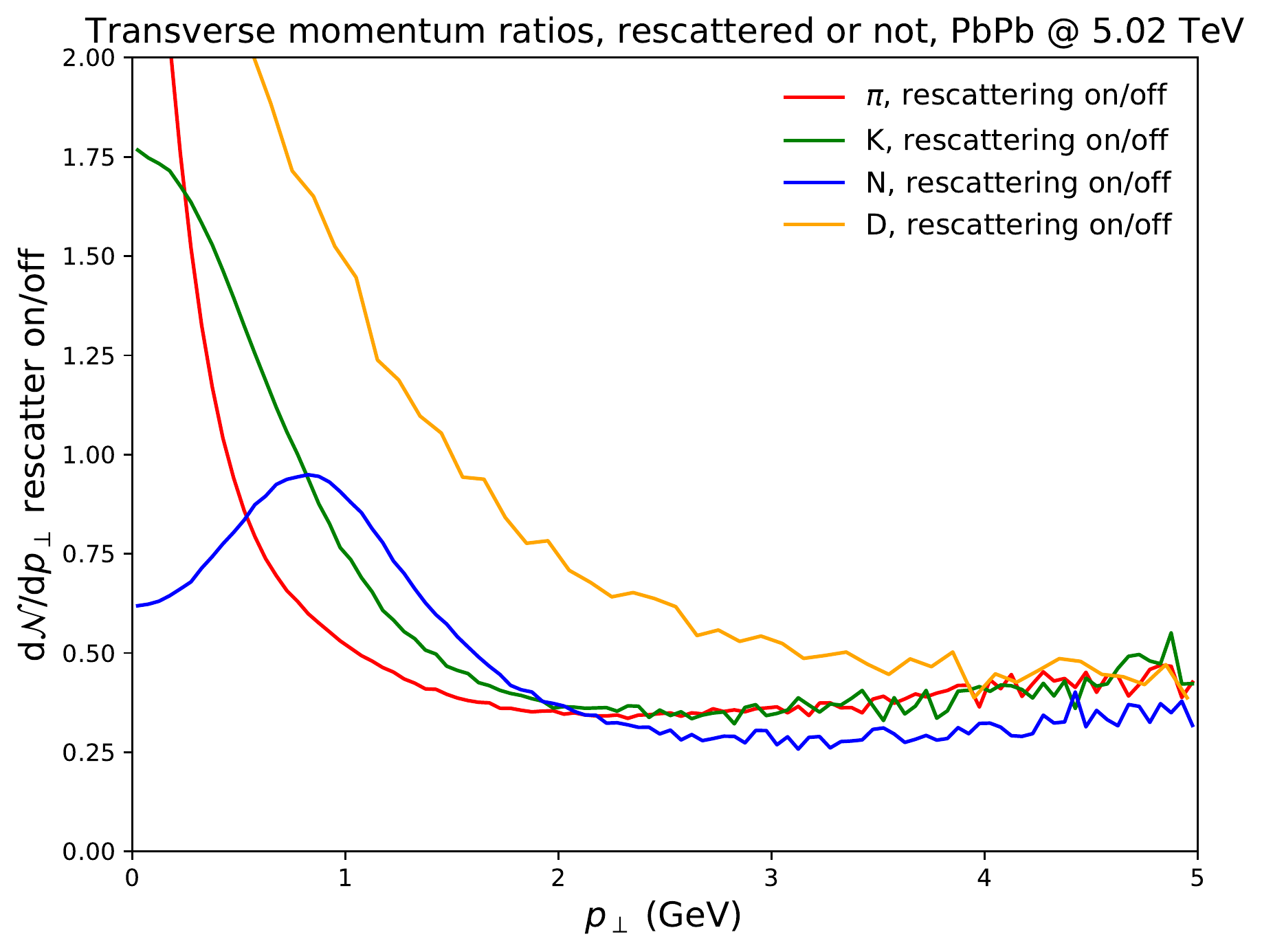}\\
(d)
\end{minipage}
\caption{$\pT$ spectra for pions, kaons, nucleons and D mesons, for (a) $\p\p$
and (b) $\PbPb$, together with ratios between the spectra with to without 
rescattering, for (c) $\p\p$ and (d) $\PbPb$.}
\label{fig:pTspectra}
\end{figure}

The $\pT$ spectra for pions, kaons, nucleons and charm mesons,
with and without rescattering, are shown in \figref{fig:pTspectra}a,b,
and the ratios with/without are shown in \figref{fig:pTspectra}c,d.
The effects are qualitatively similar for $\p\p$ and $\PbPb$, but more
prominent for the latter case. Pions get pushed to lower $\pT$, which
is consistent with the expectation that lighter particles will lose
momentum due to the ``pion wind'' phenomenon, where lighter particles
move faster than heavier and push the latter ones from behind. We remind
that all primary hadrons types are produced with the same $\pT$
distribution in string fragmentation, if the string is stretched parallel
with the collision axis. Rapid $\rho$ and $\K^*$ decays decrease the
average pion $\pT$, but initially indeed pions have the largest 
velocities.

The effect is similar for kaons, which unfortunately is inconsistent with
measurement \cite{Adam:2015qaa}. Our studies indicate that a significant
contribution to the loss of $\langle \pT \rangle$ for kaons comes from
inelastic interactions, and that the $\langle \pT \rangle$ increases if
all rescatterings are forced to be elastic. We believe this effect can be
ameliorated by implementing $3 \to 2$ and related processes.
For nucleons we note an overall loss in the rescattering scenario, which comes 
mainly from baryon--antibaryon annihilation, as already mentioned.
The $\langle \pT \rangle$ is shifted upwards by the aforementioned pion
wind phenomenon.

$\D$ mesons are enhanced at low $\pT$, all the way down to threshold.
At first glance this appears inconsistent with the pion wind phenomenon,
since $\D$ mesons are heavy. One key difference is that charm quarks
are not produced in string fragmentation, but only in perturbative
processes. Therefore $\D$ mesons start out at higher $\pT$ values than
ordinary hadrons, and can lose momentum through rescattering.
Nevertheless, the overall shift is still somewhat towards higher momenta
if only elastic rescatterings are permitted, as for kaons.

Overall we see a rather significant effect on $\pT$ spectra, and this is
to be kept in mind for other distributions. Especially for pions, where
the choice of a lower $\pT$ cut in experimental studies strongly affects
the (pseudo)rapidity spectrum deformation by rescattering, among others.

\subsection{Spacetime picture of rescattering}

\begin{figure}[t!]
\begin{minipage}[c]{\linewidth}
\centering
\includegraphics[width=0.49\linewidth]{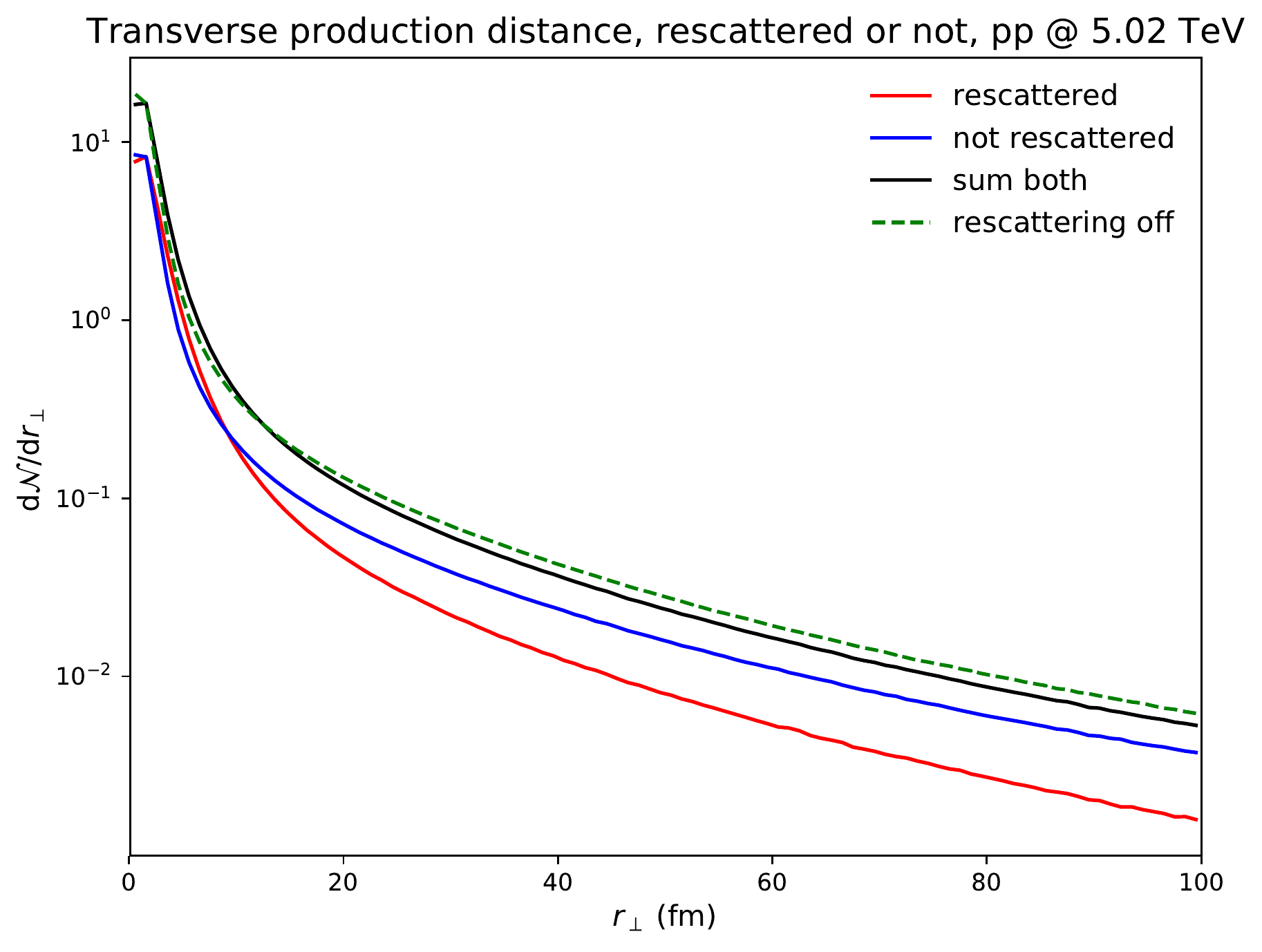}
\includegraphics[width=0.49\linewidth]{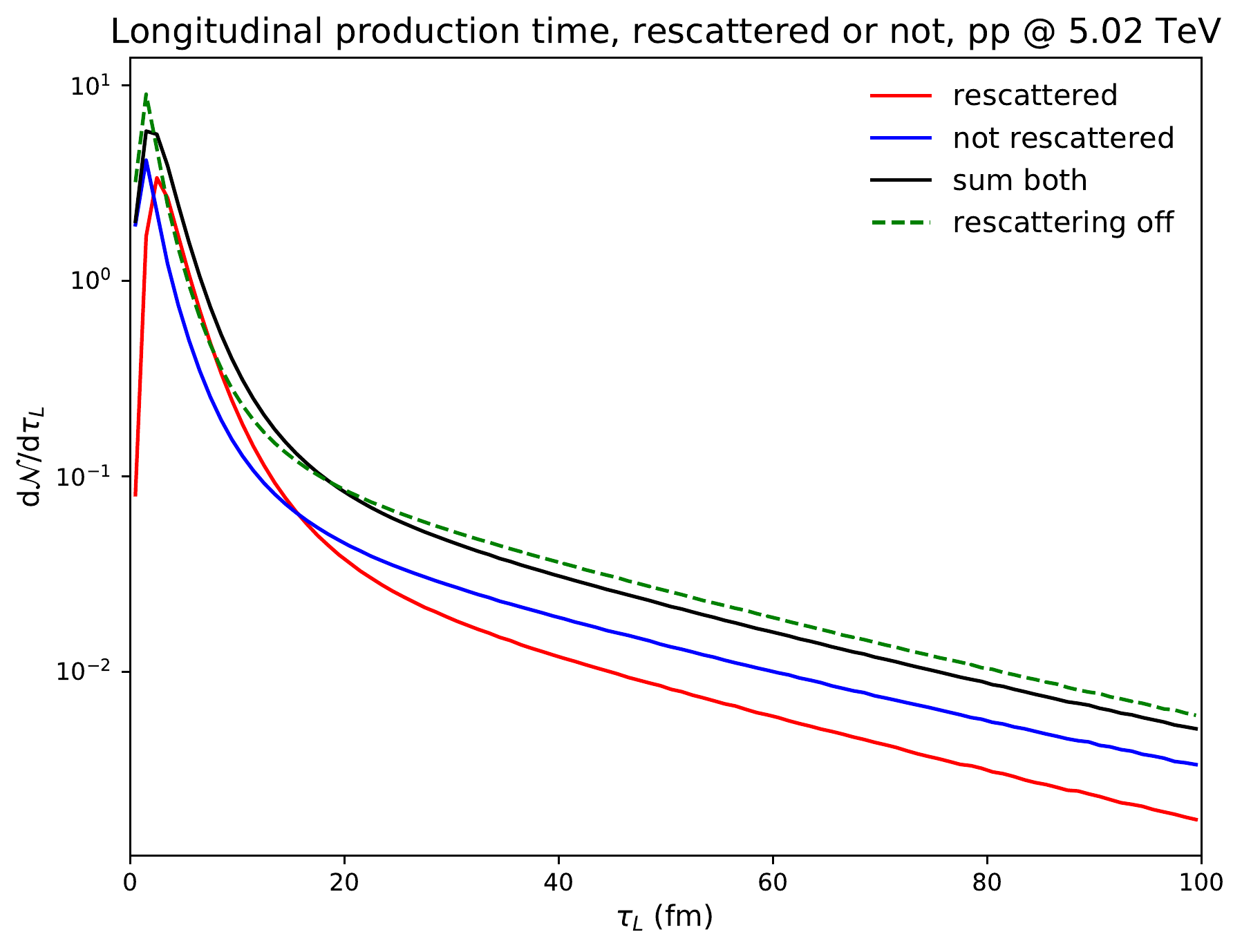}
\end{minipage}\\
\begin{minipage}[c]{\linewidth}
\centering
\includegraphics[width=0.49\linewidth]{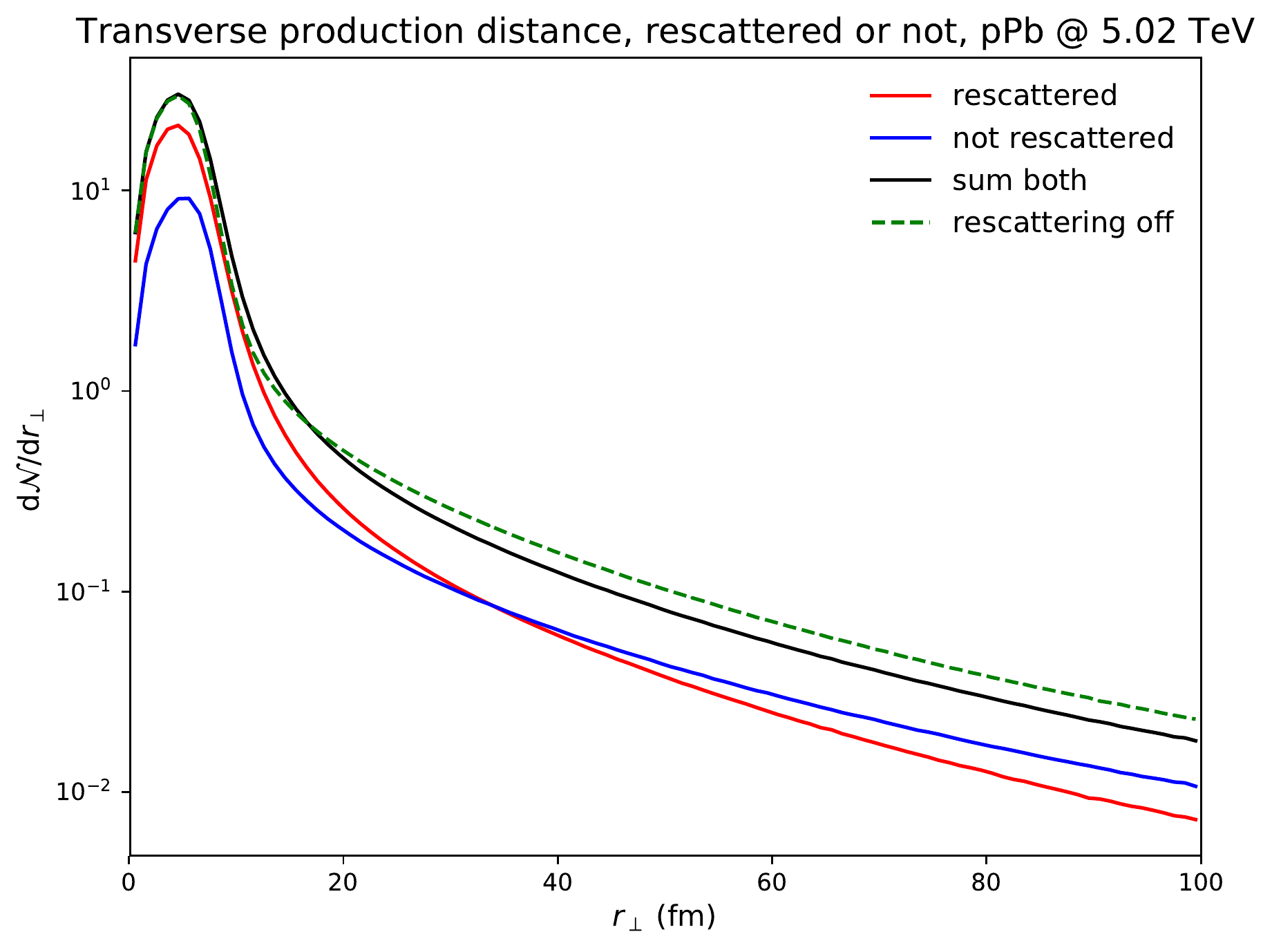}
\includegraphics[width=0.49\linewidth]{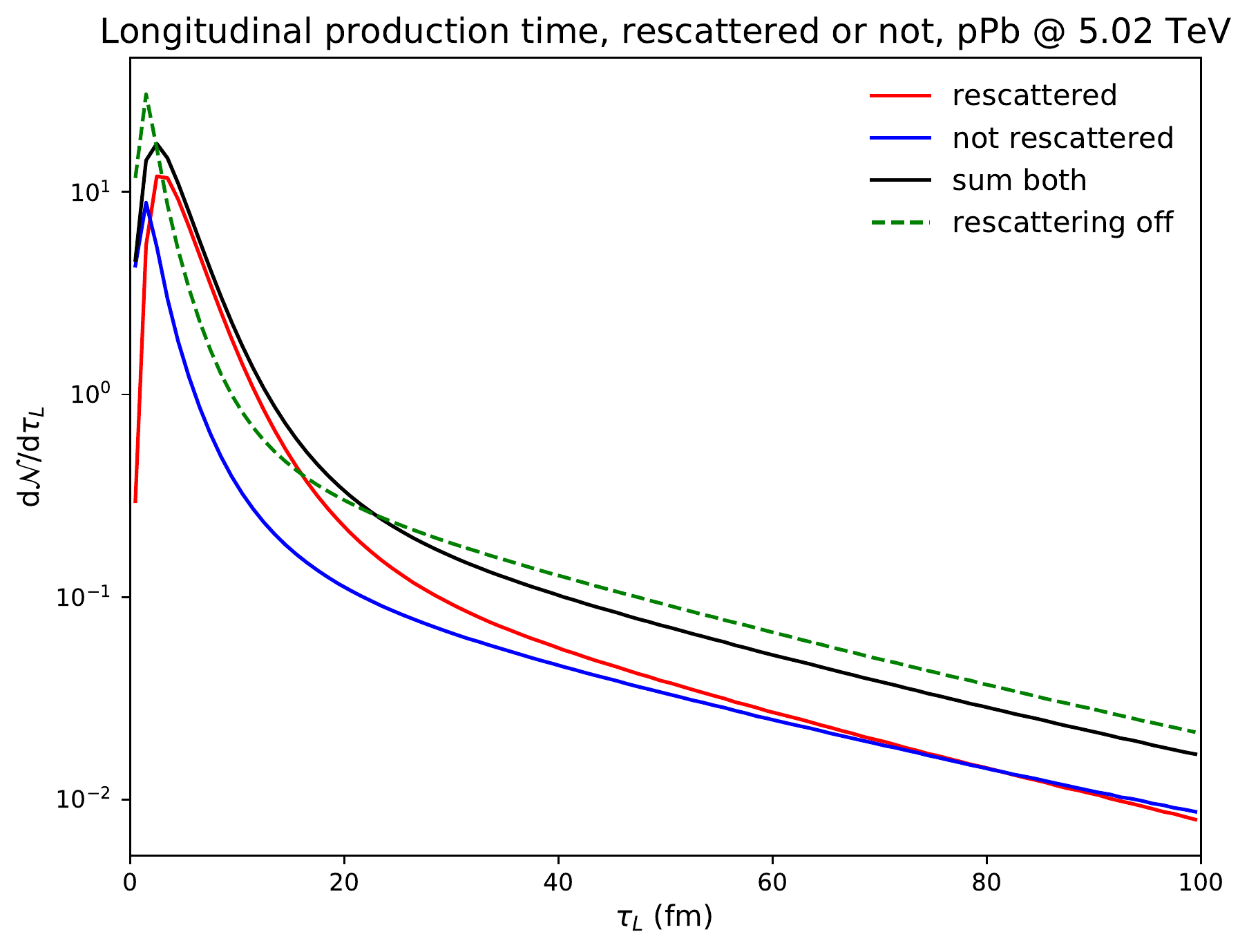}
\end{minipage}\\
\begin{minipage}[c]{\linewidth}
\centering
\includegraphics[width=0.49\linewidth]{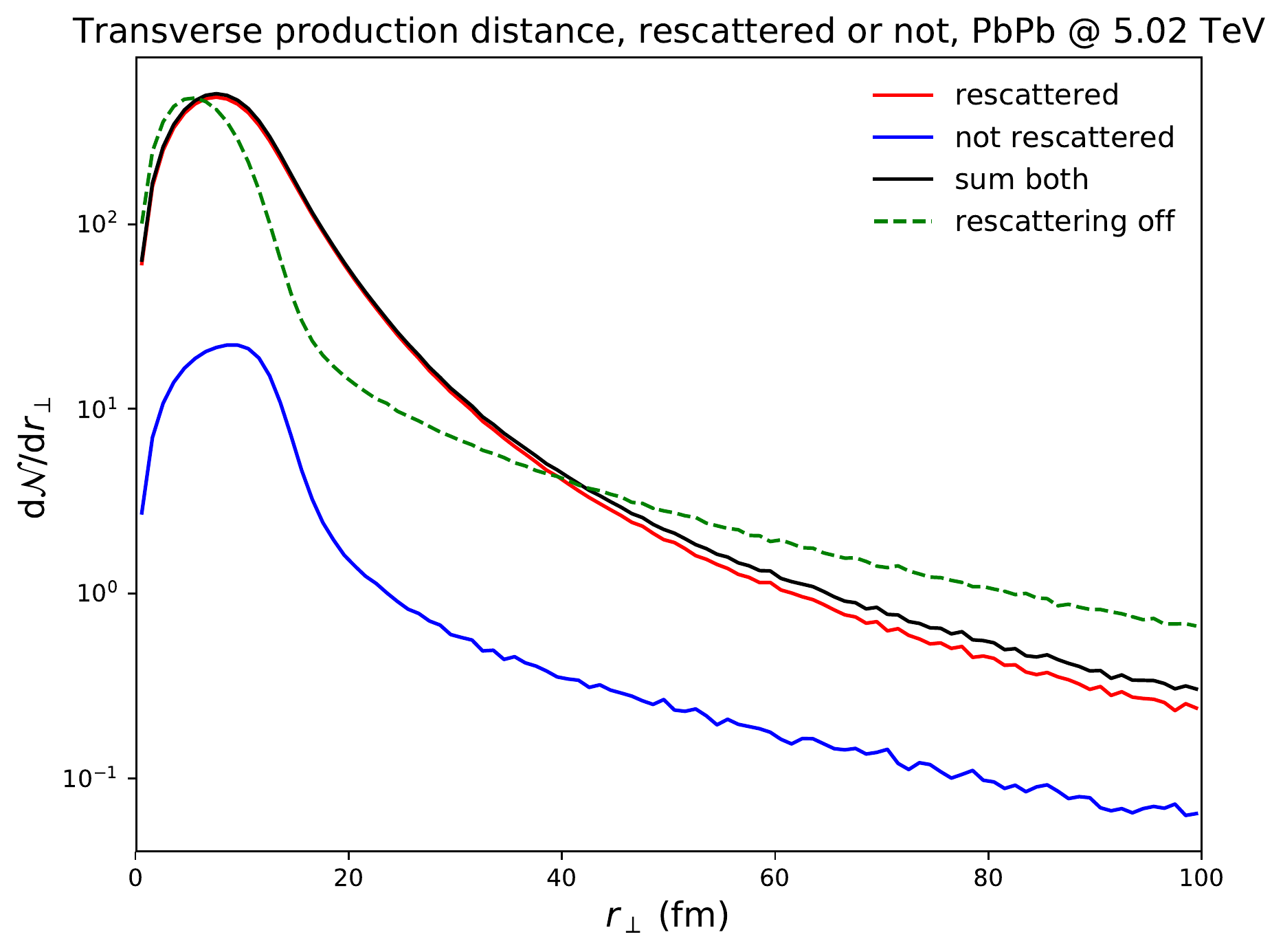}
\includegraphics[width=0.49\linewidth]{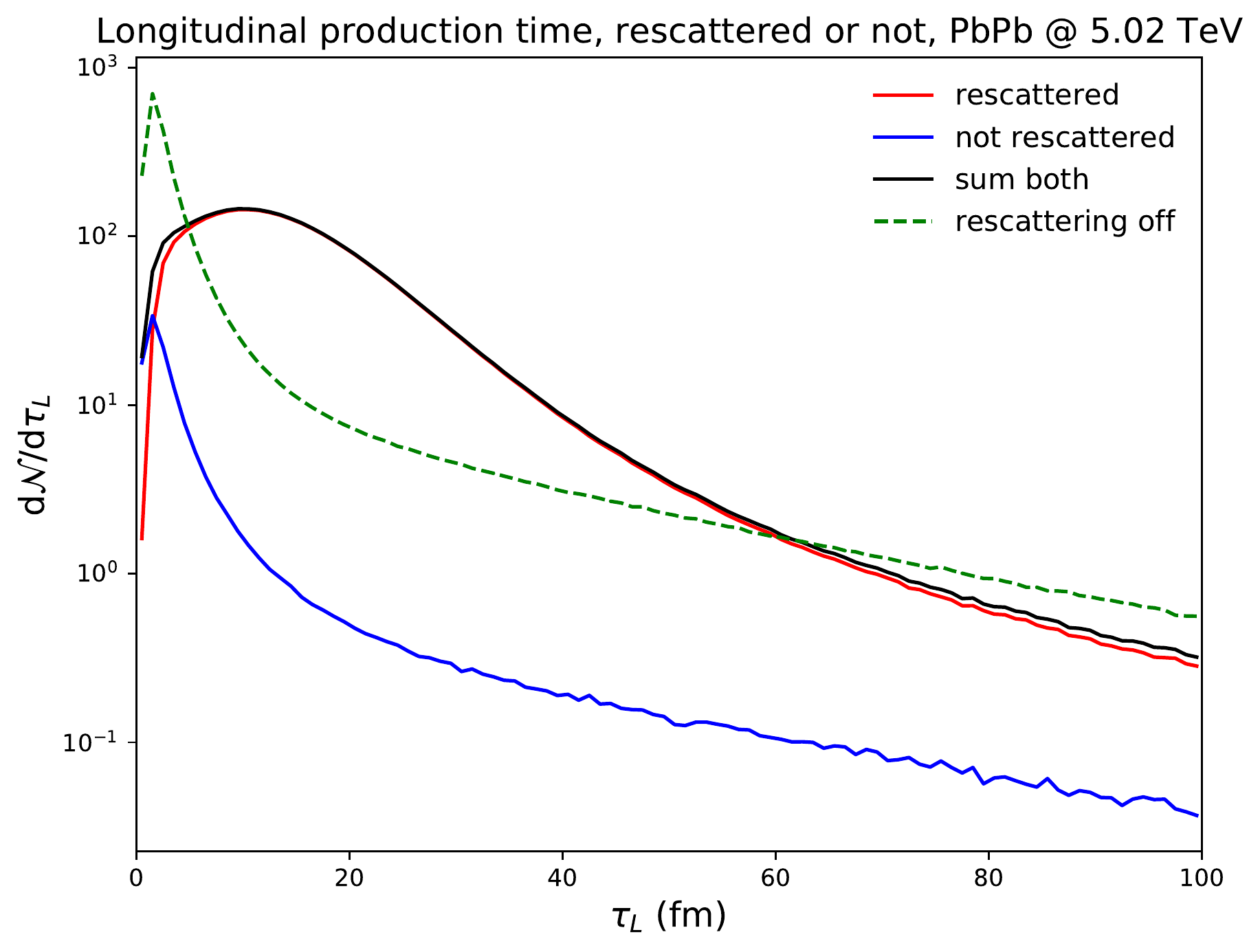}
\end{minipage}\\
\begin{minipage}[c]{0.49\linewidth}
\centering
(a)
\end{minipage}
\begin{minipage}[c]{0.49\linewidth}
\centering
(b)
\end{minipage}
\caption{(a) $r_\perp$ and (b) $\tau_L$ spectra. Note that rescattered also
refers to hadrons produced in decays of rescattered particles, even though
they themselves were not directly involved in rescattering.}
\label{fig:spacetime}
\end{figure}

In this section we study the spacetime distributions of rescatterings. 
Specifically, we consider the transverse production distance, 
$r_{\perp}^2 = x^2 + y^2$, and longitudinal invariant time, 
$\tau_L^2 = t^2 - z^2$. The two Lorentz-contracted ``pancake'' nuclei
are set to collide at $t = z = 0$, with the center of collision at
$x = y = 0$, but with sub-collisions spread all over the $(x, y)$ overlap
region. Thus the squared invariant time
$\tau^2 = t^2 - x^2 - y^2 - z^2 = \tau_L^2 - r_{\perp}^2$
tends to have a large tail out to large negative values,  
so it is not a suitable measure for heavy-ion collisions.
The $r_{\perp}$ and $\tau_L$ distributions are shown in
\figref{fig:spacetime}, separately for particles involved or not in
rescattering. For the latter it is the location of the last rescattering
that counts. Particle decays are included for particles with proper lifetimes 
$\tau_0 < 100$~fm, so that a ``final'' pion could be
bookkept at the decay vertex of for instance a $\rho$.

The overall observation is that rescattering reduces particle production at
very early and at late times, as is especially clear in the $\tau_L$
distribution for $\PbPb$. Particles produced at early times are
more likely to participate in rescattering and get assigned new
$\tau_L$ values on the way out. With this in mind, 
it may seem paradoxical that the $r_{\perp}$ distributions are comparably
broad for rescattered and unrescattered particles.
Hadrons produced in the periphery of the collision
are more likely to evade rescattering than central ones, however, so this
introduces a compensating bias towards larger $r_{\perp}$ for the
unrescattered. In this respect the $\tau_L$ distribution more follows the
expected pattern, with the unrescattered particles having comparable
average values in all three collision scenarios, whereas the rescattered
ones are shifted further out. Maybe somewhat unexpectedly, particle
production at late times and large $r_{\perp}$ is also reduced with rescattering
on. Our studies indicate that there is some rescattering activity
at late times ($\gtrsim 50$ fm), but the number of rescatterings here is
roughly a factor of three smaller than the number of decays. Now,
since rescattering produces more particles early, it tends to reduce the
average particle mass, which increases the number of stable particles
produced early and reduces the number of decaying ones in the $50 - 100$~fm
range. Furthermore, unstable particles often have lower $\pT$ and hence
smaller Lorentz factors, leading them to decay at lower $r_{\perp}$ values.

While the exact time of a rescattering cannot be measured directly, phenomena
such as resonance suppression can give an indication of the duration of the
hadronic phase \cite{Acharya:2020nyr,Acharya:2019qge}.
Experimentally, a suppression of the $\K^*/\K$ yield 
ratio at higher multiplicities has been observed, but not of the $\phi/\K$
yield ratio. The interpretation of this observation is as follows: after the
$\K^*$ decays, the outgoing $\pi$ and $\K$ are likely to participate in
rescattering because of their large cross sections, which 
disturbs their correlation and suppresses the original $\K^*$ signal. The fact
that the $\phi$ signal is not suppressed in this way indicates that they tend
to decay only after most rescattering has taken place. With the $\K^*$ and 
$\phi$ lifetimes being 3.9~fm and 46.3~fm respectively, this places bounds on
the duration of the rescattering phase. These bounds seem to be consistent 
with the spacetime distributions shown in \figref{fig:spacetime}.

With the full event history provided by \textsc{Pythia}, it is possible to
study the actual number of $\K^*$ and $\phi$ that were produced, and to trace
what happens to their decay products. A na\"ive way to approach resonance
suppression is to define a $\K^*$ or $\phi$ meson as detectable if it 
decayed and no decay product participated in rescattering. When defining
the $\K^*$ multiplicity in this way, we found that rescattering actually
increases the $\K^*/\K$ ratio for larger charged multiplicities.
This increase is not observable,
however, since it mainly comes from $\K\pi \to \K^* \to \K\pi$. That is,
some of the combinatorial background gets to be reclassified as $\K^*$,
without any change of the overall $\K\pi$ mass spectrum. To find the more
subtle effects of nontrivial processes requires a detailed fitting of the
$\K\pi$ mass spectrum. This is outside the scope of this article, but would
be interesting to study in the future. Nevertheless, the change in the
$\phi/\K$ ratio is much smaller, suggesting that qualitatively,
longer-lived resonances are indeed less affected by rescattering.

\subsection{Centrality dependent observables}
\label{sec:cent-dep-obs}

In heavy ion experiments, observables are most often characterized according to
collision centrality. The characterization is a sensible one, also for checking the
effects of hadronic rescatterings, as this will be the largest in the most central
collisions. While experiments employ a centrality definition depending on particle 
production in the forward or central regions of the experiments, we will in the 
following sections use the definition adhering to impact parameter. As such, the
centrality of a single collision is defined as
\begin{equation}
	\label{eq:cent-def}
	c = \frac{1}{\sigma_{\mathrm{inel}}}\int_0^b \d b' \frac{\d \sigma_{\mathrm{inel}}}{\d b'}.
\end{equation}

We note, however, that the results presented for nucleus-nucleus collisions can 
be transferred directly to experimental centrality measures, as the \textsc{Angantyr} model
provides a good description of \eg~forward energy, which correlates directly with
the theoretical impact parameter.

\subsubsection{Particle yields and ratios}

\begin{figure}[t!]
\begin{minipage}[c]{0.49\linewidth}
\centering
\includegraphics[width=\linewidth]{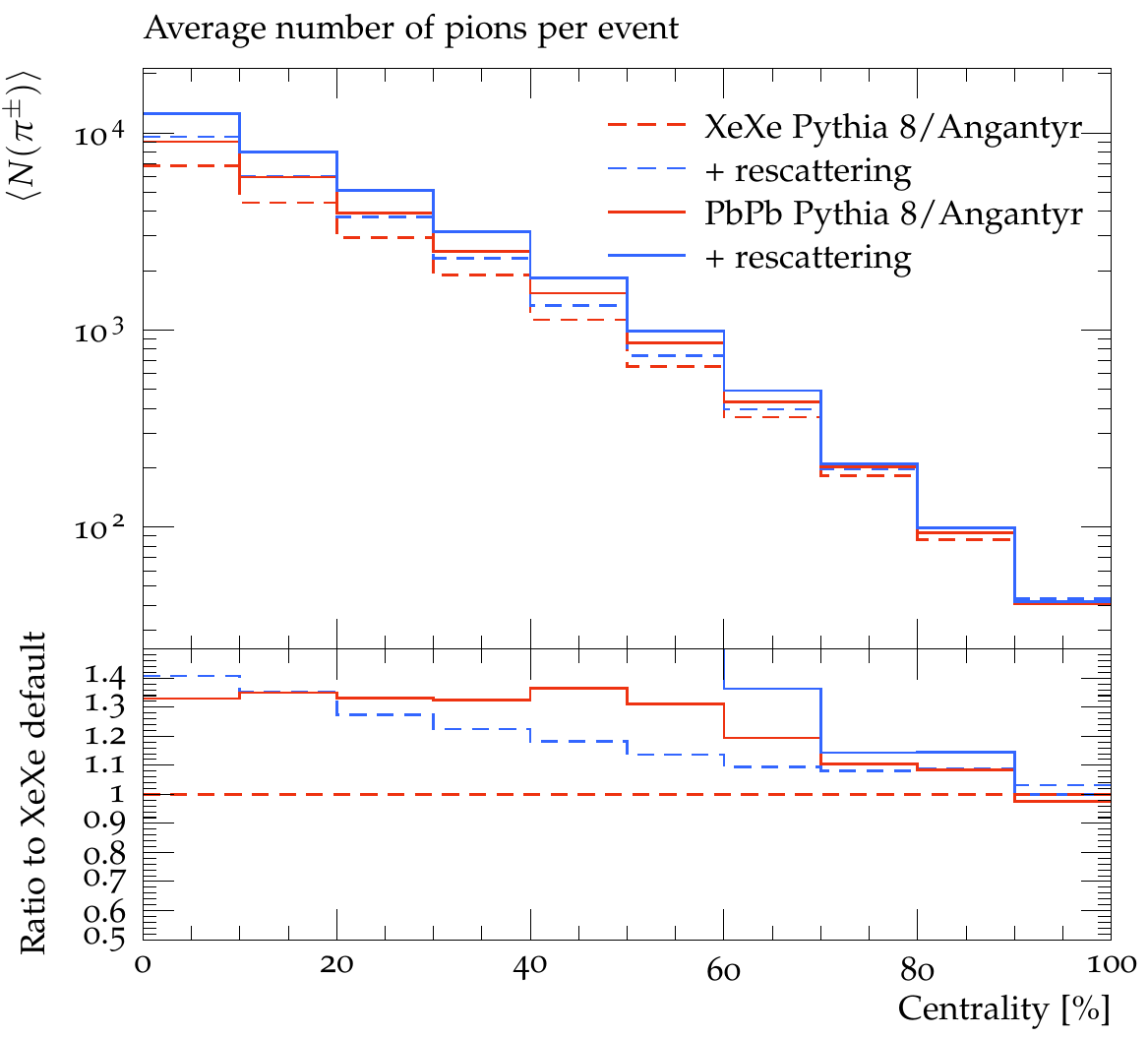}\\
(a)
\end{minipage}
\begin{minipage}[c]{0.49\linewidth}
\centering
\includegraphics[width=\linewidth]{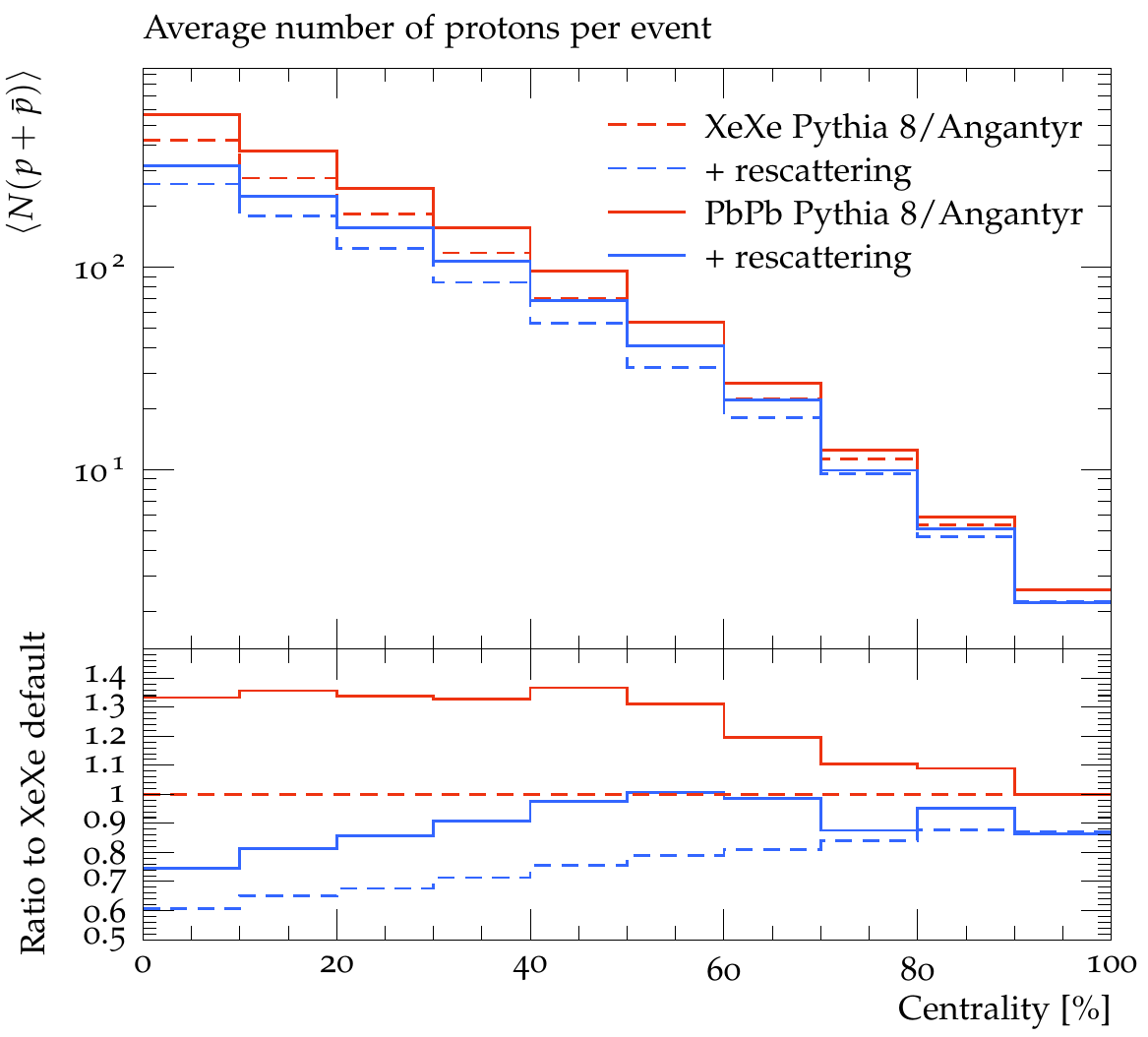}\\
(b)
\end{minipage}\\
  \caption{Average per-event yields of (a) pions $(\pi^\pm)$ and (b) protons ($\p,\pbar$)
  in $\PbPb$ and $\XeXe$ collisions at $\sqrt{s_{\N\N}} = $ 2.74 and 5.44 TeV
  respectively, as function of collision centrality.}
\label{fig:pip-yields}
\end{figure}

In the following we present the effect on identified particle yields in $|y| < 4$
(to avoid the beam region) in $\XeXe$ collisions at $\sqrt{s_{\N\N}} = 5.44$ TeV 
and $\PbPb$ collisions at $\sqrt{s_{\N\N}} = 2.76$ TeV.
Starting with light flavour mesons and baryons, we show
the average multiplicity of (a) pions ($\pi^\pm$) and (b) protons ($\p,\pbar$)
per event in \figref{fig:pip-yields} and respectively.

While the effect for pions is negligible in peripheral collisions, it grows to about 40\% in central collisions.
The effect on protons is also largest in central collisions, while in peripheral collisions it is still at a 10\%
level. This is particularly interesting in the context of recent years' introduction of microscopic models to explain
the increase of strange baryon yields with increasing multiplicity, which overestimate the amount of protons \cite{Bierlich:2015rha}.

\begin{figure}[t!]
\begin{minipage}[c]{0.49\linewidth}
\centering
\includegraphics[width=\linewidth]{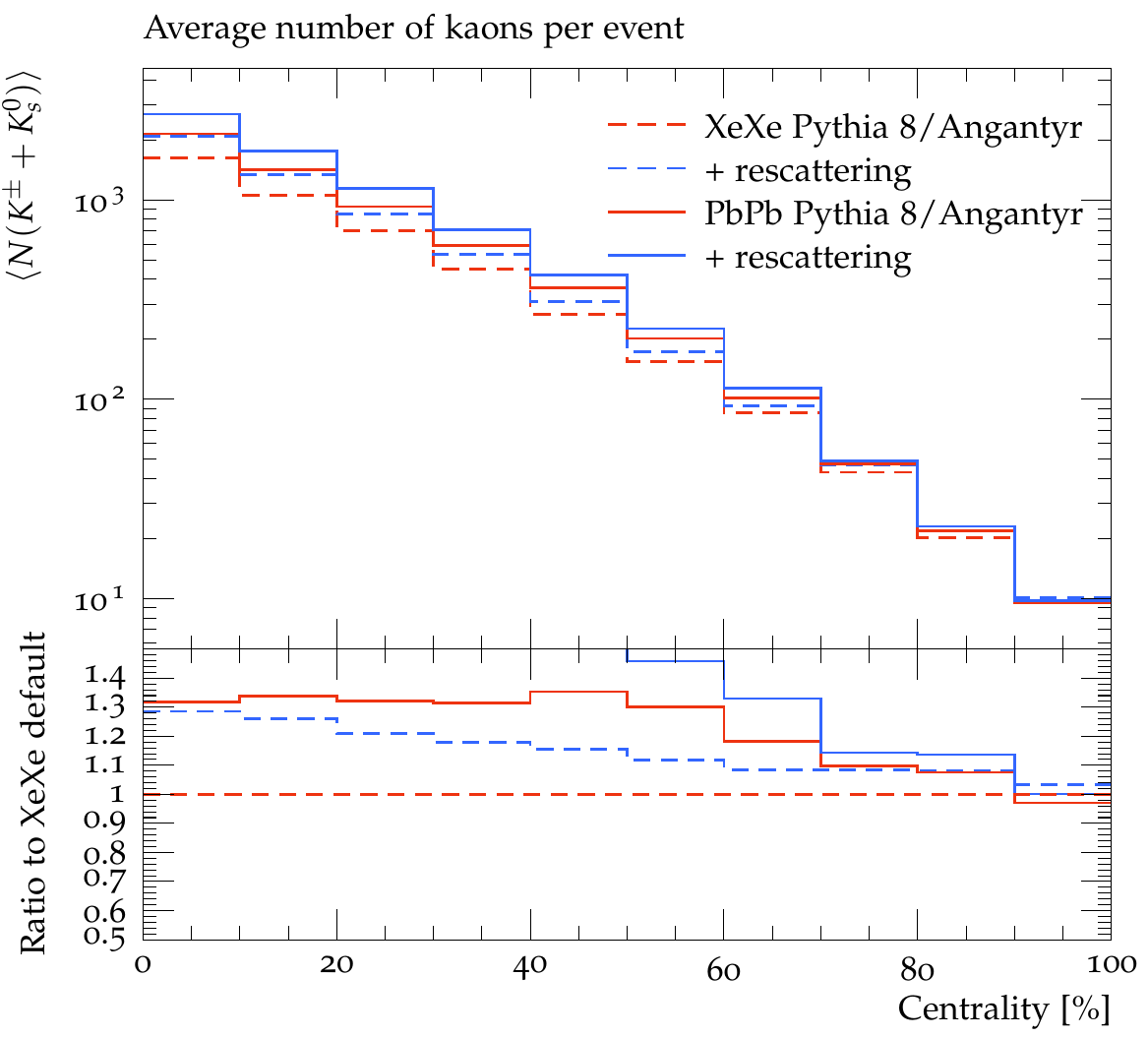}\\
(a)
\end{minipage}
\begin{minipage}[c]{0.49\linewidth}
\centering
\includegraphics[width=\linewidth]{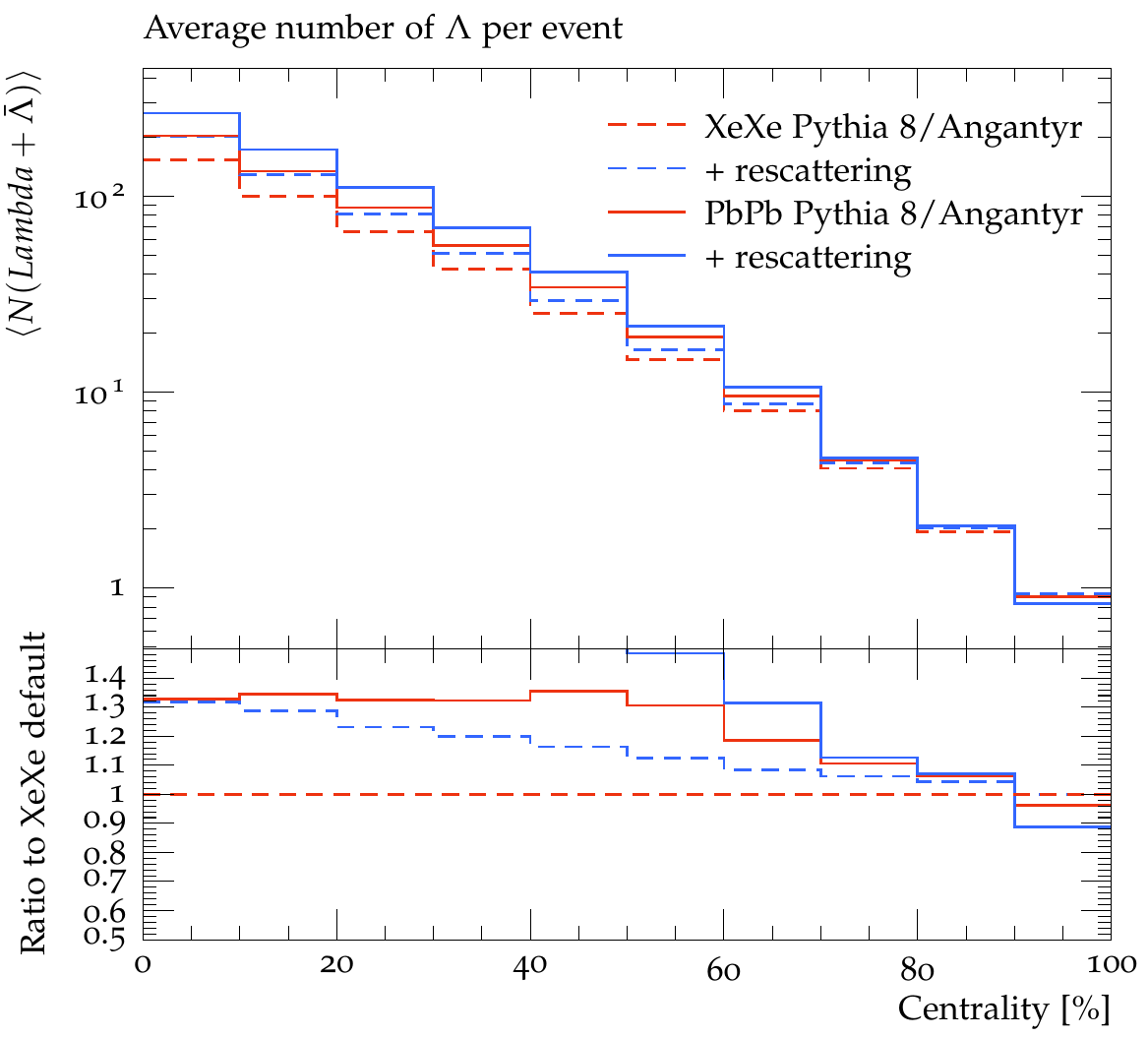}\\
(b)
\end{minipage}\\
  \caption{Average per-event yields of (a) kaons $(\K^\pm, \K^0_{\mathrm{L,S}})$
  and (b) $\Lambda$ ($\Lambda,\overline{\Lambda}$) in $\PbPb$ and $\XeXe$
  collisions at $\sqrt{s_{\N\N}} = $ 2.74 and 5.44 TeV respectively, as function
  of collision centrality.}
\label{fig:kl-yields}
\end{figure}

In \figref{fig:kl-yields} we move to strange mesons and baryons, with the total kaon
($\K^\pm$ and $\K^0_{\mathrm{L,S}}$)
and $\Lambda$ multiplicity, (a) and (b) respectively. While there is a large effect on the direct yields of both
species, it is almost identical to the change in $\pi^\pm$ in \figref{fig:pip-yields}a, leaving the $\K/\pi$ 
and $\Lambda/\pi$ ratios unchanged. 

\begin{figure}[t!]
\begin{minipage}[c]{0.49\linewidth}
\centering
\includegraphics[width=\linewidth]{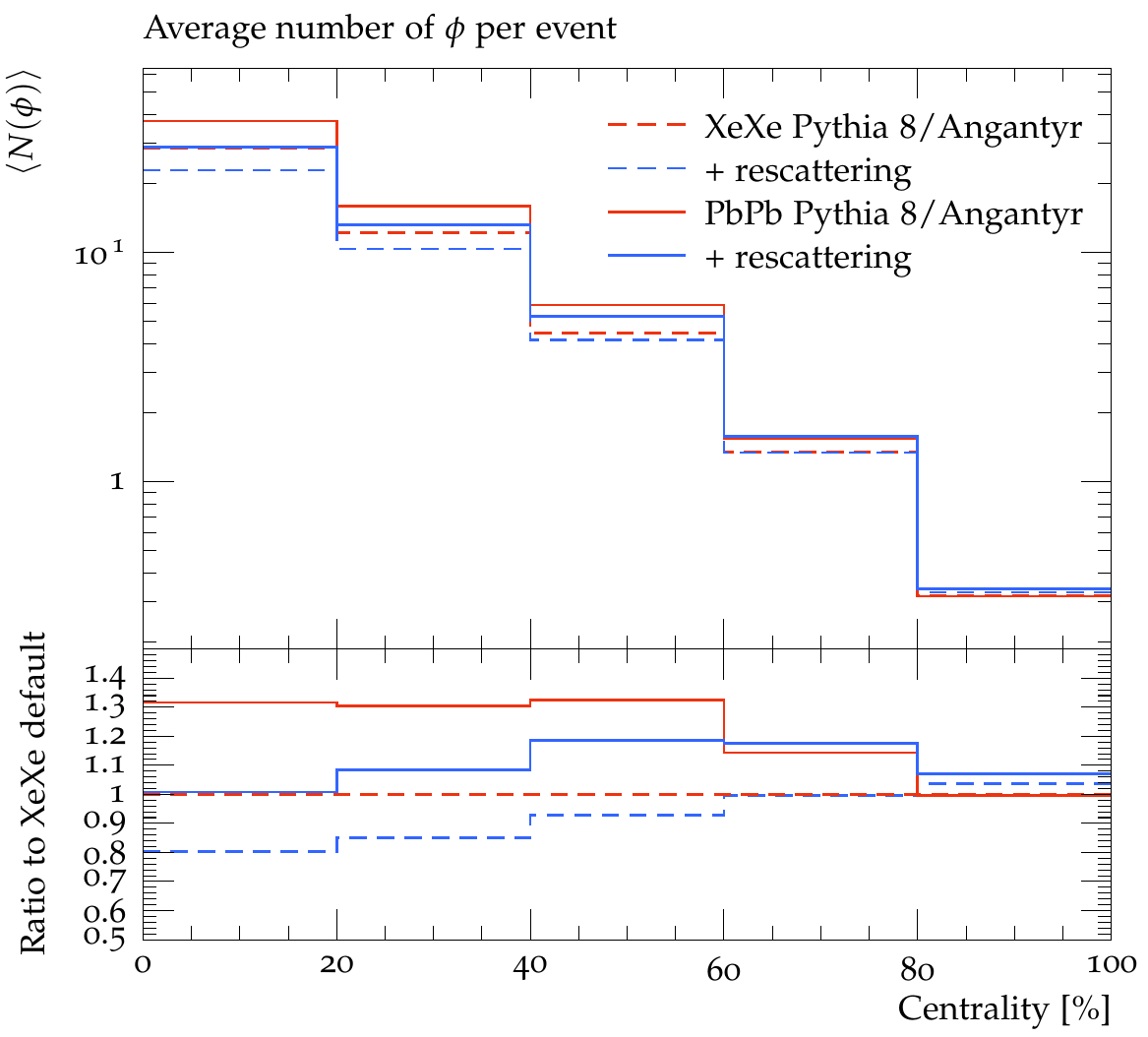}\\
(a)
\end{minipage}
\begin{minipage}[c]{0.49\linewidth}
\centering
\includegraphics[width=\linewidth]{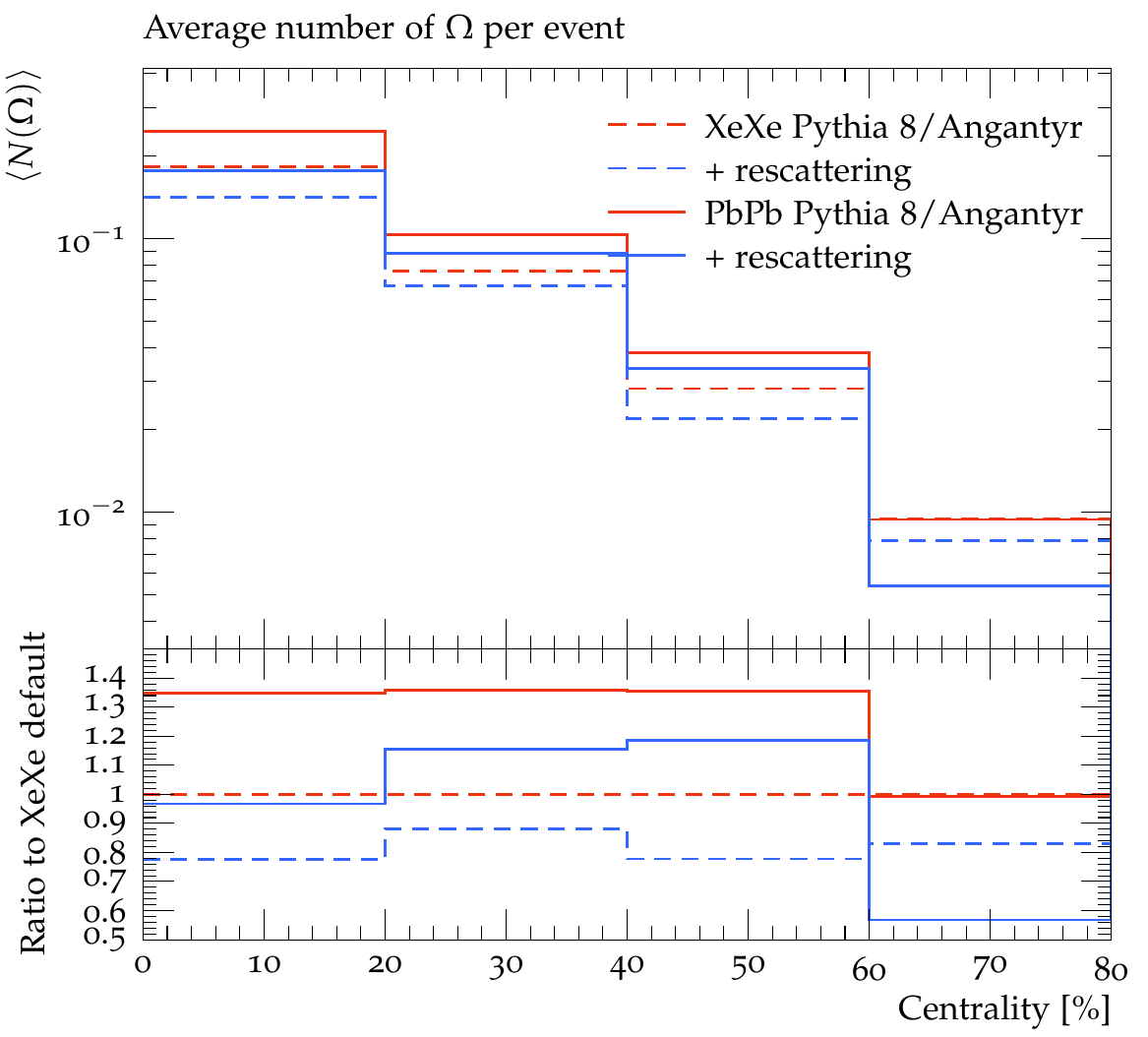}\\
(b)
\end{minipage}\\
	\caption{Average per-event yields of (a) $\phi$ and (b) $\Omega^{-}$ in $\PbPb$
	and $\XeXe$ collisions at $\sqrt{s_{\N\N}} = $ 2.74 and 5.44 TeV respectively, as function of collision centrality.}
\label{fig:phi-omega-yields}
\end{figure}

We finish the investigation of the light-flavour sector by showing the total
$\phi$ and $\Omega^{-}$ multiplicities in \figref{fig:phi-omega-yields}a 
and b respectively. The $\phi$ multiplicity decreases by about 20\% in central
events and is constant within the statistical errors in peripheral. The $\Omega$
multiplicity is decreased roughly the same amount. The decrease here, however,
is rather constant in centrality in $\XeXe$ but increases for central events in $\PbPb$.

\begin{figure}[t!]
\begin{minipage}[c]{0.49\linewidth}
\centering
\includegraphics[width=\linewidth]{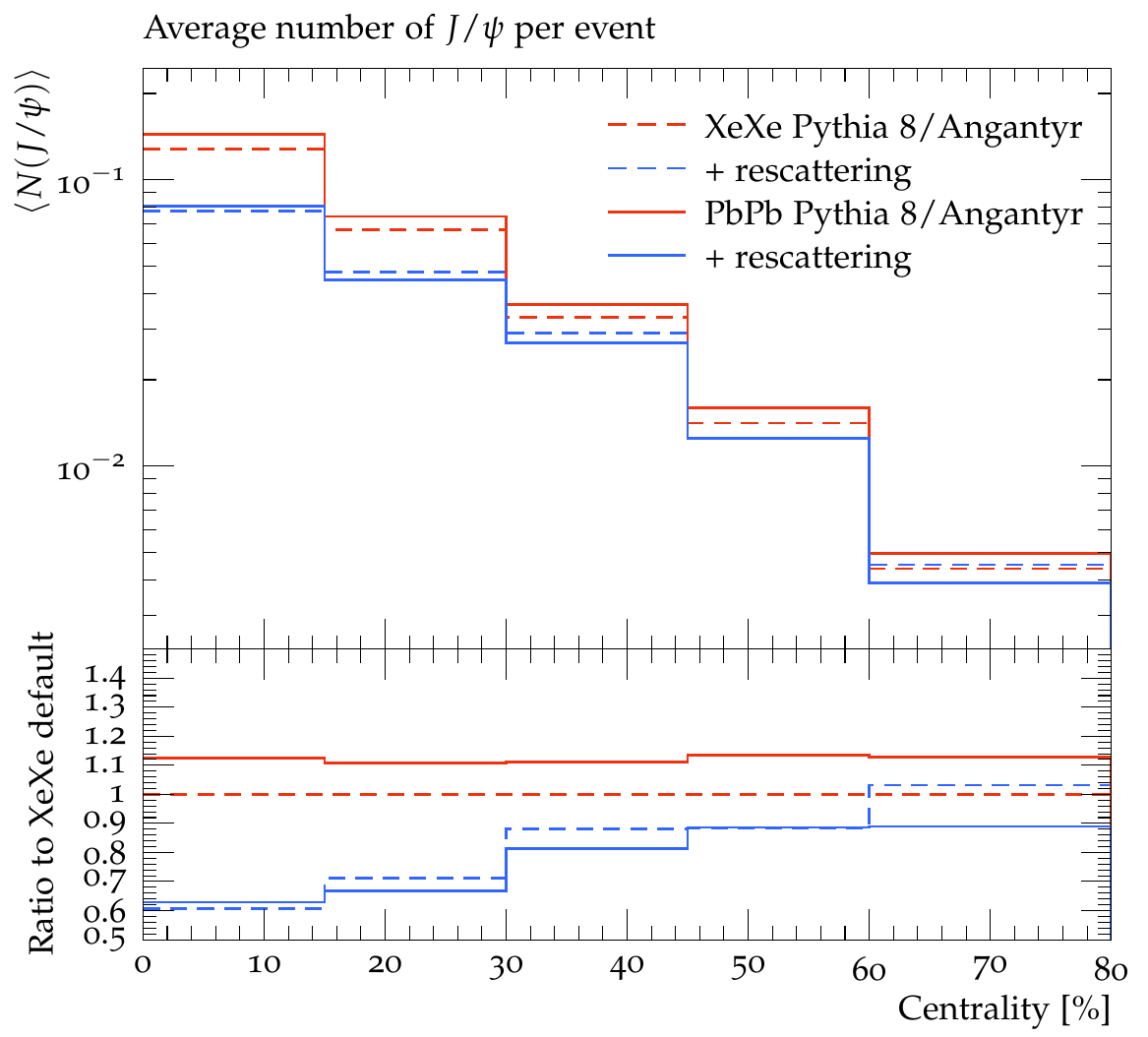}\\
(a)
\end{minipage}
\begin{minipage}[c]{0.49\linewidth}
\centering
\includegraphics[width=\linewidth]{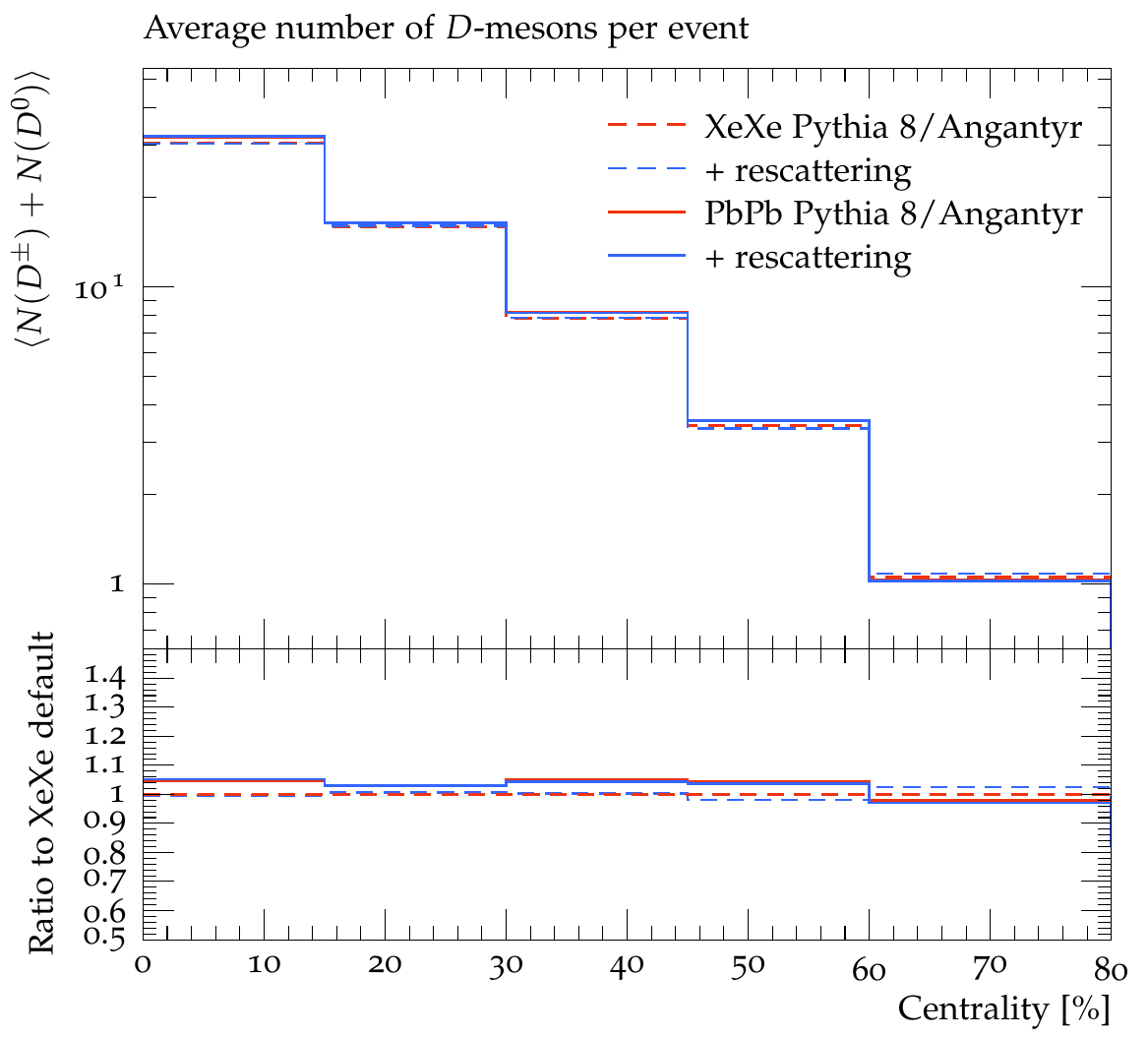}\\
(b)
\end{minipage}\\
\caption{Average per-event yields of (a) $\J/\psi$ and (b) $\D$ mesons in $\PbPb$
and $\XeXe$ collisions at $\sqrt{s_{\N\N}} = $ 2.74 and 5.44 
TeV respectively, as function of collision centrality.}
\label{fig:jpsi-d-yields}
\end{figure}

Notably (and as opposed to \eg UrQMD), the rescattering framework implemented in \textsc{Pythia},
includes cross sections for heavy flavour mesons and baryons. In \figref{fig:jpsi-d-yields}
we show the effect on (a) $\J/\psi$ and (b) $\D$ mesons ($\D^\pm$ and $\D^0$). Starting with
the $\J/\psi$ we see a significant effect in both collision systems in central events, less so 
in peripheral. While the initial $\J/\psi$ yield is roughly 10\% larger in $\PbPb$ than in $\XeXe$,
the final value after rescattering saturates at a value at roughly 60\% of the initial $\XeXe$ value,
independent of the two collision systems\footnote{This feature is clearly accidental. We have checked in smaller
collision systems to confirm.}. Whether or not this is consistent with the measured nuclear modification
factor \cite{Acharya:2019lkh} in peripheral collisions (clearly not in central collisions, where an additional
source of $\J/\psi$ production would be required) is left for future detailed comparisons to data.

In \textsc{Pythia} (rescattering or not) there is no mechanism for charm quarks to vanish from the event at early times. The
constituents of the $\J/\psi$ would therefore have to end up in other charmed hadrons. In \figref{fig:jpsi-d-yields}b 
we show the $\D$ meson yield, demonstrating that this is more than two orders of magnitude above the $\J/\psi$ one.
It is then consistent to assume that the missing charm quarks can recombine into open charm without having observable consequences. Indeed
there is no significant effect on the $\D$ meson yield from rescattering.

\subsubsection{Elliptic flow}

One of the most common ways to characterize heavy ion collisions is by the measurement of
flow coefficients ($v_n$'s), defined as the coefficients of a Fourier expansion of the single particle
azimuthal yield, with respect to the event plane $\Psi_n$ \cite{Voloshin:1994mz,Poskanzer:1998yz}:
\begin{equation}
	\label{eq:flow-def}
	E\frac{\d^3 N}{\d^3p } = \frac{1}{2\pi} \,\frac{\d^2 N}{\pT \d \pT \d y} 
	\left( 1 + 2\sum_{n=1}^\infty v_n \cos(n(\varphi - \Psi_n)) \right).
\end{equation}
The azimuthal angle is denoted $\varphi$, and $E$, $\pT$ and $y$ are the particle energy, transverse
momentum and rapidity respectively. In experiments it is not possible to utilize this definition 
directly, as the event plane is unknown. Therefore one must resort to other methods. For the purpose
of testing if a model behaves as expected, it is on the other hand preferable to measure how much
or little particles will correlate with the true event plane (when we show comparisons to experimentally
obtained values in \secref{sec:FlowCoefficients}, we will use the experimental definitions). In the following, we 
will therefore use an event plane obtained from the initial state model, defined as
\begin{equation}
	\Psi_n = \frac{1}{n} \arctan\left(\frac{\langle r^2 \sin(n\varphi)\rangle}{\langle r^2 \cos(n\varphi) \rangle}\right) + \frac{\pi}{n},
\end{equation}
for all initial state nucleons participating in collisions contributing to the final state multiplicity 
(inelastic, non-diffractive sub-collisions). The origin is shifted to the center of the sampled distribution of
nucleons, and $r$ and $\varphi$ are the usual polar coordinates. Flow coefficients can then simply
be calculated as
\begin{equation}
	\label{eq:vn-th}
	v_n = \langle \cos(n(\varphi - \Psi_n))\rangle.
\end{equation}
As in the previous section we consider all particles in $|y| < 4$ and without any lower cut on transverse momentum.

\begin{figure}[t!]
\begin{minipage}[c]{0.49\linewidth}
\centering
\includegraphics[width=\linewidth]{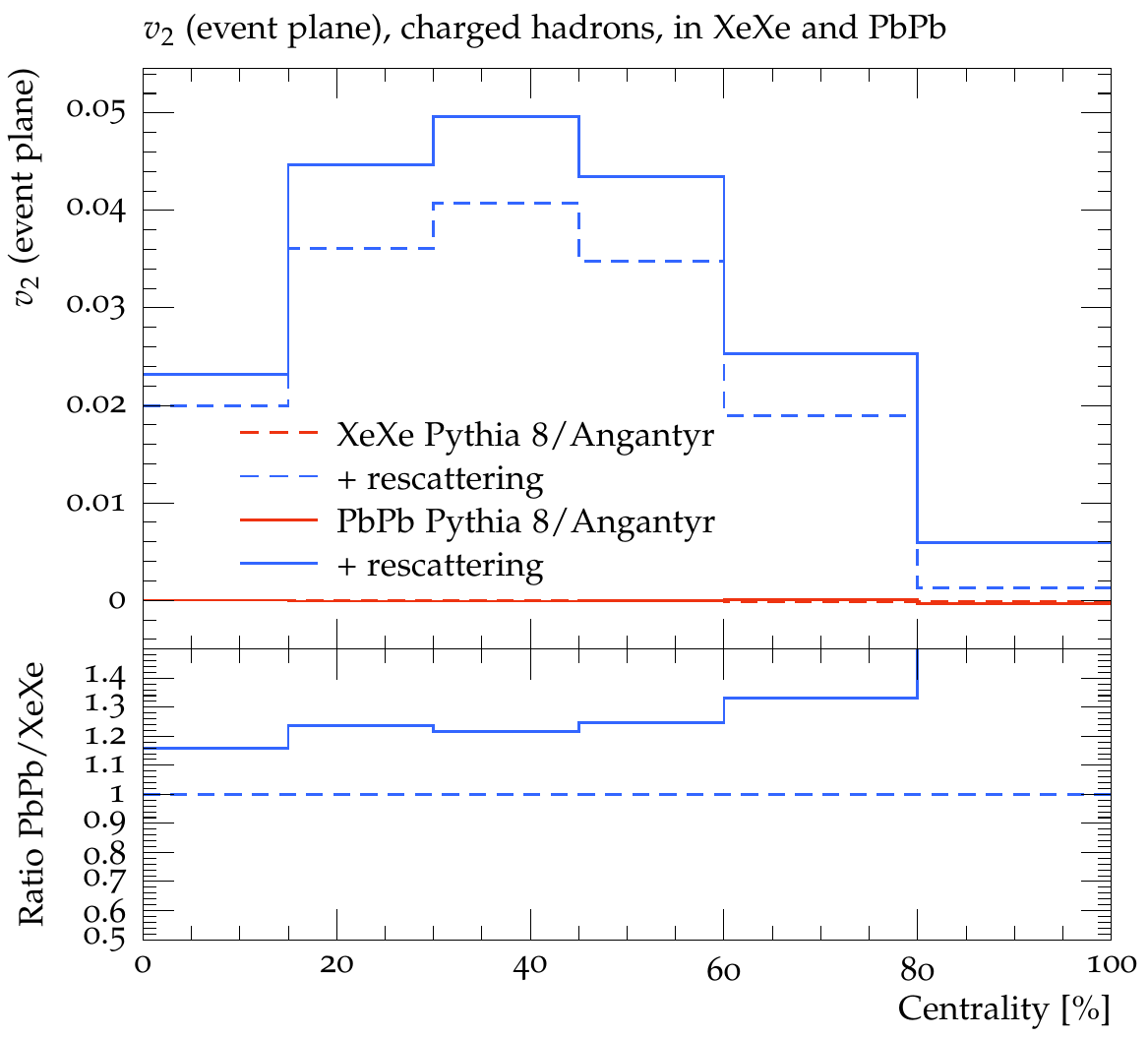}\\
(a)
\end{minipage}
\begin{minipage}[c]{0.49\linewidth}
\centering
\includegraphics[width=\linewidth]{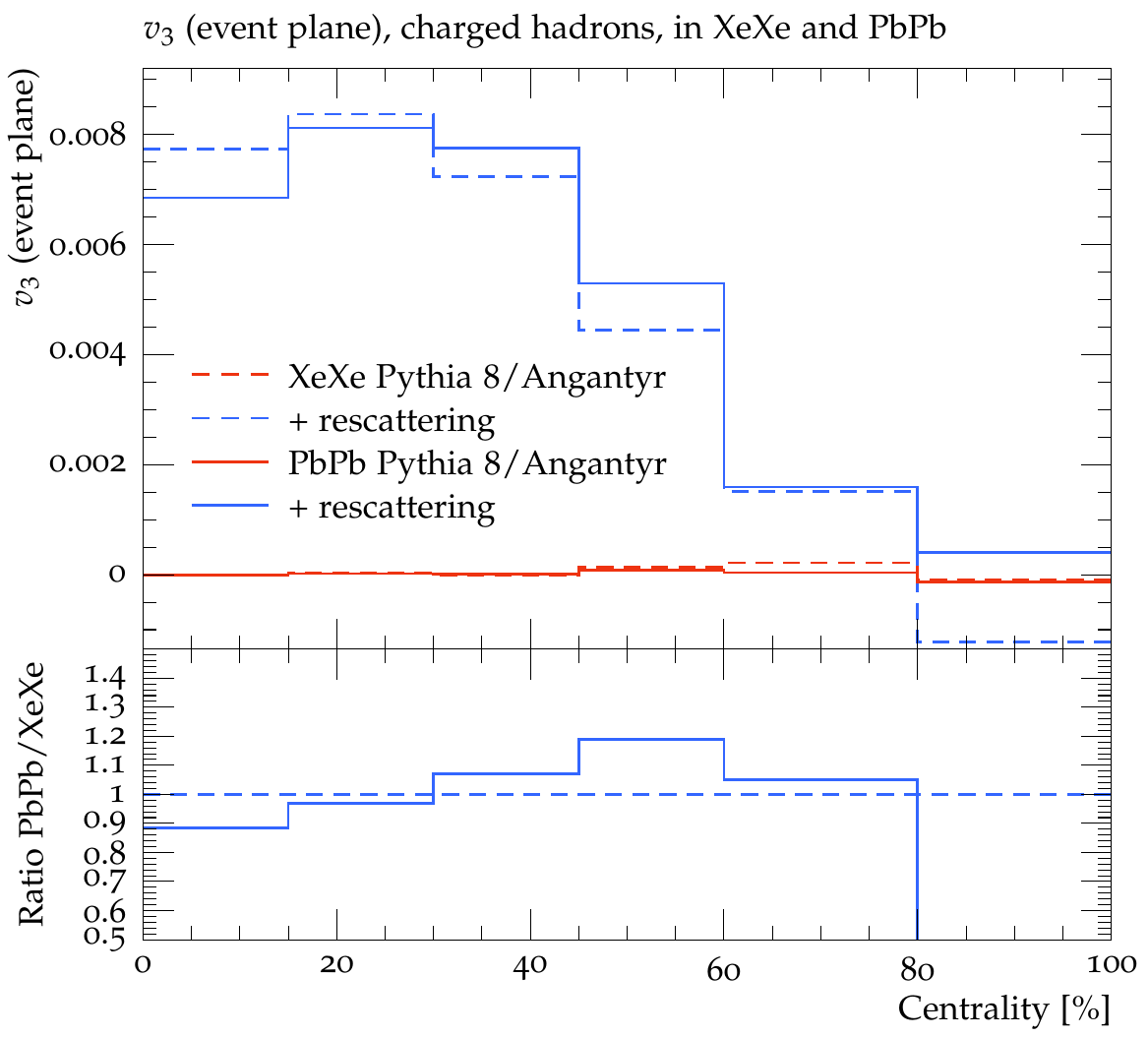}\\
(b)
\end{minipage}\\
	\caption{Flow coefficients (a) $v_2$ and (b) $v_3$ in $\PbPb$ and $\XeXe$ collisions at $\sqrt{s_{\N\N}} = $ 2.74 and 5.44 
TeV respectively. Results shown with and without rescattering, and are calculated with respect to the event plane such that the sample without rescattering
	is zero by construction.}
\label{fig:v2-nch}
\end{figure}

In \figref{fig:v2-nch} we show (a) $v_2$ and (b) $v_3$ as functions of collision
centrality for charged particles for $\XeXe$ collisions 
at $\sqrt{s_{\N\N}}=5.44$ TeV and $\PbPb$ collisions at $\sqrt{s_{\N\N}} = 2.74$ 
TeV both with and without rescattering. It is seen that $v_2$ receives a sizeable contribution from rescattering. The contribution
is larger for $\PbPb$ than for $\XeXe$, which is not surprising, given the larger density. The $v_2$ arises because particles are pushed
by rescatterings along the density gradient, which is larger along the event plane. Note that the curve without rescattering is zero, as the definition 
of $v_n$ from \eqref{eq:vn-th} ensures that no non-flow contributions enter the results.

For $v_3$ (\figref{fig:v2-nch}b) there is not much difference between $\PbPb$ and $\XeXe$. Since $v_3$ is mainly generated by initial state
shape fluctuations, this is a reasonable result.

Since different hadron species have different cross sections, hadronic rescattering will yield different flow coefficients for 
different hadron species. As an example, since the $\p\p (\pbar\p)$ cross section is larger than the average hadron-hadron cross section
(which is dominated mainly by pions), $v_2$ for protons will be higher. We note (without showing) that 
hadronic rescattering gives $v_2(\p) > v_2(\pi) > v_2(\K) \approx v_2(\Lambda) > v_2(\Omega) > v_2(\phi)$, with the
latter reaching its maximum for $v_2$ about an order of magnitude less than for protons.

For heavy flavours, the results require more explanation, due to the differing production mechanisms. In \textsc{Pythia},
$\D$ mesons are produced in string fragmentation, requiring that one of the quark ends is a charm quark. The $\J/\psi$,
on the other hand, is predominantly produced early, either by direct onium production via colour-singlet and colour-octet
mechanisms, or by an early ``collapse'' of a small $\c\cbar$ string to a $\J/\psi$. Onia are therefore
excellent candidates for hadrons mainly affected by hadronic rescattering, and not any effects of strings interacting with
each other before hadronization.

\begin{figure}[t!]
\begin{minipage}[c]{0.49\linewidth}
\centering
\includegraphics[width=\linewidth]{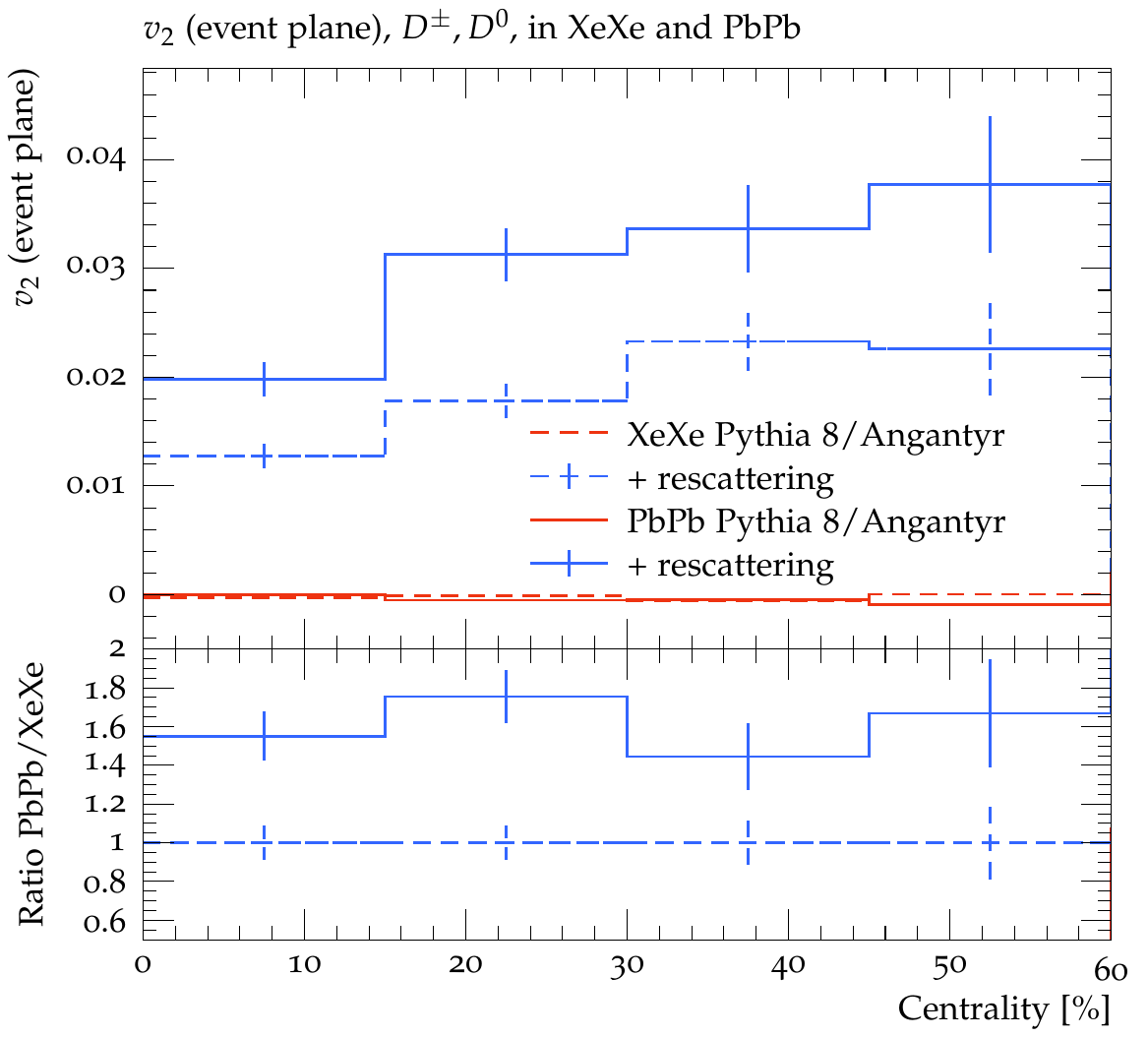}\\
(a)
\end{minipage}
\begin{minipage}[c]{0.49\linewidth}
\centering
\includegraphics[width=\linewidth]{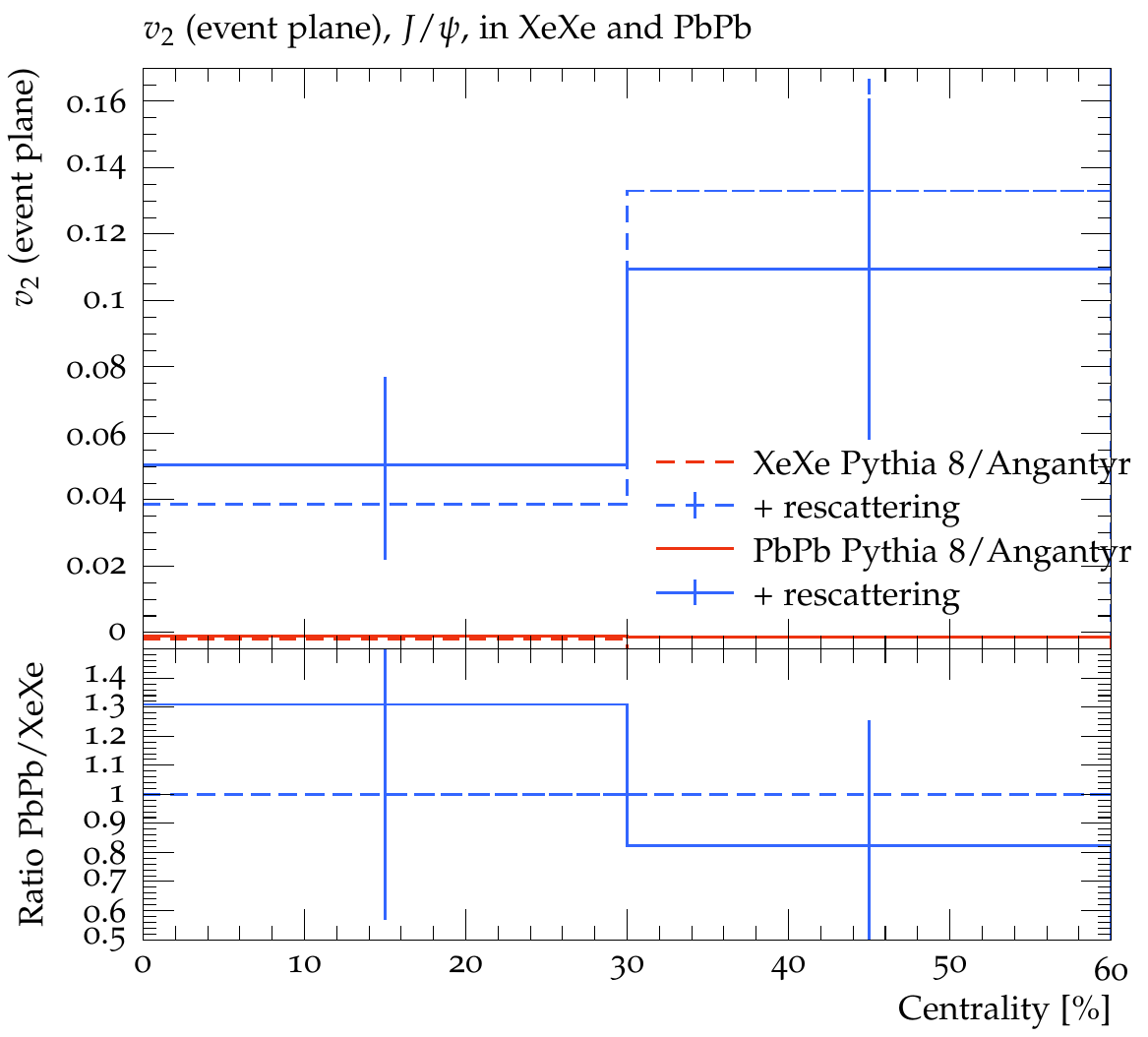}\\
(b)
\end{minipage}\\
	\caption{The $v_2$ flow coefficient for (a) $\D$ mesons ($\D^\pm, \D^0$) and (b) $\J/\psi$ as a function of centrality in 
	$\PbPb$ and $\XeXe$ collisions at $\sqrt{s_{\N\N}} = $ 2.74 and 5.44 TeV respectively. Error bars are statistical 
	errors. Results shown with and without rescattering, and are calculated with respect to the event plane such that the sample without 
	rescattering is zero by construction.}
\label{fig:v2-d-and-j}
\end{figure}

In \figref{fig:v2-d-and-j}, we show $v_2$ for (a) $\D$ mesons and (b) $\J/\psi$. Starting with $\D$ mesons we see an
appreciable $v_2$, numerically not too far from $\PbPb$ data \cite{Abelev:2014ipa}. A clear difference is observed between
$\XeXe$ and $\PbPb$. In the figure, statistical error bars are shown, as they are not negligible due long processing times
for heavy flavour hadrons.
For the $\J/\psi$, shown in \figref{fig:v2-d-and-j}b, $v_2$ for $\PbPb$ and $\XeXe$ are compatible within the statistical error.
More importantly, the result is also compatible with experimental data \cite{ALICE:2013xna}. Together with the result from 
\figref{fig:jpsi-d-yields}a, which suggests a sizeable nuclear modification to the $\J/\psi$ yield from rescattering, a detailed
comparison with available experimental data should be performed. It should be noted that the treatment of charm in the \textsc{Pythia} 
hadronic rescattering model follows the additive quark model, as introduced earlier. Thus, no distinction is made between $\J/\psi$ and
other $\c\cbar$ states. A foreseen improvement of this treatment would be to consider differences with input taken \eg~from lattice calculations.

\section{Comparison with data}
\label{sec:comparisonWithData}
In this section we go beyond the model performance plots shown in the previous section, and compare
to relevant experimental data for $\XeXe$ and $\PbPb$, in cases where Rivet \cite{Bierlich:2019rhm} implementations of the
experimental analysis procedure are available (though not in all cases validated by experiments).
We focus on observables where the rescattering effects are large, and in some cases surprising.

In all cases centrality is defined according to \eqref{eq:cent-def}, as it reduces computation time, and
the difference between centrality defined by impact parameter and forward energy flow is not large in $\A\A$ collisions.

\subsection{Charged multiplicity}

\begin{figure}
\begin{minipage}[c]{0.49\linewidth}
\centering
\includegraphics[width=\linewidth]{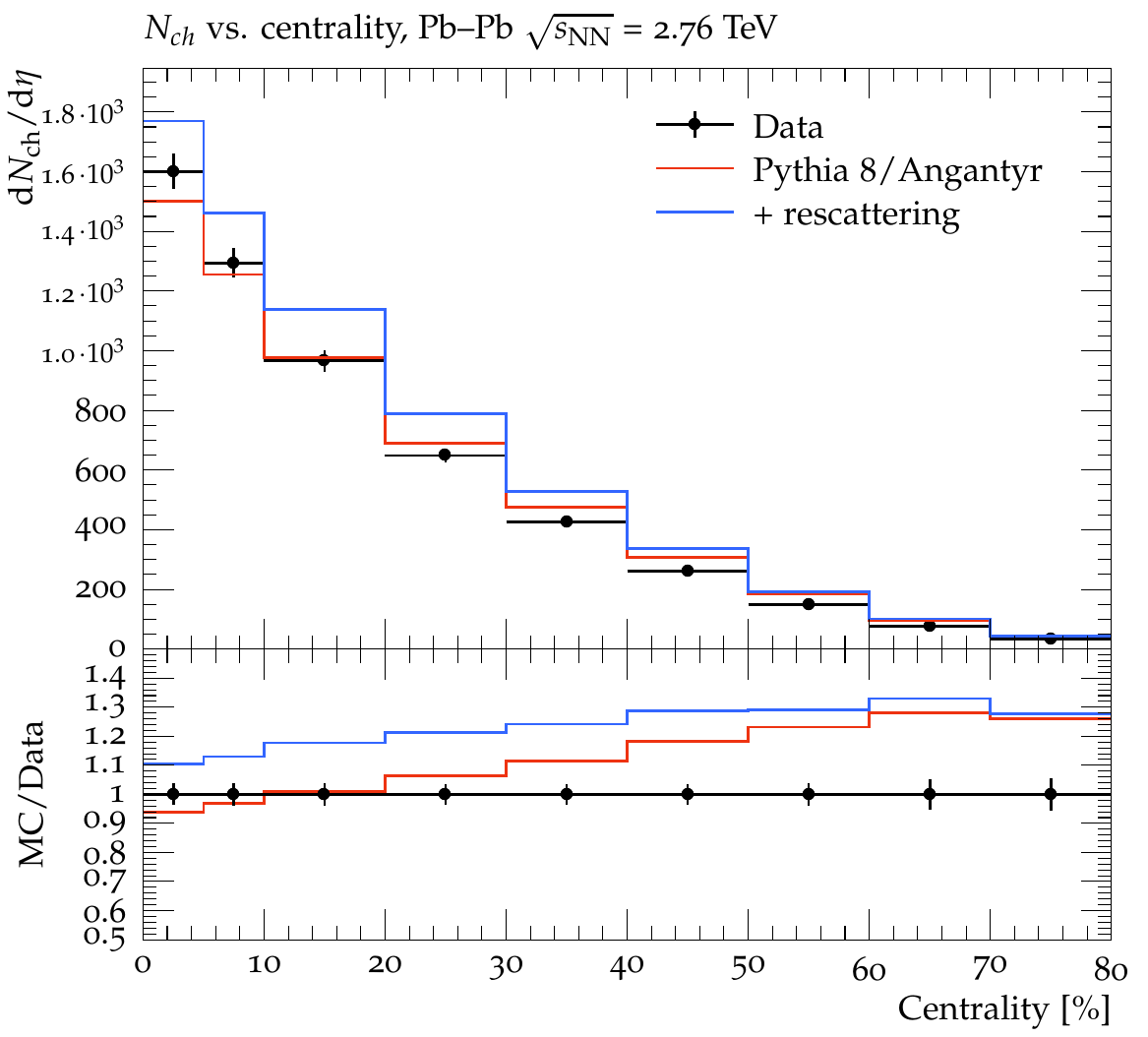}
	(a)
\end{minipage}
\begin{minipage}[c]{0.49\linewidth}
\centering
\includegraphics[width=\linewidth]{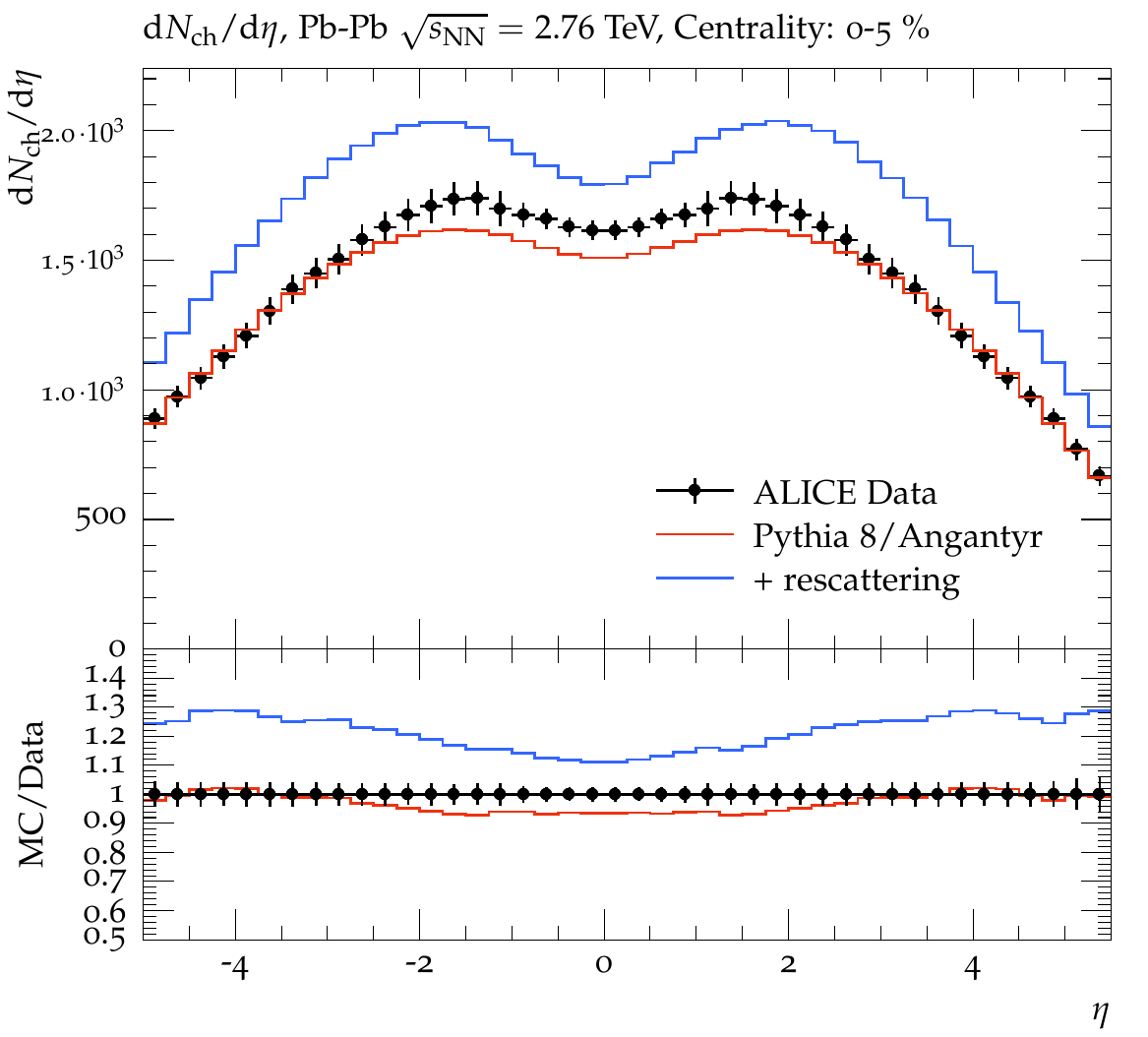}
	(b)
\end{minipage}
\caption{\label{fig:exp-mult} Charged multiplicities in $\PbPb$ collisions $\sqrt{s_{\N\N}} = 2.76$ TeV.
At mid-rapidity as (a) function of centrality, and (b) differential in $\eta$ in centrality 0-5\%.
Data from ALICE \cite{Aamodt:2010cz,Abbas:2013bpa}.}
\end{figure}

In section \ref{sec:model-multiplicities} we described how the current lack of $3 \to 2$
processes in the rescattering framework increases the total multiplicities. In \figref{fig:exp-mult}
\textsc{Angantyr} with and without rescattering is compared to experimental data \cite{Aamodt:2010cz,Abbas:2013bpa}.

In \figref{fig:exp-mult}a, $\mathrm{dN}_{ch}/\d\eta|_{\eta = 0}$ is shown as
function of centrality. It is clear that the shift in multiplicity, caused by rescattering,
is centrality dependent, with a larger effect seen in more central events. For centrality 0-5\%,
the agreement with data shifts from approximately 8\% below data to 10\% above. It is
instructive to show the differential distributions as well, and in \figref{fig:exp-mult}b,
the $\eta$-distribution out to $\pm 5$ is shown. It is seen that the shift is slightly larger
at the edges of the plateau. This effect is most pronounced in the centrality bin shown here,
and decreases for more peripheral events.

\begin{figure}
\begin{minipage}[c]{0.49\linewidth}
\centering
\includegraphics[width=\linewidth]{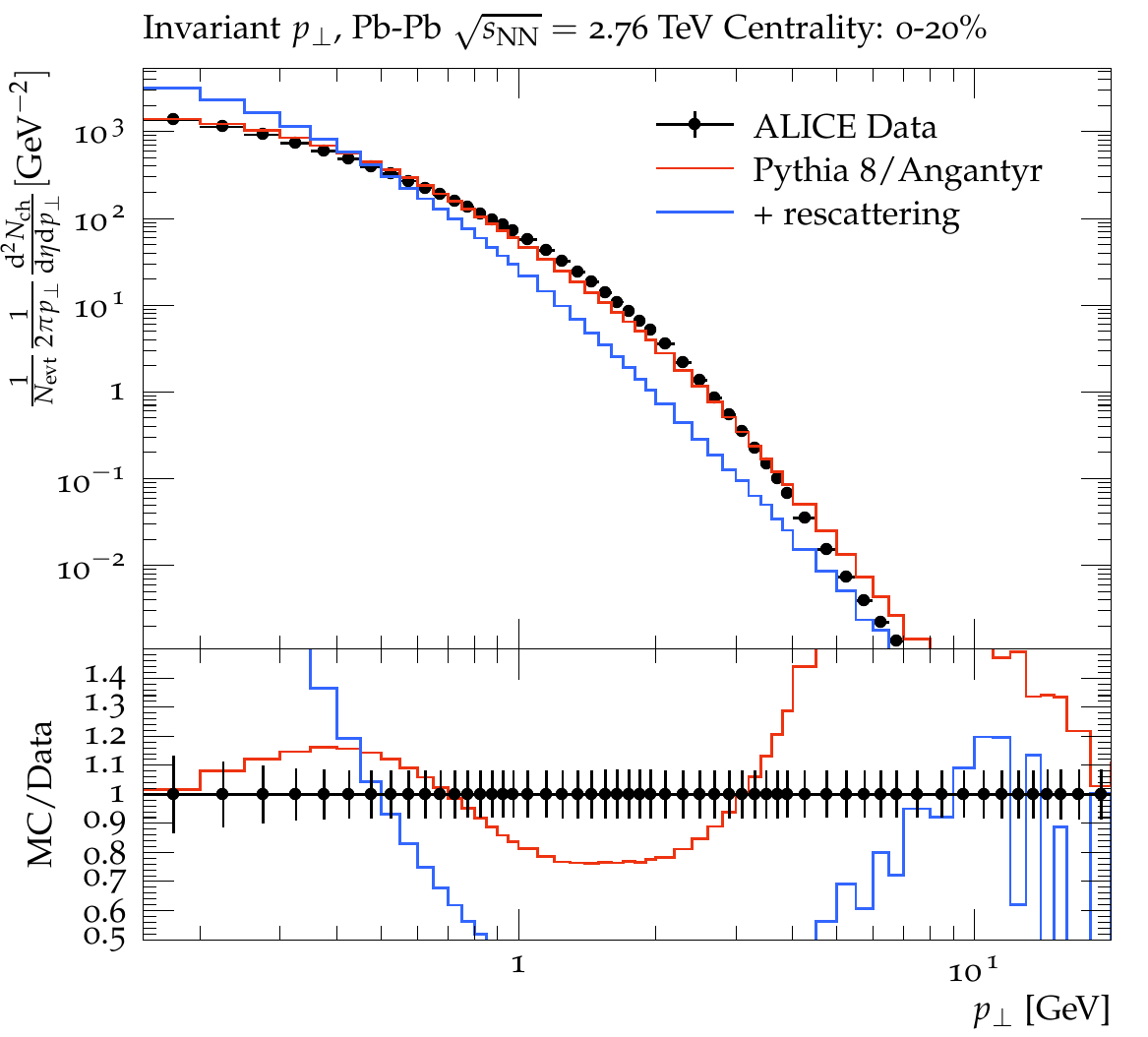}
	(a)
\end{minipage}
\begin{minipage}[c]{0.49\linewidth}
\centering
\includegraphics[width=\linewidth]{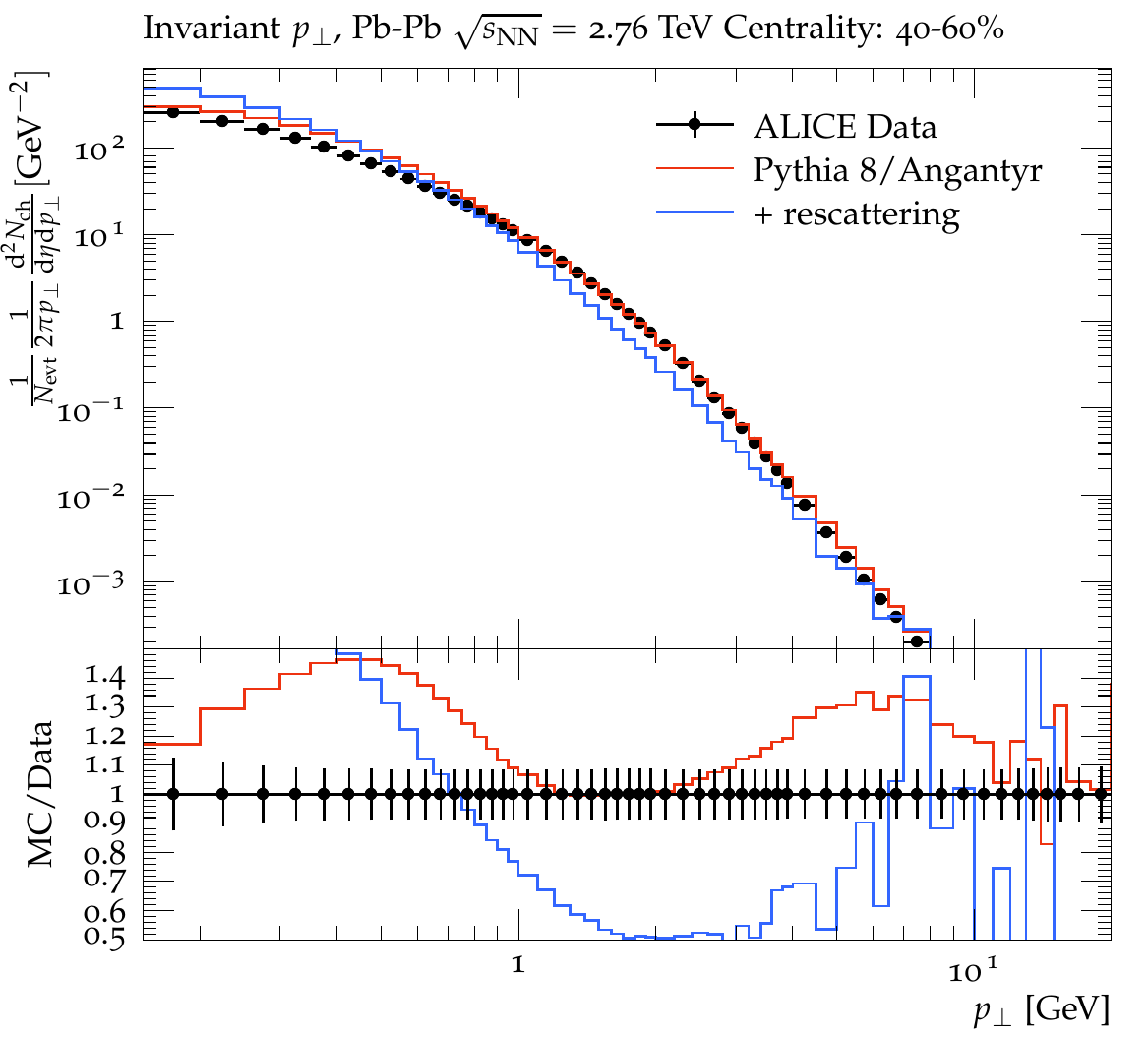}
	(b)
\end{minipage}
	\caption{\label{fig:exp-pt} Invariant $\pT$ spectra of charged particles
	in $\PbPb$ collisions $\sqrt{s_{\N\N}} = 2.76$ TeV, in $|\eta| < 0.8$. Shown
	for two different centrality intervals (a) 0-20\% and (b) 40-60\%.
	Data from ALICE \cite{Abelev:2012hxa}.}
\end{figure}

To further explore the change in charged multiplicity distributions, we show
comparisons to invariant $\pT$ distributions in the same collision system,
measured down to $\pT = 0.15$ GeV in $|\eta| < 0.8$ \cite{Abelev:2012hxa}
in \figref{fig:exp-pt}, with 0-20\% centrality shown in \figref{fig:exp-pt}a,
and 40-60\% in \figref{fig:exp-pt}b. It is seen that particles at intermediate
$\pT \approx 1-6$ GeV are pushed down to very low $\pT$ (pion wind)
in rescatterings which, due
to the lack of $3 \to 2$ processes, will generate more final state particles overall.

From this investigation of effects on basic single-particle observables from adding rescattering, it is clear that agreement with data
is decreased. Since hadronic rescattering in heavy ion collisions is physics effects which must be taken into account, this clearly points
to the need of further model improvement. Beyond re-tuning and adding $3 \to 2$ processes, the addition of string-string interactions
before hadronic rescattering will change the overall soft kinematics. This is an important next step, which will be taken in a forthcoming publication.

\subsection{Flow coefficients}
\label{sec:FlowCoefficients}

As indicated in section \ref{sec:cent-dep-obs}, rescattering has a non-trivial effect on flow observables, 
a staple measurement in heavy ion experiments. Anisotropic flow is generally understood as a clear indication
of QGP formation, as it is well described by hydrodynamic response to the anisotropy of the initial geometry \cite{Teaney:2010vd}.

The main difference between most previous investigations and this paper, of the effect of rescattering on flow, is
the early onset of the hadronic phase. Recall that with a hadronization time of $\langle \tau^2 \rangle \approx 2$ fm$^2$,
the initial hadronic state from string hadronization is much denser.

In this section we will compare to experimental data from $\XeXe$ and $\PbPb$ collisions obtained by the ALICE experiment 
\cite{Acharya:2019vdf}. When doing so, it is important to use the same definitions of flow coefficients as used by the experiment.
Since the event plane is not measurable by experiment, equations (\ref{eq:flow-def}) and (\ref{eq:vn-th}) cannot be applied directly.
Instead the flow coefficients are calculated using two- and multi-particle azimuthal correlations using the so-called generic 
framework \cite{Bilandzic:2013kga}, implemented in the Rivet framework \cite{Bierlich:2020wms}, 
including the use of sub-events \cite{Huo:2017nms}.

\begin{figure}
\begin{minipage}[c]{0.49\linewidth}
\centering
\includegraphics[width=\linewidth]{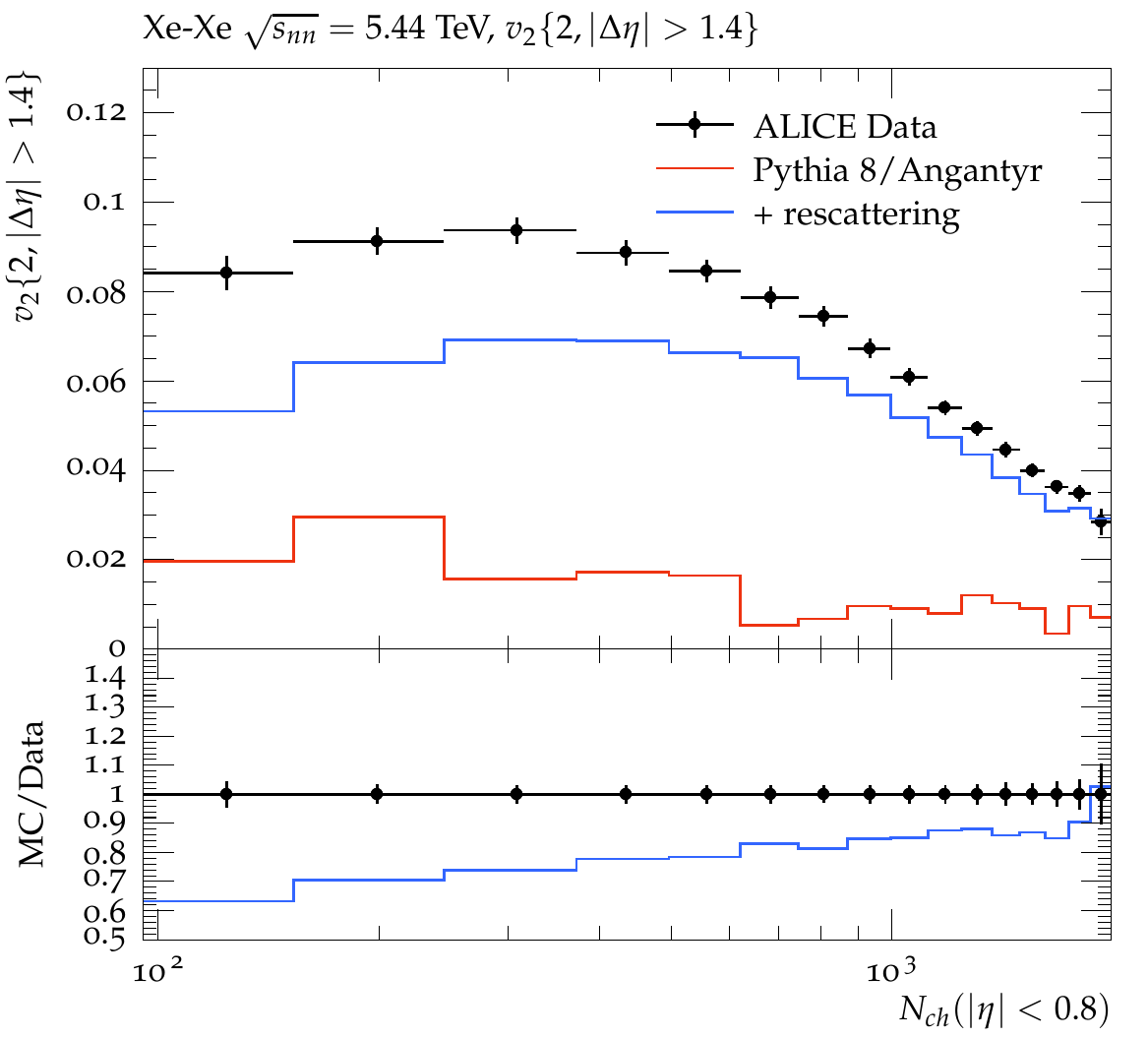}
	(a)
\end{minipage}
\begin{minipage}[c]{0.49\linewidth}
\centering
\includegraphics[width=\linewidth]{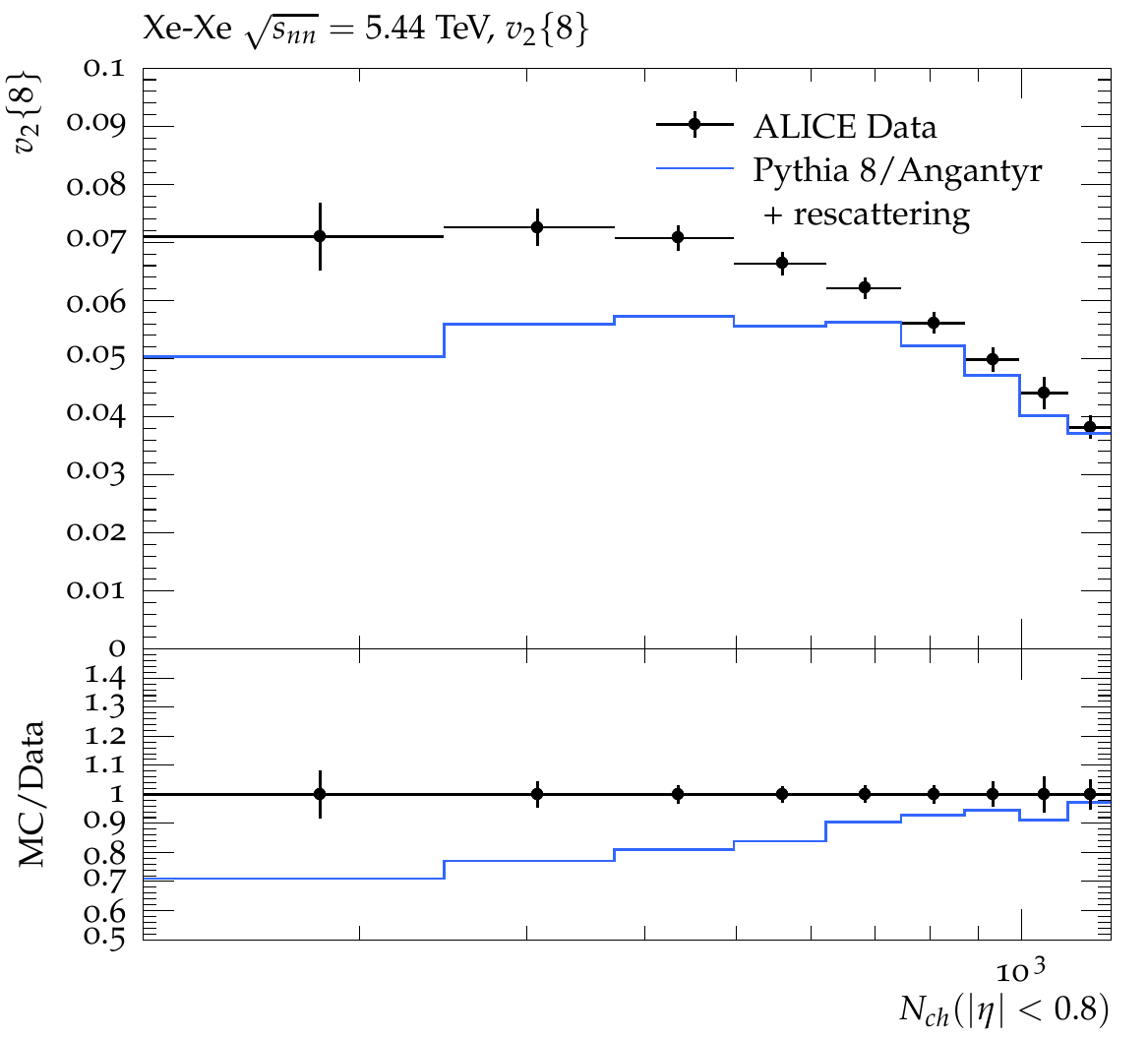}
	(b)
\end{minipage}
	\caption{\label{fig:flow-xexe}Elliptic flow in $\XeXe$ collisions at $\sqrt{s_{\N\N}} = 5.44$ TeV. (a) $v_2\{2\}$
	with $|\Delta \eta| > 1.4$ and (b) $v_2\{8\}$. The $v_2$ calculated with 4- and 6-particle correlations show a similar
	trend, but are not shown in the figure. Data from ALICE \cite{Acharya:2019vdf}.}
\end{figure}

In figure \ref{fig:flow-xexe} we show elliptic flow $v_2$ in $\XeXe$ collisions at $\sqrt{s_{\N\N}} = 5.44$ TeV calculated
with (a) two-particle correlations and $|\Delta \eta| > 1.4$, as well as (b) $v_2\{8\}$. In the former case we compare also to 
the no-rescattering option, which gives a measure of contributions from 
non-flow mechanisms such as (mini)jets and particle decays.
In both cases the data is reproduced with good (within 10\%) accuracy for very high multiplicities, but the
calculation is up to 30-40\% below data for more peripheral events. It is particularly interesting to note that even in
the case of using an 8-particle correlator, the calculation shows the same agreement as only two particles with a gap
in $\eta$ between them. This rules out the possibility that additional flow enters purely from a local increase in two-particle
correlations. This should also already be clear from the treatment in section \ref{sec:cent-dep-obs}, where it was clearly shown
that the added $v_2$ by rescattering is in the correct direction with respect to the theoretical event plane.

\begin{figure}
\begin{minipage}[c]{0.49\linewidth}
\centering
\includegraphics[width=\linewidth]{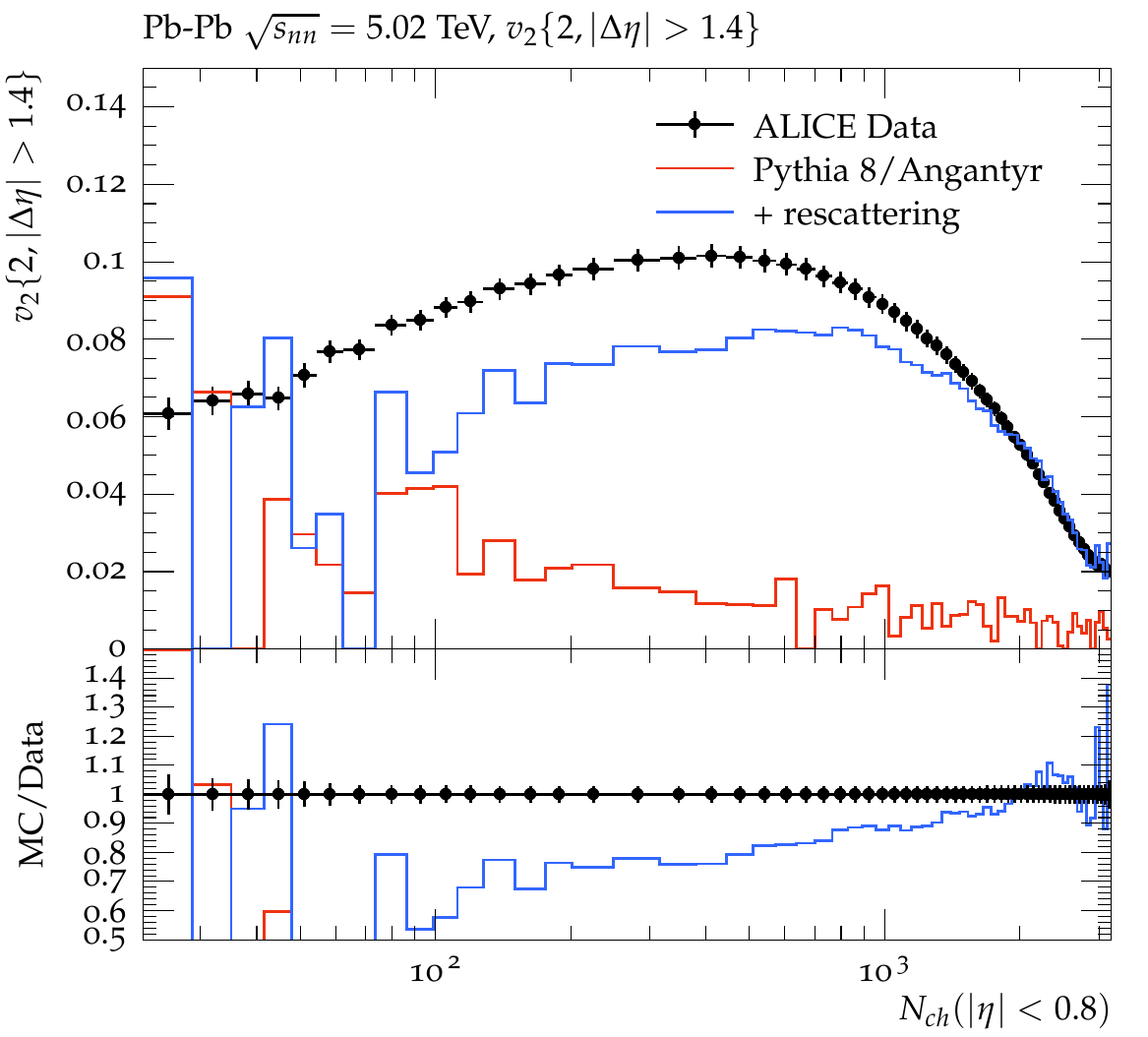}
	(a)
\end{minipage}
\begin{minipage}[c]{0.49\linewidth}
\centering
\includegraphics[width=\linewidth]{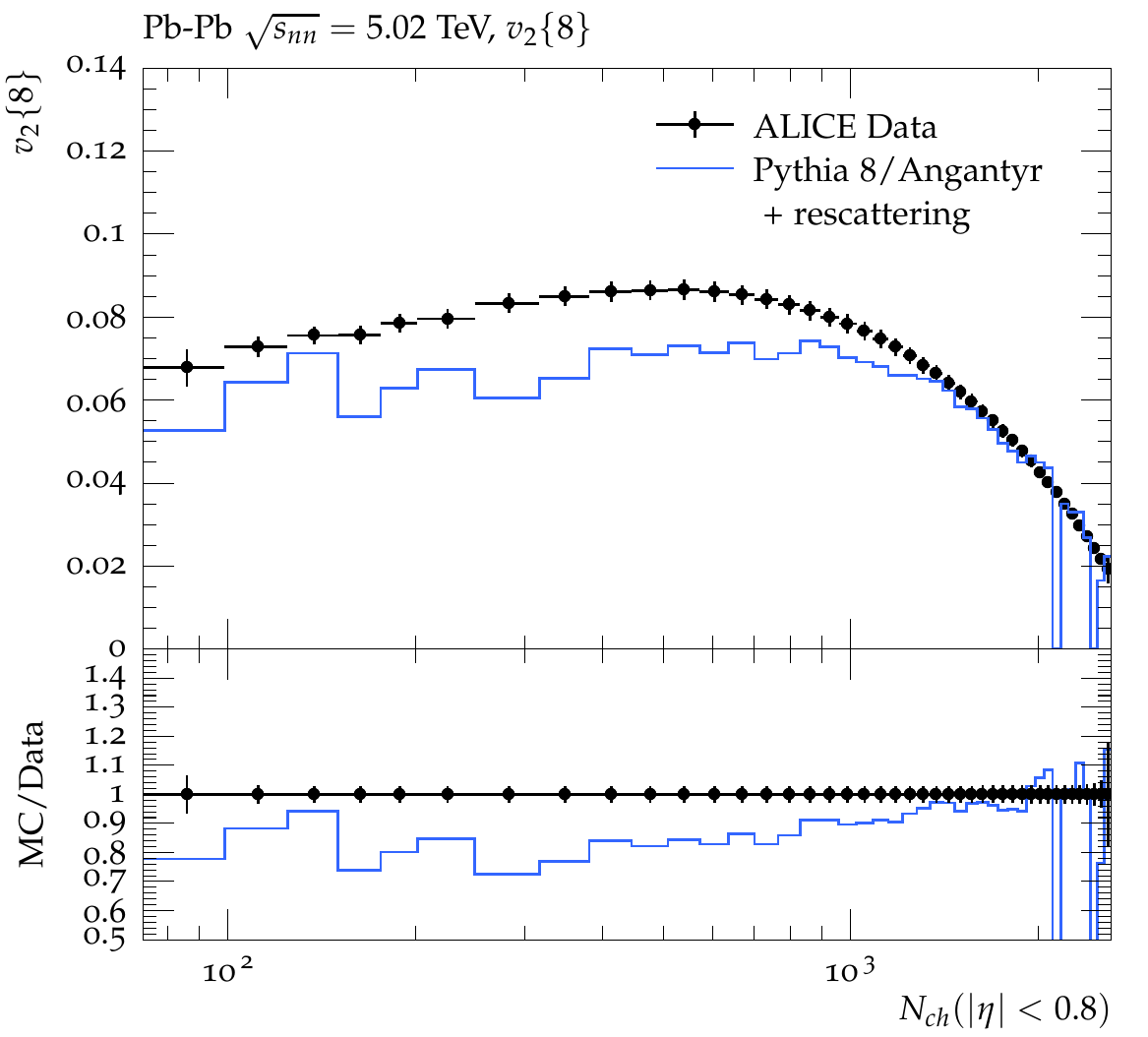}
	(b)
\end{minipage}
	\caption{\label{fig:flow-pbpb}Elliptic flow in $\PbPb$ collisions at $\sqrt{s_{\N\N}} = 5.02$ TeV. (a) $v_2\{2\}$
	with $|\Delta \eta| > 1.4$ and (b) $v_2\{8\}$. The $v_2$ calculated with 4- and 6-particle correlations show a similar
	trend, but are not shown in the figure. Data from ALICE \cite{Acharya:2019vdf}.}
\end{figure}

In figure \ref{fig:flow-pbpb} we show the same observables, $v_2\{2,~|\Delta \eta| > 1.4\}$ and $v_2\{8\}$ for $\PbPb$ collisions
at $\sqrt{s_{\N\N}} = 5.02$ TeV. While the same overall picture is repeated, it is worth noticing that the agreement at high multiplicities
is slightly better. As it was also observed in section \ref{sec:cent-dep-obs}, the effect of rescattering is in general larger in $\PbPb$ than in 
$\XeXe$, due to the larger multiplicity of primaries.

We want here to emphasize that, while hadronic rescattering can obviously not describe data for elliptic flow completely, the results here
suggest that hadronic rescattering with early hadronization has a larger effect than previously thought. This is particularly interesting seen in
the connection with recent results, that interactions between strings before hadronization in the string shoving model \cite{Bierlich:2020naj} will
also give a sizeable contribution to flow coefficients in heavy ion collisions, without fully describing data. The combination of the two frameworks, 
to test whether the combined effect is compatible with data, will be a topic for a future paper. It should be mentioned that the contributions 
from different models, acting one after the other, does not add linearly \cite{Auvinen:2013sba,daSilva:2020cyn}.

\subsection{Jet modifications from rescattering}

As shown, both in \figref{fig:pTspectra} and \figref{fig:exp-pt}, hadronic rescattering has a significant effect
on high-$\pT$ particle production. Studies of how the behaviour of hard particles changes
from $\p\p$ to $\A\A$ collisions are usually aiming at characterising the interactions between
initiator partons and the QGP. The observed phenomena are referred to as ``jet quenching'', and 
phenomenological studies usually ignore the presence of a hadronic phase. For a notable exception see 
ref. \cite{Dorau:2019ozd} for a recent exploratory study using SMASH, as well as references therein.

In this final results section, we do not wish to go into a full study on the effect of
rescattering on jet observables, but rather point to an interesting result which will be pursued further in
a future study, as well as warn potential users of the \textsc{Pythia} rescattering implementation
of a few pitfalls.

\begin{figure}
\begin{minipage}[c]{0.49\linewidth}
\centering
\includegraphics[width=\linewidth]{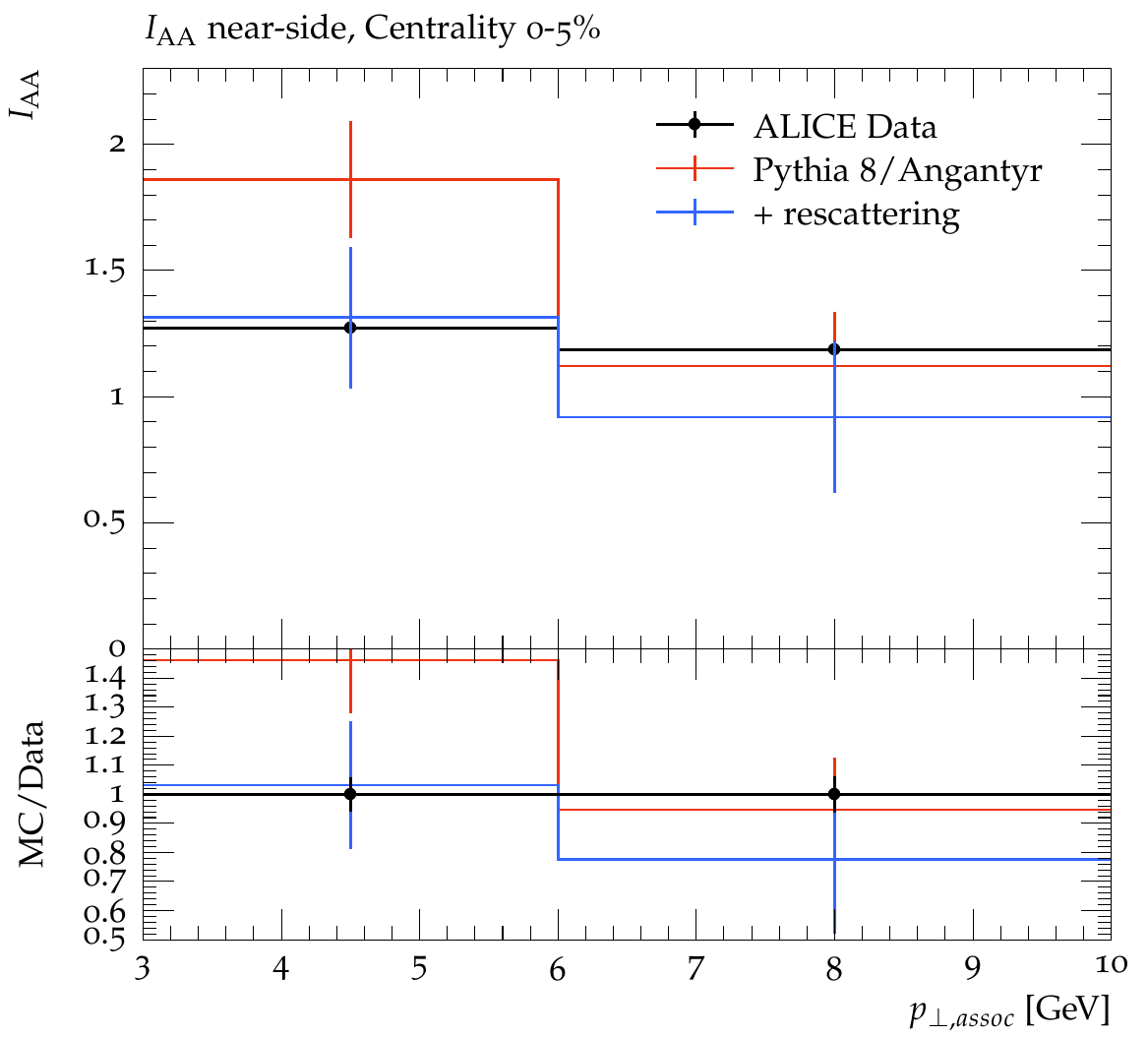}
	(a)
\end{minipage}
\begin{minipage}[c]{0.49\linewidth}
\centering
\includegraphics[width=\linewidth]{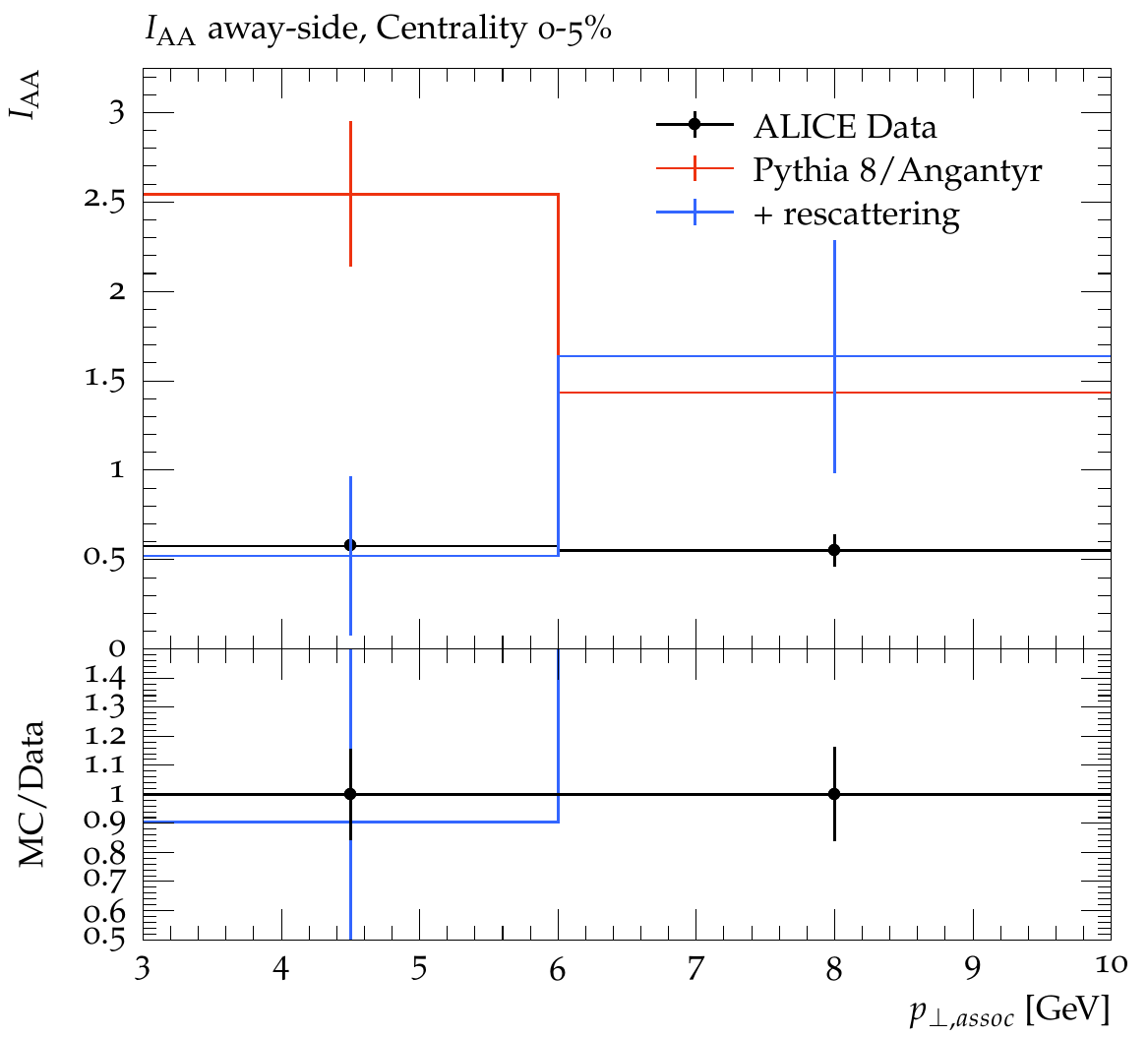}
	(b)
\end{minipage}
	\caption{\label{fig:iaa} Modification of high-$\pT$ azimuthal correlations, $I_{\A\A}$, in $\PbPb$ collisions at 
	$\sqrt{s_{\N\N}} = 2.76$ TeV on (a) the near side ($\Delta \varphi$ of $\pm0.7$) and (b) the away side ($\Delta \varphi$ of 
	$\pi\pm 0.7$), both in the $0-5$\% centrality bin. Error bars are statistical errors. Data from ALICE \cite{Aamodt:2011vg}.} 
\end{figure}

One of the early key observations of jet quenching effects was the disappearance of back-to-back high-$\pT$
hadron correlations in central $\mathrm{AuAu}$ collisions at RHIC \cite{Adler:2002tq}. Similar studies have since
also been performed at the ALICE experiment, and we compare here to data from a study of azimuthal modifications
in $\PbPb$ collisions at $\sqrt{s_\mathrm{NN}}=2.76$ TeV \cite{Aamodt:2011vg}. In this study, trigger particles
of $8$ GeV $< p_{\perp,\mathrm{trig}} < 15$ GeV are correlated in $\varphi$ with associated particles 
of $4$ GeV $< p_{\perp,\mathrm{assoc}} < p_{\perp,\mathrm{trig}}$.
The $\PbPb/\mathrm{pp}$ ratio of per-trigger yields is denoted $I_{\A\A}$, and it was noted in the study by ALICE
that the $\PbPb$ per-trigger yield is suppressed to about 60\% of $\p\p$ on the away side ($\Delta \varphi$ of $\pi\pm0.7$) and
enhanced by about 20\% on the near side ($\Delta \varphi$ of $\pm0.7$).
In \figref{fig:iaa}, $I_{\A\A}$ in $0-5$\% centrality for $\PbPb$ collisions is shown on (a) the near-side
and (b) the away side, compared to ALICE data \cite{Aamodt:2011vg}.

It is seen that, by default, \textsc{Pythia/Angantyr} overestimates the away-side $I_{\A\A}$ in the whole $p_{\perp,\mathrm{assoc}}$ range, 
while the near-side is overestimated at low $\pT < 6$ GeV. Adding rescattering brings the simulation on
par with data in all cases but the high-$\pT$ part of the away-side $I_{\A\A}$. No significant effect from rescattering was observed 
in peripheral events.

At first sight, this seems like a very significant result, but we wish to provide the reader with a word of caution. We remind 
that the current lack of $n \to 2$ processes and retuning causes a drastic shift in $\pT$ spectra as previously shown, 
incompatible with data. The depletion seen from rescattering is exactly in the region where $I_{\A\A}$
is now well reproduced. It can therefore very well be that the effect seen is mainly a token of current shortcomings.
This is of course not a statement that hadronic rescattering has no impact on jet-quenching observables, but it goes to show 
that a potential user cannot run \textsc{Pythia} to explain this or similar observables without a deeper analysis.

Finally a technical remark. Running \textsc{Pythia/Angantyr} with rescattering to reproduce an observable requiring
a high-$\pT$ trigger particle will require very long run times. The figures in this section are generated by first requiring 
that a parton--parton interaction with $\hat{p}_{\perp} > 5$ GeV takes place at all, and secondly a veto is put in place ensuring
that the time-consuming rescattering process is not performed if there are no trigger particles with the required $\pT$ present
in the considered acceptance.


\section{Summary and outlook}
\label{sec:summary}

The \textsc{Pythia} rescattering framework was first introduced in
\cite{Sjostrand:2020gyg}, which focused on validating it in the context
of $\p\p$ collisions. In this paper, the main focus has been physics
studies in $\p\Pb$ and $\PbPb$ collisions. 

Before going into finer details, it is worth to consider how rescattering
changes the bulk properties of events. Notably it increases the charged
multiplicity by about 20~\% in $\PbPb$ events. One key reason is that
we have implemented $2 \to n, n > 2$ processes but not $n \to 2$.
For a system in thermal equilibrium the two kinds should balance each
other. In the modelling of $\PbPb$ collisions, however, our original
production is not a thermal process, and the subsequent rescattering is
occurring in an expanding out-of-equilibrium system. As an example, minijet
production gives a larger rate of higher-$\pT$ particles than a thermal
spectrum would predict, and therefore a fraction of higher invariant
rescattering masses with more $2 \to n$ processes. While the inclusion
of $n \to 2$ therefore is high on the priority list of future model
developments, it is not likely to fully restore no-rescattering
multiplicities. For $\p\p$, a slight retuning of the $\pTo$ parameter
for multiparton interactions helped restore approximate agreement with
data, but for $\p\Pb$ and $\PbPb$ also some tweaks may be needed in
the \textsc{Angantyr} framework. Until that is done, users need to be
aware of such shortcomings.

The most obvious effect of rescattering is the changed shape of $\pT$
spectra, where pions lose momentum. Owing to their low mass they tend to
be produced with higher velocity, which in collisions then can be used
to speed up heavier particles. The above-mentioned $2 \to n$ processes
act to reduce the overall $\langle \pT \rangle$ values, however, and for
kaons and $\D$ mesons the result is a net slowdown.  Here it should be
remembered that the $\D$'s are produced only in perturbative processes,
and so can have large velocities to begin with. For protons indeed a
speedup can be observed, but here the significant rate of
baryon--antibaryon annihilation clouds the picture. 

Still at the basic level, the detailed event record allows us to
map out both the space--time evolution, the nature of rescatterings
and the change of particle composition. An example is resonance
suppression, which was discussed briefly, but where an analysis
paralleling the experimental one is outside the scope of the current
article. Also of interest is the converse, the formation of particular
particles in rescattering, such as $f_2(1270)$ resonances
\cite{Lebiedowicz:2020qnz} or exotic hadrons. Care must be taken in such
studies however, as particles that form resonances already are
correlated, and hence the appearance of a particle in the event record
does not necessarily translate directly to an observable signal.
We also see other future applications of space--time information,
notably for Bose--Einstein studies.
 
Amongst the physics results presented, the most remarkable is
the observation of a sizeable elliptic flow in $\A\A$ collisions,
where data is described particularly well at high multiplicities.
Flow is also visible in $\D$ meson production, at a slightly lower rate
than in the inclusive sample. The flow increases from $\XeXe$ to
$\PbPb$, \ie when moving to larger systems. This should be contrasted
with the $\p\p$ results \cite{Sjostrand:2020gyg}, where the rescattering
flow effects were tiny and far below data. Thus rescattering may be one
source of flow, but apparently not the only one. The
\textsc{Pythia/Angantyr} framework also includes other effects that
contribute to the flow, notably shoving \cite{Bierlich:2016vgw}, where
an improved modelling \cite{Bierlich:2020naj} will soon be part of the
standard code.

Another interesting observation is the suppression of $\J/\psi$
production in central collisions, by the breakup into $\D$ mesons in
rescattering. The $\D$ meson rate is hardly affected, since it is
more than two orders of magnitude larger to begin with. One should
note that the handling of charm collisions largely is based on the
Additive Quark Model, which does not distinguish between $\J/\psi$ and
$\psi'$, so again there is room for improvement.

There are also examples where rescattering, as currently implemented,
is going in the wrong direction. The $\pT$ spectrum in $\PbPb$ is
reasonably well described without rescattering, but becomes way too soft
with it. Hyperon production rates also drop, where data wants more such
production \cite{ALICE:2017jyt}. 

The most important follow-up project in a not too distant future is
to combine all the features that have been introduced on top of the basic
\textsc{Pythia/Angantyr} model, notably ropes, shoving and rescattering,
and attempt an overall tuning. It is not possible to tell where results
will land at the end, since effects tend to add nonlinearly. One can
remain optimistic that many features of the data will be described
qualitatively, if not quantitatively.

Another possible application is (anti)deuteron production, and even
heavier (anti)nuclei.
In the past this has often been modelled using coalescence of particles
close in momentum space, on the assumption that such particles also have
been produced near to each other in space--time. (One such model is even
included in \textsc{Pythia} \cite{Dal:2015sha}.) This usually is not a
bad approximation in $\e^+\e^-$ annihilation or $\p\p$, at least as
modelled by string fragmentation. But the much larger volume of
particle production and rescattering in $\A\A$ obviously requires
due consideration to the space--time proximity of (anti)nucleons,
and also that deuterons can break up by rescattering processes.

The rescattering model is made freely available, starting with
\textsc{Pythia}~8.303, with a few tiny corrections in 8.304 to allow the
extension from $\p\p$ to $\p\A$ and $\A\A$. In the past we have seen
how new \textsc{Pythia} capabilities have led to follow-up studies
by the particle physics community at large, both foreseen and unforeseen
ones, and we hope that this will be the case here as well, although
admittedly the long run times is a hurdle.


\section*{Acknowledgements}

Work supported in part by the Swedish Research Council, contract numbers
2016-05996 and 2017-003, in part by the MCnetITN3 H2020 Marie Curie Innovative 
Training Network, grant agreement 722104, and in part by the Knut and Alice Wallenberg
foundation, contract number 2017.0036.
This project has also received funding from the European Research
Council (ERC) under the European Union's Horizon 2020 research
and innovation programme, grant agreement No 668679.

\appendix
\section*{Appendix - Algorithmic complexity}
The rescattering algorithm needs to compare each hadron pair. This has an
asymptotic complexity of $\mathcal{O}(n_{\mathrm{record}}^2)$, where
$n_{\mathrm{record}}$ is the total number of particles in the event record,
including those that are not final-state particles. In practice this
asymptotic bound is never reached, since a large number of the comparisons
are trivial, \eg if one of the compared particles has already decayed or
rescattered. Instead, profiling shows that the bottlenecks are calculating the
total cross sections and rescattering vertices for pairs that can potentially
rescatter. These are calculated for a much smaller number of pairs, and give a
complexity that is less than quadratic in practice.

The average generation time is shown in \tabref{tab:runtime}. We see that
it is in the order of milliseconds for $\p\p$ and $\p\Pb$, both with and
without rescattering. The rescattering accounts for about 45~\% of the
total runtime for $\p\p$, and about 75~\% for $\p\Pb$. The situation is
radically different for $\PbPb$, where the rescattering takes more than
99.5~\% of the time, making the average generation time go from less than
a second to more than two minutes per event. Thus more careful planning
is needed for $\PbPb$ rescattering studies, since a rerun will cost.

\begin{figure}[t!]
\begin{minipage}[c]{0.49\linewidth}
\centering
\includegraphics[width=\linewidth]{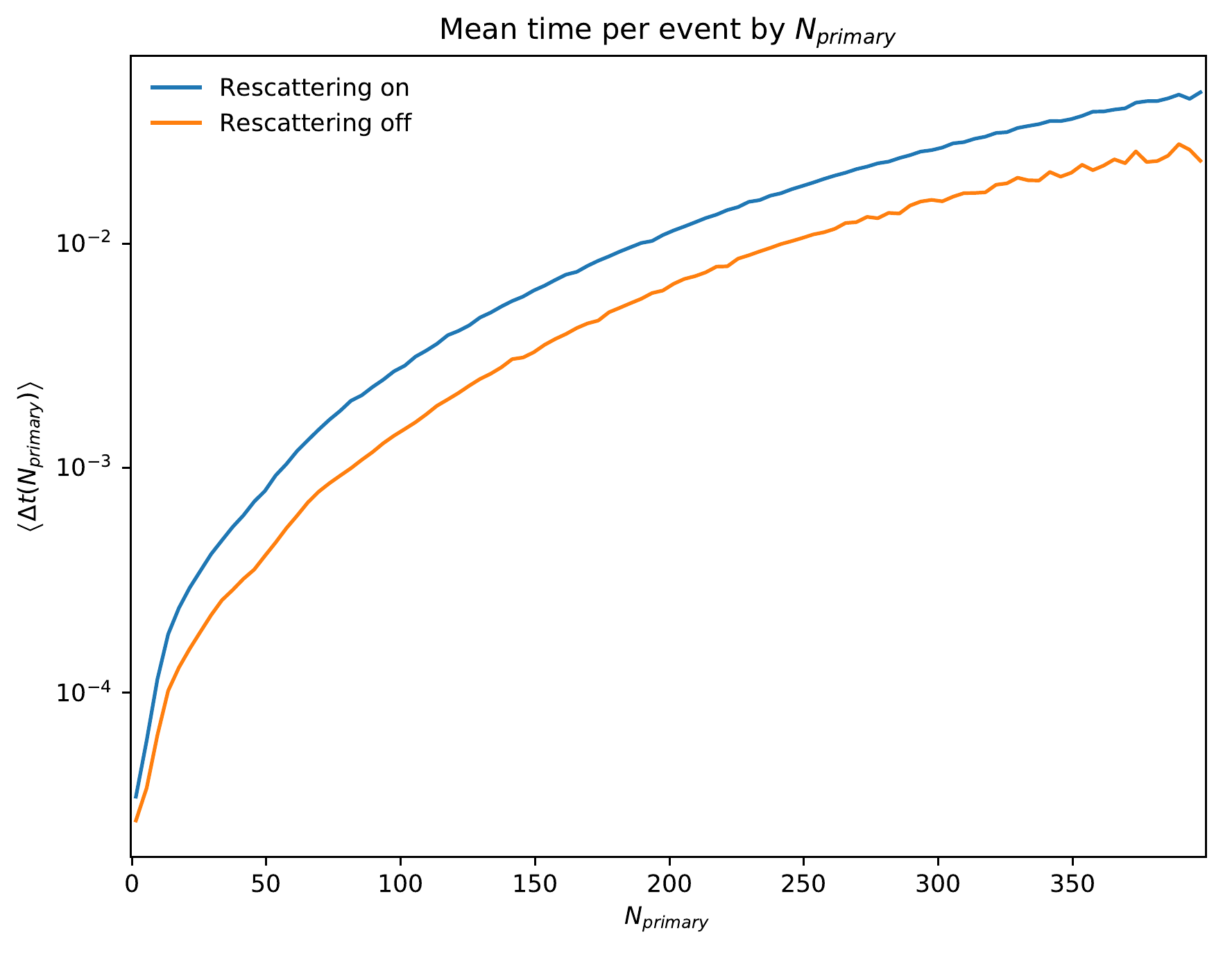} \\
(a)
\end{minipage}
\begin{minipage}[c]{0.49\linewidth}
\centering
\includegraphics[width=\linewidth]{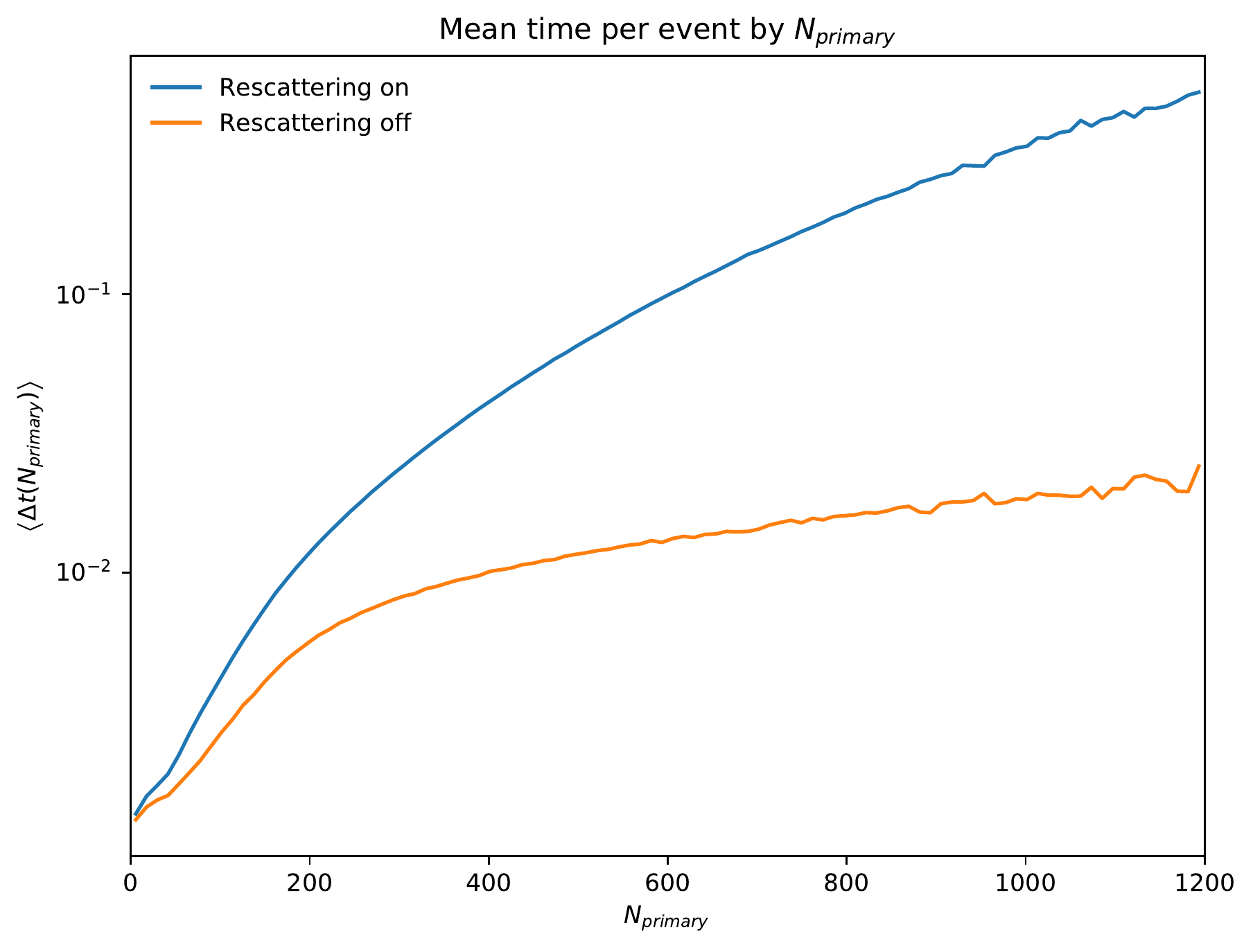} \\
(b)
\end{minipage}\\
\begin{minipage}[c]{0.49\linewidth}
\centering
\includegraphics[width=\linewidth]{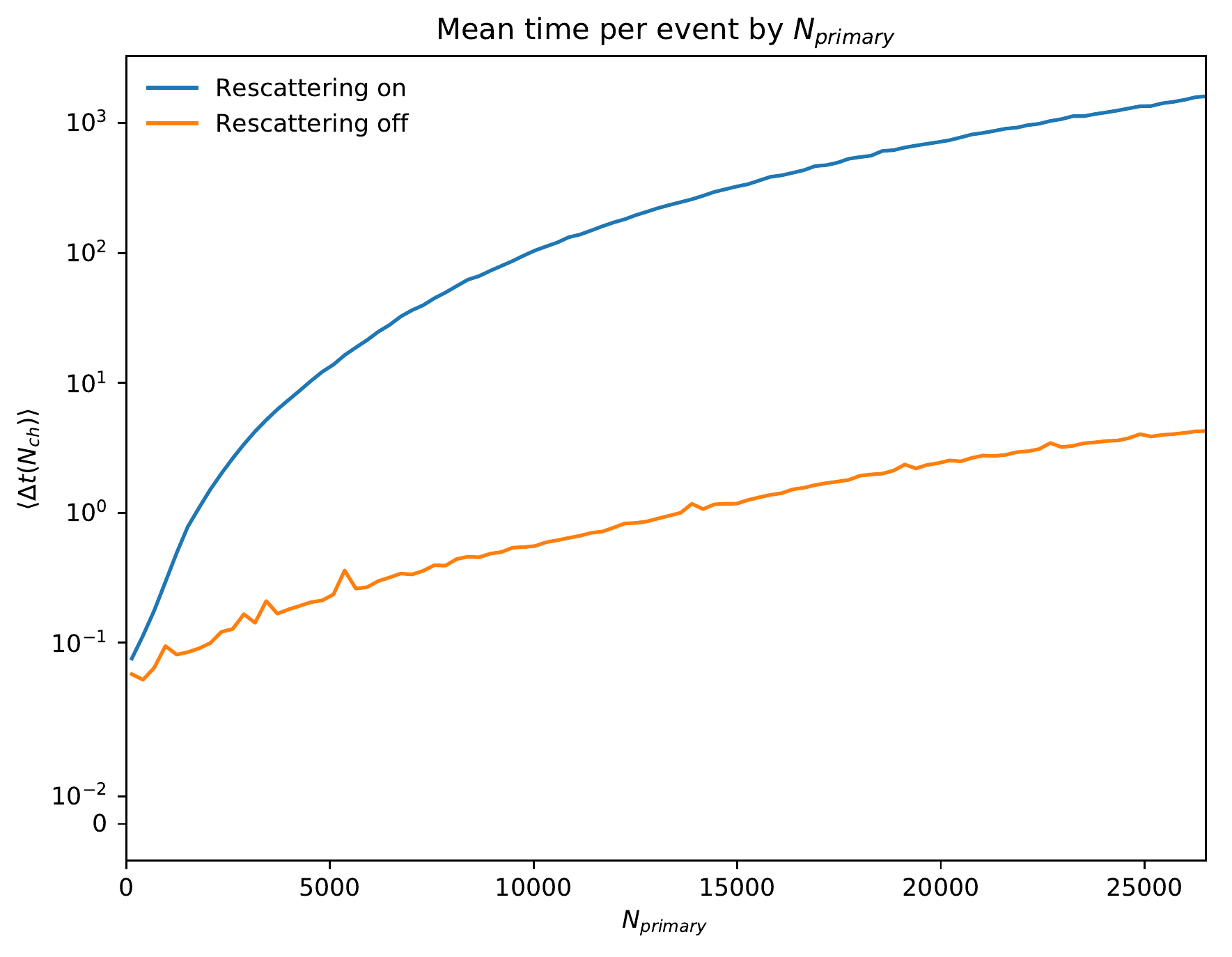} \\
(c)
\end{minipage}
\caption{
The average generation time of each event with a specified primary hadron 
multiplicity, for (a) $\p\p$, (b) $\p\Pb$ and (c) $\Pb\Pb$, with
$\sqrt{s_{\N\N}} = 5.02$ TeV.
}
\label{fig:runtime}
\end{figure}

In \figref{fig:runtime} the average generation time per event is shown
as a function of the number of primary hadrons. We do not tune the $\pTo$
parameter for this study, so that the primary hadron distribution is the
same with and without rescattering. The runtime as a function of the
primary multiplicity is essentially unchanged by this, however. In all
three processes the rescattering overhead is modest for small multiplicities.
At the tail towards larger multiplicities the slowdown is about a factor
$\sim 2$ for $\p\p$, $\sim 20$ for $\p\Pb$ and $\sim 1000$ for $\PbPb$.

In $\PbPb$ studies focused on peripheral or mid-centrality events, 
unnecessarily generating high-multiplicity events can incur a significant
slowdown. This can be mitigated by writing an impact-parameter generator
tailored to the specific needs, and passing it to \textsc{Pythia} via a
user hook. 

\begin{table}[t] 
\centering
\begin{tabular}{c|c c c}
    Case & Resc. off & Resc. on & Ratio \\
    \hline
    $\p\p$   & 2.24 ms & 4.02 ms & 1.79 \\
    $\p\Pb$  & 6.40 ms & 25.6 ms & 4.00 \\
    $\PbPb$ & 0.594 s & 150.4 s  & 253
\end{tabular}
\caption{The average generation time per event. Events were generated on 
an Intel(R) Core(TM) i7-6700K CPU at 4.00GHz.}
\label{tab:runtime}
\end{table}

\bibliographystyle{utphys}
\bibliography{bibliography}

\end{document}